\documentstyle[12pt]{article}
\newcommand{\beq}{\begin{equation}}
\newcommand{\beqr}{\begin{eqnarray}}
\newcommand{\eeqr}{\end{eqnarray}}
\newcommand{\deek}{\frac{d^4k}{(2\pi )^4}}
\newcommand{\cut}{\int_{|k| < \Lambda}}
\newcommand{\cuts}{\int_{|k|' < \Lambda}}
\newcommand{\Sid}{\mbox{$\Psi^{\dagger}$}}

\newcommand{\eeq}{\end{equation}}

\newcommand{\intom}{\int_{-\infty}^{\infty}\frac{d\omega}{2\pi}}

\newcommand{\intomi}{\int_{-\infty}^{\infty}\frac{d\omega_{i}}{2\pi}}
\newcommand{\intompi}{\int_{-\infty}^{\infty}\frac{d\omega_{i}'}{2\pi}}
\newcommand{\val}[2]{\mbox{$\vec{#1}_{#2}$}}
\newcommand{\sid}{\mbox{$\psi^{\dagger}$}}
\newcommand{\intK}{\int_{- \pi}^{\pi} \frac{dK}{2\pi}}
\newcommand{\intk}{\int_{- \Lambda}^{\Lambda} \frac{dk}{2\pi}}

\newcommand{\intki}{\int_{- \Lambda}^{\Lambda} \frac{dk_{i}}{2\pi}}
\newcommand{\intkpi}{\int_{- \Lambda}^{\Lambda} \frac{dk_{i}'}{2\pi}}
\newcommand{\sib}{\mbox{$\overline{\psi}$}}

\newcommand{\lk}{\Lambda /K_F \rightarrow 0}
\newcommand{\lf}{F (\val{\Omega}{}  \cdot  \val{\Omega'}{} )}

\newcommand{\iT}{\int_{0}^{2\pi}\frac{d\theta}{2\pi}}
\newcommand{\iTi}{\int_{0}^{2\pi}\frac{d\theta_{i}}{2\pi}}

\begin{document}
\setlength{\baselineskip}{0.375in}
\begin{center}
\Large\bf
RENORMALIZATION GROUP APPROACH TO INTERACTING
FERMIONS \\
\normalsize
\vspace{.5in}
R.Shankar \\
Sloan Laboratory of Physics \\
Yale University\\
New Haven CT 06520\\
REVISED AND EXPANDED JUNE 1993 FOR REV.MOD.PHYS.\\
\vspace{.5in}
\Large\bf
Abstract
\end{center}
\normalsize
\setlength{\baselineskip}{.3in}

The stability or lack thereof of nonrelativistic  fermionic systems to
interactions is studied within the Renormalization Group  (RG) framework, in
close analogy with the study of critical phenomena using $\phi^4$
scalar field theory.  A brief   introduction to  $\phi^4$ theory in
four dimensions and the path integral formulation for fermions is given
before turning to   the problem at hand. As for the latter, the following
procedure is used.   First, the modes on either side of  the Fermi surface
within a cut-off $\Lambda$  are chosen for study in analogy with  the modes
near the origin in $\phi^4$ theory and a path integral is written to describe
them. Next, an   RG    transformation   which eliminates a part of
these modes,  but preserves the action of  the noninteracting system
is identified. Finally the possible perturbations of this free-field
fixed point are classified  as relevant, irrelevant or marginal. A
$d=1$ warmup calculation involving a   system of fermions  shows how
, in contrast to mean-field theory, which predicts a charge density
wave   for arbitrarily weak repulsion, and superconductivity for arb
itrarily weak attraction, the renormalization group approach correct
ly yields a scale invariant system (Luttinger liquid) by  taking int
o account both instabilities. Application of the  renormalization gr
oup   in $d=2$ and $3$, for  rotationally invariant Fermi surfaces,
{\em automatically} leads to Landau's Fermi liquid theory,
which appears as a fixed point characterized by an effective mass an
d a Landau function $F$,  with the  only   relevant perturbations
being  of the superconducting (BCS) type.
  The functional flow equations for the BCS couplings are derived an
  d separated into an infinite number of flows, one for each angular
  momentum.
  It is shown that similar  results hold for rotationally non-invari
  ant (but time-reversal invariant) Fermi surfaces also, with obviou
  s loss of rotational invariance in the parametrization of the fixed
  point interactions.

   A study of a  nested Fermi surface  shows an additional  relevant
   flow  leading to charge density wave formation.  It is pointed out that
for small $\Lambda / K_F$, a $1/N$ expansion emerges, with
$N = K_F/ \Lambda$, which explains why  one is able  to solve
the narrow cut-off theory.  The  search for non-Fermi liquids
in $d=2$ using the RG is discussed. Bringing a variety of phenomena
(Landau Theory, charge density waves,  BCS instability, nesting etc.)
under the one
unifying principle of the   RG    not only allows us to better
understand and unify them , but also  paves the way for generalizations
and extensions. The  article is pedagogical in nature and  is expected
to be accessible to any serious graduate student. On the other hand its
survey of the vast literature is mostly limited to  the   RG  approach.
\newpage
\tableofcontents
\setlength{\baselineskip}{.375in}
\newpage
\section{ INTRODUCTION}
This article is an expanded version of a short paper (Shankar 1991) in
which the application of the renormalization group (  RG ) methods to
interacting nonrelativistic fermions in more than one spatial dimension
was considered. It contains  more technical details than its predecessor
and is much more pedagogical in tone. Several related topics are reviewed
here so that  readers with a variety of backgrounds  may find the article
accessible and self-contained. Consequently each reader is likely to
run into  some  familiar topics . When this happens he should go
through the section quickly to ensure that this is indeed the case and
get used to the notation and conventions. Upon reading this article,
the readers with a condensed matter background  will see how   the
RG  allows us to synthesize a variety of seemingly unrelated phenomena
in condensed matter theory such as Landau's Fermi Liquid Theory, the BCS
instability, Charge Density Wave and Spin Density Wave instabilities, ne
sting  and so on.   Readers familiar with the   RG  but not these topics
, will see  that by following a route parallel to the one that led to a
very successful treatment of critical phenomena, we are automatically le
d to many known results  in the above mentioned topics and newer ways of
understanding them.  However there are also many fascinating differences
between critical phenomena and the phenomena considered here which make
it very interesting from the point view of the   RG .  At the time of wr
iting, there  are relatively few new results and the emphasis is on the
deeper understanding and unification of the older results  the   RG  aff
ords us. However  the machinery developed here, especially for anisotrop
ic systems,  has the potential for changing this state of affairs in the
not too distant future.  The  author is working on  a few new applicatio
ns and hopes the readers will find many more.

The concept of the   RG  was first introduced by Stuckelberg and Petermann
(1953). Its implications for quantum electrodynamics were explored in a sem
inal paper by Gell-Mann and Low (1954). These concepts were extended and
ge
neralized by Callan (1970) and Symanzik (1970).

What is the RG ? When we speak of a group in quantum mechanics we are
thin
king of symmetry operations, i.e.,   transformations  that leave the physic
s invariant. What is the   transformation   here?  Let us consider quantum
electrodynamics. When we compute a physical quantity like the scattering ra
te between electrons, in a power series in the coupling constant $\alpha $,
we find that the coefficients of the series are given by integrals over par
ticle momenta $k$, and that these in turn diverge because the allowed value
s of $k$ go up to infinity. These {\em ultraviolet divergences} are at vari
ance with experiment which gives finite answers for all physical quantities
. Renormalization is the way to reconcile these two facts. In this scheme,
one first cuts off all integrals at the {\em cut-off} $\Lambda$. This gives
answers that are finite, but dependent on $\Lambda$,  which is an artifact
in continuum theory. To get around this, one asks if it is possible to choo
se for each cut-off $\Lambda$ a corresponding coupling $\alpha (\Lambda )$
so that the physical quantities   like scattering amplitudes come out $\Lambda
$-independent. (In quantum electrodynamics one must also renormalize
the mass of the electron with the cut-off. The word coupling shall mean all s
uch
parameters that define the theory.) It is by no means obvious that this can
be done in every field theory.  However in the case of quantum electrodynam
ics or any {\em renormalizable} field theory, one can prove that to any giv
en order in perturbation theory,  it is possible to choose a handful of  pa
rameters
of the model in a cut-off dependent way so as to make physics at momenta
mu
ch smaller than the cut-off independent of it. Since the cut-off is eventua
lly sent to infinity, this means physics at any finite momentum.  {\em This
change in the cut-off by a factor $s$,
accompanied by a suitable change in couplings is an invariance of the
theory.} Th
ese   transformations form a group with the composition rule that a ch
ange by a factor $s_1$ followed by a change by a factor $s_2$ should e
qual a change by a factor $s_1 s_2$. If we write
\beq
s= e^{-t}
\eeq
so that
\beq
\Lambda (t)= \Lambda_0 e^{-t}
\eeq
where $\Lambda_0$ is some fixed number,  the group composition law is
that when two   transformations are implemented in sequence the param
eters $t$ add.

A central quantity in this approach is the {\em $\beta$-function} defi
ned as follows:
\beq
\beta (g) = \frac{dg}{dt}
\eeq
where $g$ is the generic name for the coupling constant(s). Our
convention is the one used in condensed matter physics wherein  i
ncreasing $t$ decreases the cut-off. The field theorists use the
opposite convention  and differ by a sign. To avoid all confusion
let us consider the case of Yang-Mills theory in which
\beq
\frac{dg}{dt} = cg^3 + higher \ orders\ \ \  c>0.
\eeq

If we integrate this equation from $t=0$ to $t=t$, (so that the
cut-off changes from $\Lambda_0$ to $\Lambda_0 e^{-t}$)
we find
\beq
g^2(t) = \frac{g^{2}(0)}{1 - 2g^{2}(0)ct}.
\eeq
What this equation tells us is that as we send the cut-off to infinity
($t$ to $-\infty$), we must reduce the coupling to zero logarithmically:
\beq
g^2 (t) \simeq 1/|t|.
\eeq

The point to notice in all of the above is that one is interested in how to
vary the cut-off only with the intention of eventually sending it to
infinity, which is where it belongs in a continuum theory.  Had the
theory been free of ultraviolet divergences, the question of changing
the couplings and the cut-off, keeping the physics invariant,   may
never have come up. It is clear that from this vantage point
the   RG  has no place in condensed matter physics where the
degrees of freedom live on lattice and there is a natural cut-off
on all momenta: $ \Lambda \simeq 1/a$, where $a$ is the lattice constant.

This point of view was dramatically altered following the work of
Kadanoff (Kadanoff 1965) and Wilson(1971) who gave a different and
more physical interpretation of renormalization.  In this modern
view one  contemplates changing  the cut-off (and the couplings)
even in a problem where nature provides a natural cut-off such as
the inverse lattice spacing and there are no ultraviolet infiniti
es.  We will now discuss an example from statistical mechanics wh
ere
the value of such a procedure is apparent. The discussion will be
schematic
since a more detailed one follows in the next section.

Let us consider a cubic lattice (in $d$ dimensions) with a real
scalar field $\phi (\vec{n})$ at each site labeled by the vector
$\vec{n}$ with integer coefficients. The classical statistical
mechanics of this system is described by the partition function
\beq
Z = \int \prod_{\vec{n}} d\phi (\vec{n}) e^{S(\phi (\vec{n} ))}.
\eeq
This is just the usual sum over configurations with the Boltzmann
factor $e^{-\beta \varepsilon} $  written in terms of the {\em action}
$S$, also called the {\em hamiltonian. } (Both terms will be used inter
changeably to prepare the reader for what happens all the time in the
literature.) As long as the number of sites is finite, $S$ is just a r
egular function and $Z$ is just a multiple integral. In the limit of i
nfinite sites $S$ becomes a functional and $Z$ becomes a {\em function
al integral} or {\em a Feynman  path integral.} Feynman introduced his
path integral to describe quantum mechanical problems in $d$ spatial d
imensions  as a sum over classical configurations in $d+1$ dimensions.
Thus our $Z$ could very well stand for  Feynman's representation of a
quantum problem in one lower dimension and the following consideration
apply to it.
\footnote{The $d+1$-th dimension is of course time. One works with
imaginary time, with the option  to analytically continue to real
time at the end if needed. In this case one often finds that all $d
+1$ dimensions  are equivalent. In this  discussion of the scalar fiel
d we will assume that this so. } For the  problems with bosonic operator
s the derivation of the path integral can be found in  Section IIIB of
the review by Kogut ( Kogut 1979).  The derivation for the fermionic
problem will be given in section III.

A typical quantity one is interested in is the  average of the
correlation between the variables at two different sites, also
called the {\em two-point function}:
\beqr
G(\vec{n_1} , \vec{n_2}) &=& G(\vec{n_1} -\vec{n_2})  \ (assuming \
translation \ invariance)\\
&\equiv& <\phi (\vec{n_1}) \phi (\vec{n_2})> \\
	   &=& \frac{\int \prod_{\vec{n}} d\phi (\vec{n}) \phi (\vec{n_1}) \phi
(\vec{n_2})e^{S(\phi (\vec{n} ))}}{\int \prod_{\vec{n}} d\phi (\vec{n})
e^{S(\phi (\vec{n} ))}}.
	   \eeqr

	   For  long separations this {\em correlation function}
	   typically falls
off exponentially
	   as
	   \beq
	   G(\vec{n_1} - \vec{n_2}) \simeq e^{-|\vec{n_1} - \vec{n_2}|/\xi }
	   \eeq
	   where $\xi$ is the {\em correlation length}.
	   \footnote{In the fermion
problem we are going to study, the (imaginary) time and space directions are
not equivalent. The correlation length in these discussions refers to the time
direction.} The exception is
	    when the parameters are such that the system is at a {\em critical
point }, as in the case of a magnet undergoing a Curie transition from the
ferromagnetic to paramagnetic state.  In this case it falls like a power
	   \beq
	   G(\vec{n_1} - \vec{n_2}) \simeq \frac{1}{|\vec{n_1} - \vec{n_2}|^x
}
	   \eeq
where $x$ is a {\em critical exponent}. Other critical exponents characterize
other power laws at the critical point.  A remarkable feature that we will
address shortly, is that several systems with microscopically distinct
hamiltonians (or actions) have the same critical exponent.

	    In the case of quantum problems written as path integrals the
correlation length is related to $m$, the {\em mass gap}, or the lowest
excitation energy above the ground state as per
	    \beq
	    \xi = 1/m
	    \eeq
	     and the critical case corresponds to  $m=0$.

	     An equally complete description of the above system is possible in
terms of the Fourier transforms:
	     \beq
	     \phi (\vec{k}) = \frac{1}{V} \sum_{\vec{n}} e^{i\vec{k}\cdot
\vec{n}} \phi(\vec{n}),
	     \eeq
	     where $V$ is the volume of the system. The allowed  momenta
$\vec{k}$ lie within a Brillouin  cube of sides $2\pi /a$ in all directions.
The
partition function becomes
	     \beq
	     Z =  \int \prod_{|\vec{k}| \le \pi /a} d\phi (\vec{k}) e^{S(\phi
(\vec{k} ))},
	     \eeq
 and  $G(\vec{k})$, the Fourier transform of $G(\vec{n_1} -\vec{n_2})$ is
given by
\beqr
<\phi (\vec{k_1}) \phi (\vec{k_2})> &=& (2\pi )^d \delta^{(d)} (\vec{k_1} +
\vec{k_2}) G(\vec{k_1})\\
	   &=& \frac{\int \prod_{\vec{k}} d\phi (\vec{k}) \phi (\vec{k_1}) \phi
(\vec{k_2})e^{S(\phi (\vec{k} ))}}{\int \prod_{\vec{k}} d\phi (\vec{k})
e^{S(\phi (\vec{k}) )}}.
	   \eeqr
Let us now imagine that we are only interested in the physics at long
distances, (compared to the lattice spacing $a$), for example in $G(\vec{r})$
for large separations $\vec{r}$.    In momentum space this translates to small
$\vec{k}$. To be specific let us say we are interested only in correlations of
modes that lie within a tiny ball of size $\Lambda / s$ (with $s$ very large)
centered at the origin, and  not interested in the modes that
lie in the rest of
the Brillouin zone, taken to be a sphere (rather than a cube) of radius $2\pi
/a$.  (This modification makes no difference to the small $k$ asymptotics.)
We will refer to the small $k$  modes as "slow modes"  and the others as
"fast modes".  Let us define two sets of variables:
\beqr
\phi_{<} &=& \phi (k) \ for \ 0 < k < \Lambda /s \ (slow \ modes) \\
\phi_{>} &=& \phi (k) \ for \ \Lambda /s \le k \le \Lambda  \ (fast \ modes).
\eeqr
By assumption we are going to be interested only in correlations of $\phi_<$.
However at present we are computing these objects by doing an integral over
fast and slow modes. The first step in the   RG  program is to ask if there is
an effective action or Boltzmann weight $e^{S'(\phi_< )}$ such that when
integrated over just the slow modes, it will reproduce all the slow correlation
functions. We shall now see that the answer is affirmative.

Let  the action be expressed  as follows:
\beq
S(\phi_{<}, \phi_{>}) = S_0 ( \phi_{<}) + S_0 ( \phi_{>}) + S_I (\phi_< ,
\phi_> )
\eeq
where $S_0$ is a quadratic function of its arguments that separates into slow
and fast pieces and $S_I$, called the interaction, is the part which mixes the
two. Then
\beqr
Z &=& \int \prod_{0 \le k < \Lambda /s} d\phi (k)\prod_{\Lambda /s  \le k <
\Lambda } d\phi (k)  e^{S_0 (\phi_< )}e^{S_0 (\phi_> )}e^{S_I (\phi_< ,
\phi_> )} \\
& \equiv & \int [d\phi_< ] \int [d\phi_> ]  e^{S_0 (\phi_< )}e^{S_0 (\phi_>
)}e^{S_I (\phi_< , \phi_> )}\\
 &=& \int [d\phi_< ] e^{S_0 (\phi_< )}\int [d\phi_> ]e^{S_0 (\phi_>
)}e^{S_I (\phi_< , \phi_> )} \\
 &\equiv & \int [d\phi_< ] e^{S^{'} (\phi_< )}
\eeqr
which defines the {\em effective action $S^{'} (\phi_< )$}. Let us manipulate
its definition a little:
\beqr
e^{S^{'} (\phi_< )} &=& e^{S_0 (\phi_<)} \int [d\phi_> ]e^{S_0 (\phi_>
)}e^{S_I (\phi_< , \phi_> )} \nonumber \\
&=& e^{S_0 (\phi_<)} \frac{\int [d\phi_> ]e^{S_0 (\phi_> )}e^{S_I (\phi_< ,
\phi_> )}}{\int [d\phi_> ]e^{S_0 (\phi_> )}}
\underbrace{\int [d\phi_> ]e^{S_0 (\phi_> )}}_{Z_{0>}} \nonumber \\
&=& e^{S_0 (\phi_<)}<e^{S_I(\phi_< , \phi_> )}
>_{0>}\label{effectiveaction}
\eeqr
where $<>_{0>}$ denotes averages with respect to the fast modes with
action $S_0$
and where the corresponding partition function $Z_{0>}$ has been dropped
in going to the last line, since it will merely add a  constant to the
effective
action independent of $\phi_<$, which in turn will make  no difference to any
correlation function of slow modes.

Although $S'(\phi_< )$ provides a good description of the slow mode physics,
the   RG    transformation   has two more steps besides the above  {\em mode
elimination}. These steps will now be motivated.

One aim of the   RG  is to see the various parameters in the interaction evolve
or flow as the cut-off is reduced, i.e., to compute the $\beta$-function.
Suppose  before mode elimination we had
\beq
S(\phi ) = r\phi^2 + u\phi^4 + \ldots
\eeq
and after,
\beq
S'(\phi_< ) = r'\phi^{2}_{<} + u'\phi^{4}_{<} + \ldots
\eeq
 (The action above is schematic. For example $u\phi^4$ could be the
shorthand for
\beq
\int dk_1 dk_2 dk_3 dk_4 \delta (k_4+k_3 +k_2 +k_1) u(k_4, k_3,
k_2,k_1)\phi (k_4) \phi (k_3) \phi (k_2)\phi (k_1)
\eeq
where $u(k_4,...k_1)$ is a {\em coupling function}  and not just a coupling
constant.) In any event,
we are trying to compare $r$ to $r'$, $u$ to $u'$ and so on. The problem with
doing that is that we are comparing apples to oranges. The old and new
theory are defined on two different kinematical regions. For example the
coupling $u(\Lambda , \Lambda , \Lambda , \Lambda )$ has no counterpart in
the effective theory which has all its momenta below $\Lambda /s$. (In field
theory where the old and new cut-off are both sent to infinity, this point does
not come up.) To remedy this defect, we will define new momenta after mode
elimination:
\beq
k' = sk
\eeq
which run over the same range as $k$ did before elimination.

There is just one more problem. Consider two actions:
\beqr
S(\phi ) &= &r \phi^2 + u\phi^4 \\
 S'(\phi ) &= &4r \phi^2 + 16u \phi^4
 \eeqr
 which seem different. They are however physically equivalent because we
can  simply define  $2\phi = \phi^{'}$ in the second action  (and ignore the
Jacobian in the functional integral since it is a $\phi$-independent constant)
and reduce it to the first action.  In other words,
certain changes in parameters
are not of physical importance since they can be absorbed by field rescaling.
To weed these out,  we will follow  mode elimination and momentum
rescaling by a field rescaling, defining new fields:
 \beq
 \phi^{'} (k') = \zeta^{-1}   \phi_< ( k'/s)
 \eeq
 and choose $\zeta$ such that a certain coupling in the quadratic part of the
action
has a fixed coefficient. The final action $S'$ will then be expressed in
terms of this new field. Thus the three stages in the   RG    transformation
are as follows:
 \begin{itemize}
 \item Eliminate fast modes, i.e., reduce the cut-off from $\Lambda $ to
$\Lambda /s$.
 \item Introduce rescaled momenta: $k' = sk$ which now go all the way to
$\Lambda$.
 \item
 Introduce rescaled fields $ \phi^{'}(k') = \zeta^{-1}   \phi_< ( k'/s)$ and
express the effective action in terms of them. This action should have the
same coefficient for a certain quadratic term.
 \end{itemize}

 With this definition of the   RG    transformation  , we have a mapping from
hamiltonians or actions defined in a certain phase space to actions in the same
space. Thus if we represent the initial action  as a point in a coupling
constant
space, this point will flow under the   RG  transformation  to another point in
the same space.  This definition of the   RG  opens up  a possibility that did
not exist without all three steps: {\em a fixed point  $S^{*}$} of the  group
action,
that is to say the action function which reproduces itself after the three
step    RG    transformation. Geometrically this means that the point $S^{*}$
does not move or flow under the   RG  action. If the system had a correlation
length
 $\xi$
 before the   RG  (in the old units), in the new units (in which momenta
get boosted by a factor $s$) it would  decrease to $\xi /s$. On the other hand,
at the fixed point, it must remain the same under the   RG . This means that
the correlation length at a fixed point must have been  either zero or
infinite.
We shall be dealing with the latter case here. Fixed points will dominate our
analysis.

In summary,  we see   that in the modern viewpoint, the cut-off is not to be
viewed as an artifact to be sent to
infinity but as the dividing line between the
modes we are interested in and the modes we are not interested in.The
preceding discussion explains {\em how}  we may change the cut-off and the
couplings without affecting the slow mode physics even in a problem where
there were no ultraviolet infinities.  Let us now understand {\em why} we
would want to do such a thing.

Consider the  remarkable phenomenon of universality. How can  systems with
different microscopic hamiltonians have the same decay exponent $x$ in their
critical two-point functions?   The   RG  explains this as follows ( Kadanoff
1965,1977; Wilson 1971, Wilson and Kogut 1974, Fisher 1974,1983; Wilson
1975 ). Let $S_A$ and $S_B$ be two critical hamiltonians defined in the full
$k$ space.  Each is described by a set of coupling constants. Let us represent
each as a point in a space $\cal{S}$ (in the notation of Wilson and Kogut
1974, Section 12) of hamiltonians where along each axis we measure one
coupling constant. The fact that $S_A \ne S_B$ implies that they are given by
distinct points in coupling constant space and that there are many observables
that differ in the two cases. Consider however  extreme long distance physics,
in particular the long distance decay  of   two-point functions. To calculate
these we need just the slow modes. To this end let us trade each hamiltonian
for its equivalent one after renormalization down to a very small cut-off.
{\em What we will find is that they both asymptotically approach       the
same  fixed point hamiltonian $S^{*}$where the flow stops.}  This  explains
why they share the same long distance physics and in particular the exponent
$x$. Although the coupling constant space
is infinite dimensional let us consider a toy model in which it is three
dimensional. Let all critical hamiltonians (in particular $S_A$ and $S_B$) lie
in the $x-y$ plane. Under the   RG  they all flow to $S^{*}$, which lies at
say the point $(1,1,0)$. Let us shift the origin of coordinates to the fixed
point.
Any deviation $S - S^{*}$ that lies in the critical plane  is termed {\em
irrelevant} in the   RG  terminology since it renormalizes to zero and hence
makes no difference to long distance  physics. The fixed point, being a special
case of a critical point, will of course have power law decay of correlations.

What we see is that if to this fixed point an irrelevant perturbation is added,
the perturbed system will also have the same power law decay.  If the
functional integral stands for some quantum system wrtten as a path integral,
this means that a gapless system will remain gapless
if an irrelevant perturbation is added. This idea will be invoked later in this
article. By contrast, any deviation off the critical
($x-y$) plane is called {\em
relevant} and   will get amplified by the   RG    transformation. The long
distance behavior of correlations in  this problem is unclear;  it is
controlled
by the ultimate destination of this flow,  and typically (but not always)
corresponds to  exponential decay. In the general problem there can  also  be
{\em marginal } perturbations, which neither grow nor decay under the   RG
transformation . They play a major role in the nonrelativistic electron problem
to which we now turn our attention. A truly marginal perturbation does not
cause a gap.

Table I summarizes some of the above concepts.

The preceding discussions have set the stage for introducing our main topic.
Consider a system of noninteracting fermions \footnote{ We use the term
fermion instead of simply electron to accomodate {\em spinless fermions}
which do not exist in nature but simplify the analysis by obviating the need
for spin indices. While the spin of the electron
is certainly not ignorable when
comparing
theory to experiment, it will be seen that it  really is an inessential
complication in the   RG  program to be described here and maybe
incorporated readily.} at zero temperature ($T=0$) either in the continuum
with a dispersion relation
\beq
 E=K^2/2m
 \eeq
or on  a lattice with some energy function $E(\vec{K})$ defined within the
Brillouin zone. \footnote{We use upper case letters to denote momenta
measured from the origin in contrast to the preceding discussion  where lower
case symbols were used. This is a deliberate departure from convention and
reflects the different physics that emerges here.}In all cases, one particle
states with $E\le \mu$, where $\mu$ is the {\em chemical potential}, are
filled in the ground state of the many-body system. The filled states are
bounded by the {\em Fermi  surface }. In the continuum in $d=2$ or $d=3$
the Fermi  surface  is a circle or  sphere respectively of radius
  \beq
  K_F = \sqrt{2m\mu}.
  \eeq This ground state has gapless excitations corresponding to the
promotion of fermions  from just below the Fermi surface  to just above it.
The central questions we ask in this paper are the following:
  \begin{itemize}
  \item If some perturbation is added to the free theory, will the system
develop a gap at once or will it remain gapless?
  \item If it remains gapless what is the natural way to describe the low
energy
physics of the system, in particular its response to "soft probes", probes of
low frequency $\omega$ and momentum $\vec{Q}$?
  \end{itemize}
The answer to these questions are clearly  dictated by the modes near the
Fermi surface, at  least for the case of weak perturbations.   For example in
any kind of perturbation theory of the ground state, these modes will come
with the lowest energy denominators.  We will therefore focus on modes
within a bandwidth $\Lambda$ of the Fermi surface  (the slow modes of this
problem)  and get rid of all the modes out side this cut-off (the fast modes).
In
the case of  fermions in free space we define a lower case momentum
\beq
k = |\vec{K}| - K_F
\eeq
and work with modes obeying
\beq
|k| \le \Lambda .
\eeq
In the case of electrons on a lattice, the wave vector $|\vec{K}|$ is no longer
a measure of energy and we must keep those modes that whose {\em energy}
lies within some cut-off. This complication will be discussed in the sections
devoted to lattice problems. For the present let us focus on electrons in free
space and imagine  an annulus or shell (in two or three dimensions
respectively) of thickness $2\Lambda$ with mean radius $K_F$ within which
reside the slow modes of this problem.

Let us now turn to the elimination of the fast modes outside the cut-off.
This
may be done  within the operator formalism by the use of projection operators
to define an effective hamiltonian restricted to  the subspace of slow modes.
This effective quantum hamiltonian  depends  on the cut-off in such a way as
to produce cut-off independent results for  the surviving slow  modes and the
fixed point, if any, is unaffected by this   transformation.   This is the
approach used by Wilson (1975),  Anderson and Yuval (1970) , Nozieres
(1974) and Krishna-Murthy {\em et al} (1980) in their treatment of the Kondo
problem. However  this problem, which is a paradigm for how the  RG   is to
be used in quantum problems in many-body physics, is essentially one
dimensional.  \footnote{Here  one deals with a conduction band of electrons
interacting
with a single fixed impurity.  By using  spherical waves (instead of
plane waves) centered on the impurity and keeping  just the  s-wave, one
reduces it to a  quasi-one dimensional problem in the radial coordinate.}  By
contrast the problems we deal with here are truly two and three dimensional
and the application of the  RG   to these has a short history.  Although
Anderson had suggested this possibility even in his book (Anderson 1984) no
detailed analysis was carried out for some time.  Benfatto and Gallavotti
(1990) and Feldman {\em et al} (1990, 1991, 1992) then combined the RG
with  rigorous
bounds to study (to all orders in perturbations) the stability of
gapless Fermi systems  to perturbations. Shankar (1991) developed  the
method to be described here, which is less rigorous, more intuitive, covers
other instabilities like charge or spin density waves,
rotationally-noninvariant
systems   and  maybe   easier to use for people with a background in critical
phenomena or modern field theory. More recently Polchinski (1992)
employed  a very similar approach to the nonrelativistic fermion problem to
better understand the concept  of effective field theories in particle physics.
Weinberg (1993) recently derived the effective low energy action and RG
flow equations for superconductors with Fermi surfaces that obeyed time-
reversal symmetry and nothing else. All these approaches are fundamentally
different in spirit from the method used by Hertz (1976) who completely
integrated the fermions in favour of some bosonic variables. In particular he
integrated the modes at the Fermi surface . This is analogous to integrating
the $k=0$ modes in critical phenomena. The effective theory for the bosons
then has singular parameters. Hertz handled  this problem carefully and found
a way to analyze  phase transitions that can be described by the bosonic
variables.

The approach described in this paper is as follows.  To heighten the analogy
with critical phenomena one first shifts from the operator approach to a path
integral approach and writes down the path integral. First the noninteracting
problem is considered. Since it is gapless one expects it to be the fixed point
of a   RG    transformation   in which the cut-off is reduced.  Such a
transformation    is found. With respect to this   transformation
perturbations
are classified as relevant, irrelevant or marginal. In the last two cases the
system will remain gapless while in the first case one can only make
statements if one assumes that the behavior seen for small perturbations
persists at strong coupling also. In all cases considered, this corresponds to
the opening of a gap in the spectrum.
While all this sounds like critical phenomena, (and is meant to), there are
crucial difference which can be traced back to  the nature of the phase space
for slow modes. In critical phenomena all the  long distance physics comes
from    a tiny {\em ball centered at the origin of k-space} and the fixed point
is characterized by a few couplings.
\footnote{I thank Pierre Hohenberg for pointing out to me an exception:   a
problem not involving a Fermi surface, which has  nonetheless a similar phase
space after any amount of renormalization: the condensation of a liquid into a
nonuniform state, studied by Brazovskii (1975). See Swift and Hohenberg
(1977) for the study of fluctuations on an equivalent model. It is an open
question whether the methods developed here can be applied to Brezovskii's
problem.}

 The same is true for continuum field theories like quantum electrodynamics
or quantum chromodynamics: both fermion and boson momenta are
restricted to a sphere of radius $\Lambda$ centered at the origin. In the
problem at hand, we renormalize not towards a single point, the origin, but a
surface, the Fermi surface  (which may itself change under renormalization in
the nonspherical case.) Unlike in  critical phenomena where all momenta and
momentum transfers are small (bounded by the cut-off) here only $k =
|\vec{K}|- K_F$  is small and large momentum transfers of the order of
$K_F$ are possible within the  slow  modes. {\em Renormalization only
reduces the dimension normal to the Fermi surface, the tangential part
survives.}  As for the fixed point, it is  characterized  by a surface  and
{\em
coupling functions} defined on it. \footnote{All this can be stated in another
way. In field theories or in critical phenomena one also runs into coupling
functions. But these are functions of just $k$. When Taylor expanded in $k$,
only a few terms are marginal or relevant. In the present problem, the
coupling functions depend on $k$ as well as the coordinates of the limiting
Fermi surface . The latter never get eliminated and all terms in the  Taylor
series for the latter will be important. This point will be discussed further
as
we go along. } Notice that $d=1$ is special: here the Fermi surface  is a set
of
two disjoint points. Apart from this doubling (which converts nonrelativistic
fermions into Dirac fermions) we have the same situation as in a continuum
field theory in one space dimension and there are once again just a few
coupling constants. This is why there has been a lot of activity and a lot
success (Bourbonnais and Caron 1991, Solyom 1979) in applying the   RG  to
one dimensional fermion problems in condensed matter  and a lot of
resistance to going to higher dimensions.

 We now turn to the details. In Section II the reader is given a very brief
review of how the   RG  works for a scalar field theory in four dimensions.
This will serve to remind the readers familiar with the subject the
highlights
that we will recall frequently in our progress by analogy.
  As for the newcomers, it will give them the minimum required to follow this
article. References for more details will be given. Section III explains how a
path integral can be written for fermions and how one is to extract
 correlation
functions from it. This will require the
  introduction of Grassmann variables. Readers not used to these should kill
two birds with one stone by  using the pedagogical review provided here to
learn this tool which is often used  in condensed matter theory.
        In Section IV
we study  the problem of spinless fermions in one dimension at half-filling:
with one particle per every other site on the average. This section serves as a
warm up for the   RG  program, since, as explained above, it resembles
   the run of the mill field theory in one dimension. It also shows the power
of
the   RG: whereas mean field theory ( a self-consistent approximation to be
detailed later) predicts a gap for the smallest repulsion, and
superconductivity
for the smallest attraction, the exact solution
   tells us that the system remains gapless for a finite range
   of coupling of either sign. It will be seen that the   RG  gives
   results in
agreement with the exact solution.
    Sections $V$ and VI deal with  circular and spherical Fermi surfaces.
    To lowest order in a perturbative expansion  (Section V) one finds
    that there exists a fixed point described by two  marginal coupling
    function $F$ and $V$ which depend on the angles on the circle or sphere
    as the case may be. To the next order (Section VI) one finds that  $F$ is
still
    marginal while each coefficient in the angular momentum expansion of
    $V$ grows to produce the superconducting instability if attractive
    and
    renormalizes to downwards if repulsive, a result originally
    discovered by Morel and Anderson(1962). No new surprises come at
higher orders. This is explained  in the next section. The fixed point theory
which exists in the absence of $V$ is what is known as Landau's Fermi
Liquid Theory.   The Kohn-Luttinger effect which destroys the Fermi liquid at
low temperatures is derived in the RG language.  Section VII provides a new
way of understanding why it is possible to solve the fixed point theory
characterized by the interaction $F$ even though $F$ is not necessarily small.
This is tied to the fact that certain theories with a large number of
fields can
be described by an expansion in $1/N$, $N$ being the number of
components. (In other words the coupling need not be
     small as long as $1/N$ is.) It is shown that the Fermi system with
     cut-off $\Lambda$ has a $1/N$ expansion with $N = K_F /\Lambda $.
     Thus a given problem in the full momentum space can initially be
renormalized to a small  $\Lambda$ theory (without running into any
singularities) and then when $N$ is large enough, solved in  the $1/N$
approximation. Section VIII has a discussion of Landau's Fermi Liquid
Theory. Only some aspects of this extensive field are brought up. In Section
IX we consider non-circular Fermi surfaces with no special features other
than time
    reversal invariance: if $\vec{K}$ lies on it so does $-\vec{K}$.
    It is found that one must stop using $|K|$ as a measure of energy and use
actual equal-energy contours to define the fast and slow modes. The net result
is exactly as in the rotationally invariant case except for the fact that $F$
and
$V$ now depend on more variables due to the lack of rotational invariance.
Section X  deals with the very interesting case of {\em nested } Fermi
surfaces in $d=2$: surfaces such that if $\vec{K}$ lies on them,  so does
$\vec{K} + \vec{Q_N} $ where $\vec{Q_N} $ is a fixed {\em nesting}
momentum.   We choose to illustrate the ideas with spinless fermions on a
rectangular lattice in which case  the nesting vector $\vec{Q_N} $ has
components $(\pi , \pi )$. (Readers unfamiliar with nesting may wish to peek
at  Fig.17 in Section X for an example.)  What we find is that to lowest order
a third coupling function $W$ insinuates itself at the fixed point. At next
order it begins to flow. One can show that there are definitely some relevant
directions if this force is repulsive
    and these tend to produce charge-density waves: the ground state has a
nonuniform charge density which oscillates with momentum $\vec{Q_N} $.
In Section XI we use the methods developed here to look for non-Fermi
liquids in two dimensions. Regrettably the results are negative for the case of
weakly coupled problems with a circular Fermi surface .
 Section XII contains the summary and outlook. Many
  of the remarks made in this preview will be repeated there, and the reader
will have a clearer picture of their  significance. The Appendix deals with
two
special topics: Coulomb screening and the Kohn-Luttinger effect as they
appear within the   RG  framework.
\section{ AN EXAMPLE OF THE   RG  FROM $d=4$}
The problem chosen to illustrate the   RG  at work involves a complex scalar
field in $d=4$. The functional integral can be viewed either
as describing the quantum field theory of a charged scalar field in three
space dimensions or as describing the classical statistical mechanics of
a system with one  complex field or a two real fields at each point
on a lattice. The Ising model which is described by a real field
is not chosen here since the fermion problem we will study later involves
charged
Fermi fields. Readers new to the problem should be aware that this section
has the very limited objective of making the rest of the paper comprehensible
to them. For a deeper introduction to critical phenomena, the reader is
directed to any of the excellent reviews (Wilson and Kogut 1974, Fisher
1974, Kadanoff 1977) or books (Goldenfeld (1992), Itzykson and
Drouffe(1989),  Le Bellac(1991), Ma (1976) Plischke and Bergersen(1989),
Zinn-Justin(1989)).
Readers familiar with the subject are still urged to skim through
this section to get acquainted with the notation and as well as to
refresh their memory,  since our approach to  the interacting Fermi problem
will  rely heavily on analogy to this problem where the
  RG  approach has been spectacularly  successful.

The partition function for this problem is
\beqr
Z &=& \int_{|k| < \Lambda} [d\phi d\phi^{*}]e^{S( \phi , \phi^{*})}, \ \ \ \
where \\
\left[ d\phi d\phi^{*} \right] &=& \prod_{|k| < \Lambda } \frac{dRe \phi
(\vec{k}) dIm \phi  (\vec{k} )}{\pi} \\
S(\phi , \phi^{*} ) &=& \int_{|k| < \Lambda}\phi^{*} (\vec{k}) J(k) \phi
(\vec{k}) \frac{d^4k}{(2\pi )^4},\   \ \ \ \ and \\
J(k) &=& 2 \left[ (\cos k_x -1 ) +  (\cos k_y -1 ) +  (\cos k_z -1 ) +
 (\cos k_t
-1 ) \right].
\eeqr
 This action is obtained by Fourier   transformation   of
   the following  nearest
neighbor  interaction in coordinate space:
 \beq
 S( \phi , \phi^{*}) = -\frac{1}{2} \sum_{\vec{n} , \vec{i}} | \phi (\vec{n}) -
\phi (\vec{n} + \vec{i} )|^2
 \eeq
 where $\vec{n}$ is the vector with integer coordinates used to label
 the sides of the hypercubic lattice and $\vec{i}$ is any of the eight
 unit vectors in the direction of  increasing or decreasing coordinates.
 Notice that this action favors the alignment of neighboring fields, i.e., is
 ferromagnetic.

 Since we are interested in small $k$ physics, let us hereafter approximate
$J(k)$ by its leading term in the Taylor series and write
 \beqr
 S_0(\phi , \phi^{*} ) &=& -\int_{|k| < \Lambda} \phi^{*} (\vec{k}) k^2 \phi
(\vec{k}) \frac{d^4k}{(2\pi )^4}\\
 Z &=& \int_{|k| < \Lambda} [d\phi d\phi^{*}] e^{-\int_{|k| < \Lambda}
\phi^{*} (\vec{k}) k^2 \phi (\vec{k}) \frac{d^4k}{(2\pi )^4}} \\
\left[ d\phi d\phi^{*} \right] &=& \prod_{|k| \le \Lambda}\frac{dRe \phi
(\vec{k} )dIm \phi  (\vec{k} )}{\pi} =  \prod_{|k| \le \Lambda}
\frac{d\phi^{*}  (\vec{k} )d \phi  (\vec{k} )}{2\pi i}
 \eeqr
This is called the {\em gaussian model}.  The corresponding functional
integrals is product of ordinary gaussian integrals, one for each $\vec{k}$.
This makes it possible to express  all the correlation functions in terms of
averages involving a single gaussian integral.
{\em  The only averages that do not vanish are products of  an even number
of variables, wherein each $\phi (\vec{k} )$ is accompanied by its complex
conjugate}. This is because  the action and measure are  invariant under
\beq
  \phi (\vec{k} ) \rightarrow \phi (\vec{k} )e^{i\theta } \ \ \phi^{*}
   (\vec{k}
)\rightarrow \phi^{*} (\vec{k} )e^{-i\theta}.
  \eeq
  where $\theta$ can be different for different  $\vec{k}$'s.
 Thus the only
integral we will ever need follows from the simple problem involving just a
pair of
complex conjugate variables $z$ and $z^{*}$:
\beq
 <zz^{*}> = \frac{\int_{-\infty}^{\infty} \frac{dzdz^{*}}{2\pi i}zz^{*}
 e^{-azz^{*}}}{\int_{-\infty}^{\infty} \frac{dzdz^{*}}{2\pi i}
 e^{-azz^{*}}}=\frac{1}{a}.
 \eeq

 The other two bilinears have zero average:
  \beq
  <zz> = <z^{*}z^{*}>=0
  \eeq  because the action and measure are invariant under
  \beq
  z \rightarrow z e^{i\theta } \ \ z^{*} \rightarrow z^{*} e^{-i\theta}
  \eeq
  while the bilinears are not.
  The reader wishing to verify the above results is asked to switch to
  $x$ and
  $y$, the real and imaginary  parts of the integration variables and to use
  \beq
  \frac{dzdz^{*}}{2\pi i} \rightarrow \frac{dxdy}{\pi}.
  \eeq

  If there are two sets of variables we have
  \beq
 <z^{*}_{i}z_j> = \frac{\int_{-\infty}^{\infty} \frac{dz_1dz^{*}_{1}}{2\pi
i} \frac{dz_2dz^{*}_{2}}{2\pi i}z^{*}_{i}z_j
 e^{-a_1z_1z^{*}_{1}  -a_2 z_2 z^{*}_{2} }}{\int_{-\infty}^{\infty}
\frac{dz_{1}dz^{*}_{1}}{2\pi i}
 \frac{dz_2dz^{*}_{2}}{2\pi i}e^{  -a_1z_1z^{*}_{1}  -a_2 z_2 z^{*}_{2}
}}=\frac{\delta_{ij}}{a_i} \equiv <\overline{i}j>.
 \eeq

As for the four point function, the reader may verify that
\beq
< z^{*}_{i}z_j z^{*}_{k}z_l > =
<\overline{i}j><\overline{k}l><\overline{i}l><\overline{k}j>
\label{smallwick2}
\eeq
This result makes sense: it demands that for the answer to be nonzero,
 the fields must come in complex conjugate pairs.  Since this  can happen
 in two ways, the result is a sum of two terms. The generlization to more
variables and longer strings is obvious.

 In view of the above, the reader will not be surprised
 that the {\em two-point function} in our gaussian model   is
\beqr
<\phi^{*}(\vec{k}) \phi (\vec{k}')> &=& \frac{(2\pi )^4 \delta^4 ( \vec{k} -
\vec{k}' )}{k^2} \\
						     &\equiv&
                                                      (2\pi )^4 \delta^4 (
\vec{k} - \vec{k}' ) G(k)\\
						     &\equiv& <\overline{2}1>.
						     \eeqr
 and likewise
 \beq
 <\phi^{*}(\vec{k}_4)  \phi^{*}(\vec{k}_3) \phi^{}(\vec{k}_2)
 \phi^{}(\vec{k}_1) > = <\overline{4}2><\overline{3}1> +
<\overline{4}1><\overline{3}2>.
 \eeq
 This is a case of {\em Wick'sTheorem } for bosons. For the case of $2n$
fields, the answer is a sum over all possible pairings, each term in the sum
being a product of $n$ 2-point functions.                             The
result follows from  the preceding discussion, upon making the change  from
Kronecker deltas to Dirac delta functions in Eqns.(\ref{smallwick2})  to take
into account the fact that the action in the gaussian model is an integral
 (over
$\vec{k}$) rather than sum over variable labels.

 Note that $G$ has power law behavior in momentum space ( $1/k^2$)
 and hence will do so in coordinate  space ( $1/r^2$).  Thus the action
 of the
gaussian model is critical and must flow to a fixed point under the action of
the   RG. We will now see that it is itself a fixed point.

 In the first stage of the   RG  transformation, we integrate out $\phi_>$.
Since $S_I =0$ here, we see from Eqn.(\ref{effectiveaction}) that
 \beq
 S'(\phi_< ) =  -\int_{|k| < \Lambda  /s} \phi^{*}_{<}
 (\vec{k}) k^2 \phi_{<} (\vec{k}) \frac{d^4k}{(2\pi )^4}.
\eeq
 We now carry out the last two steps by rewriting the action in terms of
 \beq
 \phi^{'}(\vec{k}' ) = \zeta^{-1}\phi_{<} (\vec{k}' /s)
 \eeq
 and obtain
 \beqr
  S'(\phi^{'}_{<} ) &=&  -s^{-6}\int_{|k'| < \Lambda  } \phi^{*}_{<}
(\vec{k}' /s) k^{'2} \phi_{<} (\vec{k}' /s) \frac{d^4k'}{(2\pi )^4} \\
  &=& -\frac{\zeta^2}{s^6}\int_{|k|' < \Lambda } \phi^{'*} (\vec{k}') k^{'2}
\phi' (\vec{k}') \frac{d^4k'}{(2\pi )^4}\\
  &=& S'(\phi' ).
  \eeqr
If we now make the choice
\beq
\zeta = s^3
\eeq
we find that the gaussian action is the fixed point:
\beq
S'(\phi' ) = S(\phi ) = S^{*}.
\eeq
Having found the fixed point, we next classify its perturbations as relevant,
irrelevant or marginal.
 We will only consider perturbations involving an even number of fields. Let
us start with the quadratic case:
 \beq
 \delta S = -  \int_{|k| < \Lambda} \phi^{*} (\vec{k}) r(k) \phi (\vec{k})
\frac{d^4k}{(2\pi )^4} \label{r}
 \eeq
 where the {\em coupling function $r$} is assumed to have a Taylor
expansion
 \beq
 r(k) = r_0 + r_2 k^2 + \ldots
 \eeq
 which reflects the short range nature of the perturbation in coordinate
 space.
One often writes
 \beq
 r_0 = m^{2}_{0}
 \eeq
 and refers to $m^{2}_{0}$ as the {\em mass term} since in the quantum
field theory interpretation of the functional integral, adding this term
to the
gaussian model
 yields the quantum field theory of a particle of mass $m_0$.

 Since this perturbation does not mix slow and fast modes, all we have to do
is replace $\phi$ by $\phi_<$ everywhere and reexpress the result in terms of
new momenta and fields. This gives
 \beqr
 \delta S' (\phi' (\vec{k}' )) &=& -s^2 \cuts \phi^{'*}
  (\vec{k}') r(k'/s) \phi'
(\vec{k}') \frac{d^4k'}{(2\pi )^4}\\
&\equiv & -\cuts \phi^{'*} (\vec{k}') r'(k') \phi' (\vec{k}') \frac{d^4k'}
{(2\pi
)^4}
\eeqr
where in the last equation we have invoked the definition of the renormalized
coupling $r'(k')$.
By comparison, we find
\beq
r'(k') = s^2 r(k'/s)  \label{rgforr}
\eeq
which implies that the Taylor coefficients obey
\beqr
r_{0}^{'} &=& s^2 r_0 \\
r_{2}^{'} &=& r_2 \\
r_{4}^{'} &=& s^{-2} r_4
\eeqr
and so on. Thus we find that $r_0$ is relevant, $r_2$ is marginal and the rest
are
irrelevant. This is a concrete example of how in
the low energy physics the coupling function $r(k)$ reduces
to a few coupling constants. (In fact $r_2$ makes
no difference since it can be absorbed by field rescaling.)
In quantum field theory, where we send the cut-off to infinity, all
momenta are small compared to the cut-off and the
theory is defined by  a few coupling constants. We
shall see that the same thing will happen for the quartic
interaction: a coupling function of four different momenta will
reduce to a single coupling constant. We may understand all this
as follows. In the original Brillouin zone,
of size $1/a$, all these functions are nontrivial and we need them
in their  entirety. As we eliminate modes, we need their behavior
in a smaller and smaller  ball near the origin, see Eqn.(\ref{rgforr}).
Not surprisingly, the function is well described by
a few terms in the Taylor series. This is the picture in fixed or
"laboratory units". In the   RG  one uses sliding units that
constantly change
to keep the
 cut-off (ball size) fixed at $\Lambda$ and  the same phenomenon appears as
the rapid shrinkage of higher coefficients in the Taylor series.
 (Of course as we renormalize, we are not just rewriting the original
 coupling function in new units, the function itself changes due
 to eliminated modes.  But it is expected nonetheless
 to be smooth in $k$. This is one of the  points
 emphasized in the modern   RG  theory: elimination of modes does
 not introduce new singularites into the couplings. As we
 shall see, this is because the effect of mode elimination
 may be expressed  in terms of  integrals which are convergent at both ends.)

Let us now consider the quartic perturbation
\beqr
\delta S &=& - \frac{1}{2!2!} \cut \phi^{*} (\vec{k}_4)\phi^{*}
(\vec{k}_3)\phi^{} (\vec{k}_2)\phi^{} (\vec{k}_1)
u(\vec{k}_4\vec{k}_3\vec{k}_2\vec{k}_1) (2\pi )^4 \delta^{4} (\vec{k}_4
+ \vec{k}_3 - \vec{k}_2 - \vec{k}_1 )
\prod_{i=1}^{4}\frac{d^4k_i}{(2\pi )^4} \nonumber \\
    &\equiv & -\int_{\Lambda}\phi^{*} (4)\phi^{*} (3)\phi^{} (2)\phi^{} (1)
u(4321)\label{short}
\eeqr
  where the coupling function obeys the symmetry condition
  \beq
  u(4321) = u(3421) = u(4312).
  \eeq
  In other words the coupling function is invariant under the exchange of the
first two or last two arguments. Even if we started with a
function that did not
have this symmetry, the invariance of the measure and the rest of the
integrand under this symmetry would automatically project out the symmetric
part and annihilate the antisymmetric part. The factorials up front are
conventional and are put there to prevent  similar  factors from arising  in
subsequent calculations.

  The renormalization of the quartic interaction is complicated by
  the fact that
unlike the quartic perturbations, it mixes up the slow and fast modes.
Thus we
have to use the formula
  \beqr
  e^{S'(\phi_< )}  &=& e^{S_0 (\phi_<) }\left< e^{\delta S
   (\phi_< , \phi_> )}
\right>_{0>} \\
  &\equiv & e^{S_0 + \delta S'}
  \eeqr
   Next we invoke the {\em cumulant expansion} which relates the mean of
the exponential to the exponential of means:
   \beq
   \left< e^{\Omega } \right> = e^{\left[ <\Omega > + \frac{1}{2}
(<\Omega^2> - <\Omega >^2 )+ \ldots \right] }.
   \eeq
   The reader may wish to verify the correctness of this expansion to
    the order
shown. Using this expansion we find
   \beq
   \delta S' = <\delta S > + \frac{1}{2} (<\delta S^2 >  - <\delta S >^2 ) +
\ldots
   \eeq
   Since $\delta S$ is linear in $u$, this a weak coupling expansion.
   It is now clear what has to be done. Each term in the series contains
   some monomials in fast and slow modes. The former  have to be averaged
   with respect to the Boltzmann weight $S_0 (\phi_> )$ by the use of Wick's
theorem. The result of each integration will be to give a monomial
 in the slow
modes. When reexpressed in terms of the rescaled fields and momenta, each
will renormalize the corresponding coupling.
   In principle the reader has been given information to carry
   out this process. There is however no need to reinvent the
   wheel. There is a procedure involving Feynman diagrams which
   automates this process. These rules will not be discussed here
   since they may be found, for example in Sections 3-5 of Wilson
   and Kogut (1974) or in any good field theory book
   (Itzykson and Zuber 1980, Zinn-Justin 1989).
   Instead we will go over just the first term
   in the series in some detail and comment on some aspects of the
   second term. Readers familiar with  Feynman diagrams should
   note that while these diagrams have the same multiplicity and
   topology as the  field theory diagrams, the momenta being integrated
   out are limited to the  shell being eliminated, i.e.,
   $\Lambda /s < k < \Lambda $.

   The leading term has the form
   \beq
   <\delta S > = - \frac{1}{2!2!} \left< \cut (\phi_> + \phi_<)^{*}_{4}
   (\phi_> + \phi_<)^{*}_{3}(\phi_> + \phi_<)^{}_{2}(\phi_> + \phi_<)_1
u(4321) \right>_{0>}.
   \eeq
   The sixteen possible monomials fall into four groups:
   \begin{itemize}
      \item  8 terms with an odd number of fast  fields.
       \item  1 term with all fast modes.
   \item  1 term with all slow modes.
      \item  6 terms with two slow and two fast modes.
   \end{itemize}
   We have no interest in the first two items: the first  since it vanishes
   by
symmetry and the second
    since  it makes a constant contribution, independent of  $\phi_<$
   to the effective action.
   Consider next the third term with all slow modes,  distinguished by the
   fact that it requires no integration (or averaging) over fast modes.
   This is called the {\em tree level term} in field theory.
   The tree level term is obtained from the original perturbation by simply
   setting $\phi = \phi_<$. Rewriting it in terms of new momenta
   and fields we find it leads to the following quartic renormalized
   interaction
    \beq
   \delta S^{'}_{4,tree} =   - \frac{1}{2!2!} \cut \phi^{'*}
   (k^{'}_{4})\phi^{'*} (k^{'}_{3})\phi^{'} (k^{'}_{2})\phi^{'}
   (k^{'}_{1}) u(k^{'}_{4}/s,\ldots k^{'}_{1}/s) (2\pi )^4
   \delta^{4} (\vec{k}^{'}_{4} + \vec{k}^{'}_{3} - \vec{k}^{'}_{2}
   - \vec{k}^{'}_{1} )
\prod_{i=1}^{4}\frac{d^4k^{'}_{i}}{(2\pi )^4}.\label{bony}
\eeq
The reader should note that the field rescaling factor $s^{12}$ has been
exactly cancelled
    by rewriting the delta function and integration measure in terms of new
moments. (Note that the delta function scales oppositely to the momenta.)

   It is evident that the renormalized four point coupling is given by
   \beq
   u^{'}(k^{'}_{4},\ldots k^{'}_{1}) = u(k^{'}_{4}/s,\ldots k^{'}_{1}/s).
   \eeq
      Carrying out the Taylor expansion
      \beq u= u_0 + O(k) \eeq
      we see that the constant term is marginal
      \beq
      u^{'}_{0} = u_0
      \eeq
      and the rest are irrelevant. This is why the scalar field
 theory in four dimensions is described by a coupling constant and not a
coupling function. Hereafter we will replace the coupling function by the
coupling constant.
 The effect will be irrelevant in the   RG  sense.

 We now pass from the tree level term to the six terms which have two slow
and two fast modes in them. Of these, two with $\phi_> \phi_>$ or their
conjugates are zero.    The others clearly renormalize the quadratic
interaction:
 \beq
 \delta S^{'}_{2} (\phi_< ) = - \frac{1}{2!2!} u_0 \left< \cut (\phi^{*}_{>}
(4) \phi^{*}_{<} (3) +
 \phi^{*}_{<} (4) \phi^{*}_{>} (3) )(\phi^{}_{>} (2) \phi^{}_{<} (1) +
\phi^{}_{<} (2) \phi^{}_{>} (1)) \right>_{0>} .
 \eeq
 If we now evaluate the averages of the fast modes we will find that all four
terms give the same contribution (which takes care of the factorials in front)
and we end up with
 \beqr
 \delta S^{'}_{2} (\phi_< ) &=& - u_0 \int_{|k|<\Lambda /s} \deek
\phi^{*}_{<} (\vec{k}) \phi_< (\vec{k}) \int_{\Lambda /s}^{\Lambda}\deek
\frac{1}{k^2} \label{oneloopquad}\\
 \delta S^{'}_{2} (\phi^{'} (\vec{k}') ) &=& - u_0 s^2 \cut \deek
\phi^{'*}(\vec{k}') \phi^{'} (\vec{k}' ) \Lambda^2 \frac{1}{2} (1-
\frac{1}{s^2}) \frac{2\pi^2}{(2\pi )^4} \label{43}
 \eeqr
 where in the last step we
 have used the fact that the area of a unit sphere in
four dimensions is $2\pi^2$.

 Eqn.(\ref{43}) gives us the change  in $r_0$:
 \beq
 \delta r_0 = \frac{u_0 \Lambda^2}{16\pi^2} (s^2 -1).
 \eeq
   Let us
   agree to measure $r_0$ in units of the cut-off  squared and drop the
$\Lambda^2$ from now on.

   Notice that the quartic coupling has renormalized the quadratic coupling.
   This is more the rule than the exception. The quadratic perturbations
   were special in that they did not generate new couplings. In view of this,
   we must really study the problem in which both $r_0$ and
   $u_0$ are present from the outset. This amounts to
   replacing the propagator $1/k^2$ by $1/(k^2+r_0) $ in
   Eqn.(\ref{oneloopquad}). However  this only modifies the result
   to higher  order in the expansion in $r_0$ and $u_0$.
   The flow to this order is
   \beqr
   r^{'}_{0} &=& s^2 (r_0 + \frac{u_0}{16\pi^2 } (1- 1/s^2)) \\
   u^{'}_{0} &=& u_0.
   \eeqr
   If we take $s=1+t$, with $t$ infinitesimal, we find the differential
   equations
   \beqr
   \frac{dr_0}{dt} &=& 2r_0 + \frac{u_0}{8\pi^2}  \label{rflow1}\\
   \frac{du_0}{dt} &=& 0\label{uflow1}.
   \eeqr
This completes our analysis of the first term in the cumulant expansion. Let
us see briefly how the above results follow in the diagrammatic approach.
First
we associate with each quartic perturbation $\delta S$ a four pronged $X$ as
in
Fig.Ia. The incoming arrows correspond to $\phi$ and the outgoing ones to
$\phi^{*}$. Each prong can stand for a $\phi_<$ or a $\phi_>$. Next we do
the average
over the fast modes. The prongs corresponding to the
matching pairs that give a nonzero contribution are  joined and correspond to
the
propagator. The diagrams in Fig.I
tell us what happens. The first one corresponds to all slow modes and there is
nothing to average, i.e., no lines to join.  Figure Ib corresponds to the eight
terms with an odd number of
 fast lines. These average to zero. Figure Ic describes
the case with two fast and two slow lines with both sets coming in complex
conjugate
pairs. The two fast lines are joined by the averaging and the line joining them
is the propagator and this corresponds to the renormalization of the quadratic
term as per Eqn.(\ref{oneloopquad}). This is called the {\em tadpole
diagram}.  Finally Fig.Id describes the case where
all lines are fast and come in pairs. We now have two propagators. We did
not consider
this above since it is a constant independent of $\phi_<$.

Notice that although all terms are of order $u_0$, they have very different
topologies.
The tree level term has no loops or sum over fast modes. Figure Ic has one
loop
and Figure Id has two loops. {\em Now the correct way to organize the
cumulant expansion is
by counting loops.} The reason is best seen in the language of
quantum field theory where the action has the prefactor $1/\hbar $
and the number of loops measures the powers of $\hbar$. In critical
phenomena
this fact becomes very clear when one works in $4-\varepsilon$ dimensions
(Wilson and Fisher 1972).  One finds then that
the loop expansion is an expansion in $\varepsilon$. The reader who wants to
know more
should consult references given at the beginning of this section.

At zero loops, or tree level, the equations are
\beqr
 r^{'}_{0}&= & s^2 r_0  \Rightarrow \frac{dr}{dt} = 2r \\
 u^{'}_{0} &=& u_0  \Rightarrow \frac{du}{dt} = 0.
 \eeqr
 Equations (\ref{rflow1}-\ref{uflow1}) are halfway between zero and one
loops: they are good to
 one loop for $r_0$ and to tree level for $u_0$.
To be consistent, we must evaluate the flow of $u_0$ to one loop also, which
means going to
second order in $u_0$ via the next term in the cumulant expansion namely
$$\frac{1}{2}\left[ <(\delta S)^2> - <(\delta S)>^2 \right] . $$
Here we draw two crosses and do the usual pairing. All diagrams in which
no line runs from one cross to the other, i.e., the disconnected diagrams may
be
dropped since they get cancelled by $-<(\delta S)>^2$. Of the rest the only
graphs
that affect $u$ are shown in Figure 2 and correspond to the following
analytical
expression:

\beqr
u'(\vec{k_4}', \ldots, \vec{k_1}')&=& u_0 - u^{2}_{0} [
\underbrace{ \int_{d\Lambda} \deek \frac{1}{k^2 |\vec{k} - \vec{k'}_3 /s +
\vec{k'}_1 /s |^2}}_{ZS graph}
+
 \underbrace{ \int_{d\Lambda} \deek \frac{1}{k^2 |\vec{k} - \vec{k'}_4 /s +
\vec{k'}_1 /s |^2}}_{ZS' graph}   \nonumber \\
 & &+
\underbrace{\frac{1}{2} \int_{d\Lambda} \deek \frac{1}{k^2 |-\vec{k} +
\vec{k'}_2 /s + \vec{k'}_1 /s |^2}}_{BCS graph}
]
\eeqr
 Several remarks are in order. First note that even though we started with a
constant $u=u_0$, the renormalized coupling has acquired momentum
dependence.   If we expand the renormalized coupling in a Taylor series,
keeping just the lowest term, we will get the renormalized $u_0$. This is
what we will do, and ignore the irrelevant higher terms in the series. This in
turn means that we can set all external momenta to zero. Before so doing,  let
us look at the three one loop  diagrams. Since we need to refer them
individually many times we need a system of nomenclature. The one  used
here is by no means standard. Consider the first diagram labeled ZS which
stands for "zero sound." In this  diagram lines labeled $1$ and $3$ meet  at a
vertex. In Fermi liquid theory a graph with the {\em same topology} occurs
and is very important when $\vec{Q} = \vec{k_1} - \vec{k_3}$ is small. The
{\em physics} of the present problem couldn't be more different: the lines
here stand for bosons and unlike in Fermi liquid theory, the internal loop
momenta are restricted to lie at the cut-off rather than take all values within
the cut-off. In the second ZS' diagram lines $1$ and $4$ meet at a vertex.
Usually when $\vec{Q}$ is small, $\vec{Q'}= \vec{k_1} - \vec{k_4}$ is
large and this diagram is not very important in Fermi liquid theory, and does
not have a name. However in problems with nesting this diagram can be
important if $\vec{Q'}$ is the nesting momentum. The BCS diagram with
lines $1$ and $2$ meeting at a vertex has the topology as one  that will
appear later in our  description of the superconducting instability.
The reader is once again cautioned that the names of these diagrams are
based solely on the topology and do not generally imply the corresponding
physics.

Readers familiar with Feynman diagrams could have easily written them
down. They must however pay attention to the symbol $\int_{d\Lambda}$
which reminds us that all internal propagator momenta (corresponding to
integrated fast modes) are summed only over the band being eliminated,
which we take to be a shell of thickness $d\Lambda$ at the cut-off.

Readers  new to the subject are strongly urged to work out the combinatorics
and derive  this result. They will then see why the factorials were included in
the definition of the perturbation and why an extra factor of $1/2$ appears in
the BCS diagrams.

 All readers should note that the one loop correction has a minus sign in front
of it, reflecting the decrease of the interaction strength as we go to the
infrared modes. (Although the one loop graphs have a positive value, they
reduce $u_0$ since the latter is defined to occur in the action with a negative
sign, see Eqn.(\ref{bony})

 Let us now set all external momenta to zero, since we are interested in just
$u_0$.  We are now assured that  if the loop momentum $\vec{k}$ lies in
$d\Lambda$, so does the other momentum which either equals $\vec{k}$ in
the ZS and ZS' diagrams, or equals $-\vec{k}$ in the BCS case. All the
integrals are now equal and we get
 \beqr
 u^{'}_{0} &=& u_0 - \frac{5u^{2}_{0}}{2} \int_{d\Lambda}\frac{ k^3 dk
d\Omega}{(2\pi )^4 k^4}\\
 \frac{du_0}{dt} &=& -\frac{5u_{0}^{2}}{16\pi^2} .\label{oneloopu}
 \eeqr
 where in the last step we have recalled
 \beq
 \frac{|d\Lambda |}{\Lambda} = dt
 \eeq
 and the area of a unit sphere in four dimensions ($2\pi^2$).

 To one loop accuracy we have the following flow:
 \beqr
 \frac{dr_0}{dt} &=& 2r_0 + au_0  \label{95}\\
 \frac{du_0}{dt} &=& -bu^{2}_{0}
 \eeqr
 where $a$ and $b$ are positive constants whose precise values we are no
longer interested in.

 We shall now analyze these equations. First observe that besides the
gaussian fixed point  at the origin, there are  no other points where both
derivatives vanish. Next, the equation for $u_0$ is readily integrated to give
 \beq
 u_0(t) = \frac{u_0(0)}{1+bu_0(0)t}.
 \eeq
 This means that if we start with a positive coupling $u_0(0)$ and
renormalize, the effective coupling renormalizes to zero as $1/t = 1/(\ln
(\Lambda_0 /\Lambda ))$. One says {\em $u_0$ is
 marginally irrelevant.} In the case of bosons a negative $u_0$ is unphysical
since the functional integral over fields will then diverge for large fields.
In
some fermion  problems one gets the same equation and negative $u_0$ is
allowed. In that case the coupling is {\em marginally relevant } and grows.
The above equation, derived for weak coupling, will soon have to be
abandoned, in contrast to the postive $u_0$ case, where it gets more and
more reliable at larger and larger $t$. Notice that the fate of marginal
couplings (unlike relevant or irrelevant couplings)
  depends on the sign.

  The statement that $u_0$ is marginally irrelevant at the gaussian
  fixed point
needs to be understood properly. {\em In particular, it does not mean that if
we add a small positive $u_0$  to the gaussian fixed point, we will
renormalize  back to the gaussian fixed point.} This is because the small
$u_0$ will generate an $r_0$ and that will quickly grow under
renormalization. What is true is that ultimately $u_0$ will decrease to zero,
but $r_0$ can be large. In fact to flow to the gaussian fixed point, we must
start with a particular combinations of $r_0$ and $u_0$ which describes the
critical surface. All this comes out of the equation (\ref{95} )
for $r_0$ which
is integrated to give
  \beq
  r_0(t) = e^{2t} \left[ r_0(0) + \int_{0}^{t} e^{-2t'} \frac{au_0(0)}{1+
bu_0(0)t}dt' \right] .
  \eeq
 Let us consider large $t$. Typically $r_0$ will flow to infinity exponentially
fast due to the exponential prefactor, unless we choose
 $r_0$ such that the object in brackets vanishes:
 \beq
 r_0 (0)+ \int_{0}^{\infty} e^{-2t'} \frac{au_0(0)}{1+ bu_0(0)t}dt' = 0
 \eeq
 which, for very small $u_0(0)$ translates to
 \beq
 r_0(0) = -\frac{au_0(0)}{2}\label{CS}
 \eeq
 which defines the critical surface  (a line in this case) in the $r_0 -u_0$
plane. Any point on this approaches the origin as follows:
 \beqr
 u_0(t) &\simeq & a/t \\
r_0(t) &\simeq &  e^{2t} \left[ - \int_{t}^{\infty} e^{-2t'}
\frac{au_0(0)}{1+
bu_0(0)t}dt' \right] \simeq - \frac{a}{2bt}\label{flow2}
\eeqr
Figure 3 depicts the state of affairs. Table II summarizes the
 results for the
gaussian fixed point and its leading perturbations.

The analysis of couplings with more powers of the fields is similar. All of
them are irrelevant even at the tree level and higher loops cannot
 change that.
For example the constant part of the sextic $(\phi^{*} \phi )^3$ coupling
falls like $1/s^2$.

 \subsection{The Field Theory Approach to the $\beta$-function}
 We just derived  the flows in the modern approach,  which is intuitively very
appealing   and consists of integrating out fast modes. We will now rederive
the one loop flow of $u_0$   the old way, where the aim is to banish all cut-
off dependence from physical quantities. The two approaches will then be
compared and contrasted. The reason we even bring up the field theory
method is that at higher loops it is more tractable than the modern approach.
In the Appendix  we perform two calculations involving  interacting fermions
for which the field theory method proves more convenient. The present
discussion  will rather succinct and readers new to diagrams will have to
work that much harder.

 Consider a field theory with two coupling constants: a mass term $r_0$,  a
quartic coupling $u_0$ and a cut-off $\Lambda.$ The physical quantity we
wish to hold fixed is $\Gamma (k_4 \ldots k_1)$, called the {\em irreducible
four point vertex or four point function.} (Arrows on vectors will be
suppressed.) It is defined as follows. Let us define the action of a massive
free field $S_0$
 \beq
 S_0 = -\int_{0}^{\Lambda}\deek   \phi^{*} (k) (k^2 + r_0 )\phi (k) .
 \eeq
 Consider now  (suppressing the momentum integration measure for variables
labeled $5-8$ in the quartic coupling)
 \beqr
 -<\phi^{*}(k_4) \phi^{*} (k_3 ) \phi (k_2) \phi (k_1)> &=& \frac{\int
[d\phi^{*} d\phi ] (-\phi^{*}(k_4) \phi^{*} (k_3 ) \phi (k_2) \phi
(k_1))e^{S_0}
   e^{-\frac{u_0}{2!2!} \int_{\Lambda }\phi^{*}  (k_8) \phi^{*}  (k_7)
\phi^{}  (k_6) \phi^{}  (k_5) }}{\int [d\phi^{*} d\phi ] e^{S_0} e^{-
\frac{u_0}{2!2!} \int_{\Lambda }\phi^{*}  (k_8) \phi^{*}  (k_7) \phi^{}
(k_6) \phi^{}  (k_5)}} \nonumber \\
 &\equiv & \frac{\left< (-\phi^{*}(k_4) \phi^{*} (k_3 ) \phi (k_2) \phi (k_1))
e^{-\frac{u_0}{2!2!} \int_{\Lambda} \phi^{*}  (k_8) \phi^{*}  (k_7) \phi^{}
(k_6) \phi^{}  (k_5)} \right>_{0} }{\left< e^{-\frac{u_0}{2!2!}
\int_{\Lambda }\phi^{*}  (k_8) \phi^{*}  (k_7) \phi^{}  (k_6) \phi^{}
(k_5)} \right>_{0} }
 \eeqr
 where in going to the last equation, we have multiplied and divided
 by the partition function with action $S_0$ and $<>_0$  stands for
 averages with respect to this measure. Notice that all momentum integrals
 go from $0$ to $\Lambda$. This is because we are not eliminating modes,
we
 are carrying out a calculation of some correlation function in a given
 theory. We now calculate the answer in a power series in $u_0$ by
expanding the exponential. We then
 throw out all disconnected diagrams (diagrams in which some lines are not
 connected to the others) and delete the four propagators that link the
 external fields (whose momenta are labeled $1-4$) to the vertices that come
from the exponential, and
 the $\delta$-function for overall momentum conservation. This defines
 $\Gamma (4321)$ which is the object we want to be cut-off independent.
 In field theory $\Gamma(4321)$ is the scattering amplitude for the
 process in
 which $1 +2 \rightarrow 3 +4$,
 and is a
 measure of the interaction between particles. In the gaussian
 model it will vanish since all diagrams will be disconnected, the disconnected
diagrams
 describing the independent propagation of noninteracting particles.

 Let us now calculate $\Gamma$ to order $u_0$.

 If we expand the exponential in the numerator to first order, we get a
connected piece in which the external fields numbered $1$ to $4$ get paired
with the quartic interaction fields numbered
 $5$ to $8$. The factorials get neutralized by the number of ways to pair and
the propagators get dropped and the net result is that to this order
 \beq
 \Gamma ( 4321) = u_0.
 \eeq
 (The denominator is set equal to unity since expanding it to order $u_0$ will
change the answer to order $u^{2}_{0}$.) The reader new to this subject is
very strongly urged to carry out the steps using Wick's theorem and paying
attention to the combinatorics.
 Since to lowest order in perturbation theory
 $\Gamma (4321) = u(4321)$, we will
 sometimes refer to $u$ as the amplitude for scattering. To be exact it is just
a
 coupling constant in the theory which equals $\Gamma$ in the weak
coupling
 limit.  It does however have the all (anti)symmetries of $\Gamma$ under the
exchange of
 momentum labels of the external  (fermions) bosons.

 The above answer for $\Gamma$ is clearly cut-off independent and one may
choose $u_0$ to match scattering experiments. Notice that it is also
independent of external momenta.

 To next order we must expand the numerator to order $u^{2}_{0}$ and the
denominator to order
 $u_0$ (since $\Gamma$ starts out at order $u_0$ in the numerator.) The
Feynman diagrams are exactly as before except for the fact that all loop
momenta go up to the cut-off. The result is
 \beqr
 \Gamma (4321) &=& u_0 - u^{2}_{0} \left[ \int_{0}^{\Lambda} \deek
\frac{1}{(k^2 +r_0) (|k + k_3 - k_1 |^2 + r_0)} +
 \int_{0}^{\Lambda} \deek \frac{1}{(k^2 +r_0)( |k + k_4 - k_1 |^2 +r_0)}
\right. \nonumber \\
 & & + \frac{1}{2}
\left. \int_{0}^{\Lambda} \deek \frac{1}{(k^2 +r_0)( |-k + k_2 + k_1 |^2
+r_0)} \right] .
 \eeqr

 Let us now demand that $\Gamma (0000)$ be independent of cut-off as the
latter goes to infinity. In this limit we find
 \beq
 \Gamma (0000) = u_0  - u^{2}_{0} \frac{5}{32\pi^2}\ln
\frac{\Lambda^2}{r_0} .\label{div}
 \eeq
 Let us now act on  both sides with $d/dt = -\Lambda d/d\Lambda $ and
demand that they vanish.
 This gives us
 \beq
 0 = \frac{du_0}{dt}  + \frac{5}{16\pi^2}u^{2}_{0}
 \eeq
 from which we find
 \beq
 \beta (u_0) =  \frac{du_0}{dt}  = - \frac{5}{16\pi^2}u^{2}_{0}
 \eeq
 Two points need clarification here. First: why did we  not take the implicit
$t$ derivative of the $u^{2}_{0}$ term in Eqn.(\ref{div})? The answer is that
 $\beta$ is of second order in the coupling and this will give a third order
term.  Next one may wonder about
 $\Gamma$ with all $k$'s not equal to zero.  Will they also be cut-off
independent if we choose $u_0(\Lambda )$ as above to make $\Gamma
(0000)$ cut-off independent? The answer is yes. If we expand the above
integrals in the external momenta, the integrals will become convergent and
cut-of independent if we send the cut-off to infinity. Thus the external
momenta must be much smaller than the cut-off for the field theroy
renormalization to work. If we want the physics to be cut-off independent for
external momenta comparable to $\Lambda$, we will need to introduce new
couplings besides $u_0$. )

 In the same way one can derive the flow for $r_0$ be demanding that the
pole in
the full propagator  (the two-point function in the theory with $u_0 \ne
0$) have a certain cut-off independent location.

 Suppose we add just the quartic coupling but no mass term $r_0$ to the
gaussian model.
 Then we will find that $\Gamma (0000)$ has an {\em infrared} logarithmic
divergence. (Send $r_0$ to zero in Eqn.(\ref{div}). )This is a physical
divergence in a massless theory, analogous the infinite cross section for
Rutherford scattering in electrodynamics. However the $\beta$-function,
which involves the derivative with respect to the upper limit of momentum
integration  is still  well defined and has the same value quoted above.
In the
modern approach, even if $r_0 = 0$, we will never see any infrared
divergence in the calculation of the $\beta$-function  since the loop
momentum  will now go from $\Lambda /s$ to $\Lambda$. This was the
meaning of the earlier statement that mode elimination does not produce
singularities in the parameters that appear in the effective action because
the
flow is given by integrals that are well behaved at both ends.

  Although the two methods gave the same answer, this is a fact that needs
some explanation since the methods are very different.  In the modern
approach  a change in cut-off is compensated by a change in an infinite
number of couplings while in the latter one tries to compensate by changing
just $r_0$ and $u_0$.  How can this be possible? The answer is that in the
field theory approach one always sends the cut-off to infinity (or equivalently
looks at correlation functions  with external momenta very small compared to
the cut-off), while in the modern approach we can ask for quantities involving
momenta right up to the cut-off. If in the modern approach we limit ourselves
to momenta much smaller than the cut-off, we could trade the complicated
hamiltonian
  for a simpler one at low momenta dominated by a few marginal and relevant
couplings.

  At a graphical level there are  differences in the range of integration in
the
loop graphs that contribute to the flow. In the modern approach we demand
that each internal line lie in the shell  of width $d\Lambda$ near the cut-off.
In the field theory approach, where we take the $\Lambda$-derivative of
momentum integrals going up to $\Lambda$, the answer is a sum of terms in
which
  one of the internal propagator momenta is at the cut-off and the others go up
to the cut-off. {\em In our flow equation for $u_0$ this  difference was
suppressed.} This  was because we argued that only the lowest term in the
Taylor series for the coupling was marginal and the rest were irrelevant,
allowing us to set all external momenta equal to zero. This meant that if one
line in the loop was at the cut-off, the other being either equal to it (ZS and
ZS') or opposite
to it (BCS) also had to be at the cut-off. {\em Thus both lines
were at the cut-off in both approaches to the flow. }
Had we been interested in the renormalization of irrelevant operators, we
would have had to consider non-zero external momenta and the two schemes
would have  yielded different answers.

Notice that the two schemes do not have to give the same flows, they just
have to give the same physics (at momenta much smaller than the cut-off).
The book-keeping can be very different. Consider a more general graph in the
field theory
approach with many internal lines and 4 external lines, so that it
contributes to the renormalization of the 4-point coupling. Some of these
internal momenta may be at the cut-off and the rest below it. Such a graph is
forbidden in the modern approach. The effect of these graphs (with slow and
fast momenta in the loops) will appear as follows in the modern approach.
First  all internal lines (propagators) with slow momenta are snipped and the
dangling lines are made  into external lines. This graph is then used   to
renormalize a higher point function with that many more external lines, say 6
lines in all  if two new external lines were produced by snipping an internal
line. Suppose  we now stop renormalizing and    compute an object, say the
4-point vertex  of slow modes,  using these couplings.  The answer will be
given as  integrals over slow momenta. The 6-point coupling that was
generated by snipping will contribute to the 4-point function when  the two
lines that
got snipped  get joined again.    In the  field theory approach such a
contribution would  already be sitting inside the effective 4-point coupling
which got renormalized by graphs with slow and fast lines in the loop.
\section{PATH INTEGRALS FOR FERMIONS}
In this section the reader is introduced to the path integral
representation of
fermion problems. Some elementary  problems involving dynamics and
thermodynamics will be first solved by operator methods and then the same
results will be rederived using  path integral methods
reviewed here. For a more detailed treatment, the reader is asked to turn to
the standard references (Berezin 1966, Itzykson and Drouffe 1989,
Schwinger 1970, Negle and Orland (1988)).
\subsection{The fermionic oscillator: dynamics and thermodynamics via
operators}
Let $\Psi$ and $\Psi^{\dagger}$ be two fermionic operators obeying {\em
anti}commutation relations:
\beqr
\{ \Psi^{\dagger} , \Psi \} &=&  \Psi^{\dagger} \Psi + \Psi \Psi^{\dagger}
=1\\
\{ \Psi , \Psi \} & =& \{ \Psi^{\dagger} , \Psi^{\dagger} \} =0.
\eeqr
Note that the last equation tells us
\beq
{\Psi^{\dagger}}^2 = \Psi^2 =0.
\eeq
This equation will be used all the time without explicit warning.
The number operator
\beq
 N = \Psi^{\dagger} \Psi
 \eeq
 obeys
 \beq
 N^2 =   \Psi^{\dagger} \Psi \Psi^{\dagger} \Psi = \Psi^{\dagger} ( 1-
\Psi^{\dagger} \Psi ) \Psi =  \Psi^{\dagger} \Psi =N.
 \eeq
 Thus the eigenvalues of $N$ can only be $0$ or $1$.
 The corresponding normalized eigenstates obey
 \beqr
  N|0> &=& 0 |0> \\
  N|1> &=& 1 |1>.
  \eeqr
  We will now prove that
  \beqr
  \Psi^{\dagger} |0> &=& |1> \\
  \Psi |1> &=& |0>.
  \eeqr
  As for the first,
  \beq
  N\Psi^{\dagger} |0> =  \Psi^{\dagger} \Psi \Psi^{\dagger} |0> =
\Psi^{\dagger} (1- \Psi^{\dagger} \Psi )|0> = \Psi^{\dagger} |0> ,
  \eeq
  which shows that $\Psi^{\dagger} |0>$ has $N=1$. Its norm is unity:
  \beq
  || \Psi^{\dagger} |0> ||^2 = <0|\Psi \Psi^{\dagger} |0>=
  <0|(1- \Psi^{\dagger}
\Psi)|0> = <0|0> =1.
  \eeq
  It can be similarly shown that
  $\Psi |1> = |0>$ after first verifying that $ \Psi
|1>$ is not a null vector, that it has unit norm.

  There are no other vectors in the Hilbert space: any attempts to produce
more states are thwarted by $\Psi^2 = {\Psi^{\dagger}}^2 =0$.

  Consider now a Fermi oscillator with hamiltonian
  \beq
  H_0 = \Omega_0  \Psi^{\dagger} \Psi
  \eeq
  whose eigenvalues are clearly $0$ and $\Omega_0$.

  We will work not with $H_0$ but with
 \beq
 H = H_0 - \mu N
 \eeq
 where $\mu$ is the {\em chemical potential.} For the oscillator, since
 \beq
 H = (\Omega_0 - \mu ) \Psi^{\dagger} \Psi
 \eeq
  this merely amounts to measuring all energies relative to the chemical
potential.
  \footnote{The eigenvalues of $H$ are $T=0$ free energies rather than
energies. We shall however often refer to $H$ as the hamiltonian.}

  Let us now turn to thermodynamics. The grand partition function is
  \beq
  Z = Tr e^{-\beta (H_0 - \mu N)} = e^{-\beta A(\mu , \beta )}
  \eeq
  where the trace is over any complete set of eigenstates, $\beta$ is the
inverse temperature $1/T$ and $A$ is the free energy. The latter is clearly a
function of $\mu$ and $\beta$ and its differential is
  \beq
  dA = - <N> d\mu - S dT
  \eeq
  where $S$ is the entropy and $<N>$ stands for the average particle number.
Let us verify that
  \beq
  <N>  = - \frac{\partial A}{\partial \mu}
  \eeq
  as follows:
  \beqr
- \frac{\partial A}{\partial \mu} &=& \frac{\partial}{\partial \mu} \frac{\ln
Z}{\beta}\\
&=& \frac{1}{\beta} \frac{Tr \beta N e^{-\beta (H_0 - \mu N)} }{Z} \\
&=&\frac{1}{\beta} <\beta N> = <N>.
\eeqr
The free energy $A$ is the double Legendre transform of the internal energy
$E(S,<N>)$:
\beq
A(\mu , \beta ) = E(S,<N>) - ST - \mu <N>.
\eeq
Thus $E$ must equal $<H_0>$. This is indeed so:
\beqr
E &=& A - T \frac{\partial A}{\partial T} - \mu \frac{\partial A}{\partial
\mu}\\
&=& A + \beta \frac{\partial A}{\partial \beta} - \mu  \frac{\partial
A}{\partial \mu}\\
&=& \frac{\partial A\beta}{\partial \beta} - \mu  \frac{\partial A}{\partial
\mu}\\
&=& \frac{\partial (-\ln Z)}{\partial \beta} +\frac{ \mu }
{\beta} \frac{\partial
\ln Z}{\partial \mu}\\
&=& <H_0>.
\eeqr
where the steps leading to the last line are left to the reader.
 It also follows
from the definition of the Legendre transform that
\beq
\mu = \frac{\partial E}{\partial <N>}
\eeq
so that $\mu$ is the (minimum) energy needed to add an extra particle.

The partition function of the Fermi oscillator is easily found (by doing the
trace over eigenstates of $N$) to be
\beq
Z = 1 + e^{-\beta (\Omega_0 - \mu )}
\eeq
from which it follows that
\beq
A = -\frac{1}{\beta} \ln (1 + e^{-\beta (\Omega_0 -\mu )}),
\eeq
which in turn implies
\beq
<N> = \frac{1}{1+ e^{\beta (\Omega_0 -\mu )}}.
\eeq
We shall be interested in the limit  $\beta \rightarrow \infty$ in which case
\beq
<N> = \theta (\mu - \Omega_0 )
\eeq
which means the fermion is present if its energy is negative (relative to the
chemical potential) and absent if it is positive.
This is to be expected since at
$T=0$, $A= E-\mu <N>$ and minimizing the free energy is the same as
minimizing $<H>$.

We now consider the dynamics.  From the Schr\"{o}dinger operators we can
form Heisenberg operators:
  \beqr
  \Psi (t) &=& e^{iHt} \Psi (0) e^{-iHt} =\Psi (0) e^{-i(\Omega_0 -\mu )t}\\
  \Psi^{\dagger} (t) &=& e^{iHt} \Psi^{\dagger} (0) e^{-iHt} =
\Psi^{\dagger} (0) e^{i(\Omega_0 - \mu )t}.
  \eeqr
  We will study {\em imaginary time quantum  mechanics} for which the time
evolution operator is
  \beq
  U(\tau ) = e^{-H\tau}
  \eeq
  and in which
  \beqr
  \Psi (\tau ) &=& \Psi (0) e^{-(\Omega_0 - \mu) \tau} \\
  \Psi^{\dagger} (\tau) &=& \Psi^{\dagger} (0) e^{(\Omega_0 - \mu )\tau} .
  \eeqr
  Note that despite
  the notation, $\Psi (\tau )$ and $\Psi^{\dagger} (\tau )$ are
are not adjoints except
  at $\tau =0$ owing to the fact that $U(\tau )$ is not unitary.

 Readers  not familiar with imaginary time quantum mechanics merely have to
observe how  the functional formalism
 reproduces the results of the operator formalism. They may also wish to learn
about imaginary time quantum mechanics  using this simple example.

 Next consider the {\em time-ordering symbol T} whose action on a pair of
fermionic  Heisenberg operators is:
 \beqr
 T(\Psi (\tau )
 \Psi^{\dagger} (0) ) &=& \Psi (\tau ) \Psi^{\dagger} (0) \ \ \ \tau
>0\\
 &=& -\Psi^{\dagger} (0) \Psi (\tau ) \ \ \ \tau <0.
 \eeqr
 Note that
 \beq
 N = \lim_{\tau \rightarrow 0^{-}} -T(\Psi (\tau ) \Psi^{\dagger} (0) )
\label{NfromT}.
 \eeq
 In field theory and many-body physics one is interested in the Green's
function:
 \beq
 G(\tau ) = <T(\Psi (\tau ) \Psi^{\dagger} (0) ) >
 \eeq
 where $<>$ denotes the average with respect to $Z$. For our problem we
find
 \beqr
 G(\tau ) &=&
 \frac{<0|T(\Psi (\tau ) \Psi^{\dagger} (0) ) |0> + <1|T(\Psi (\tau
) \Psi^{\dagger} (0) ) |1>e^{-\beta (\Omega_0 -\mu )}}{1 + e^{-\beta
(\Omega_0 - \mu )}}\\
 &=& \frac{\theta (\tau )e^{- (\Omega_0 -\mu )\tau } - \theta (-\tau ) e^{-
(\Omega_0 -\mu )(\tau + \beta )}}{1 + e^{-\beta (\Omega_0 - \mu )}}
 \eeqr
 In the zero temperature limit this reduces to
 \beqr
 G(\tau ) & =& \theta (\tau )e^{- (\Omega_0 -\mu )\tau }\ \  \ \mu < \Omega_0
\\
 &=& -\theta (-\tau )e^{- (\Omega_0 -\mu )\tau } \ \ \ \mu > \Omega_0 .
 \eeqr
 Let us define the pair of transforms:
 \beqr
 G(\omega ) &=& \int_{-\infty}^{\infty} G(\tau ) e^{i\omega \tau } d\tau ,\\
 G(\tau ) &=&  \int_{-\infty}^{\infty} G(\omega ) e^{-i\omega \tau }
\frac{d\omega}{2\pi}.
 \eeqr
 We find that
 \beq
 G(\omega ) = \frac{1}{\Omega_0 - \mu - i\omega }
 \eeq
 independent of which of $\Omega_0$ or $\mu$ is greater.

 Let us calculate $<N>$ using Eqn.(\ref{NfromT}) and the above Green's
function:
 \beqr
 <N> &=& -G(0^{-}) \\
 &=& \intom \frac{e^{i\omega 0^{+}}}{i\omega - \mu -\Omega_0}\\
 &=& \theta (\mu - \Omega_0 )
 \eeqr
 where in the last step we have argued that unless the $\theta$-function is
satisfied, the contour must  be closed in the upper half plane (as dictated by
the exponential) will be  free of singularities.

 Consider finally  a toy "Hubbard model"  with two fermions and a repulsive
interaction $U$:
 \beqr
	H_0 &=& \Omega_0 (\Psi^{\dagger}_{1} \Psi_{1} +
\Psi^{\dagger}_{2} \Psi_{2}) + U \Psi^{\dagger}_{1}
\Psi_{1}\Psi^{\dagger}_{2} \Psi_{2}\\
	&=& \Omega_0 N + \frac{U}{2} N(N-1)
	\eeqr
	where
	\beq
	N = N_1 + N_2 = \Psi^{\dagger}_{1} \Psi_{1} + \Psi^{\dagger}_{2}
\Psi_{2}.
	\eeq
(Readers wishing to fill in the missing steps should note that they have to use
$N_{1}^{2} = N_1$ and likewise for $N_2$. In any event they should check
the correctness of the final result for various choices of $N_1$ and $N_2$.)
Each  fermion has the usual anticommutator of unity  with its adjoint and
anticommutes with everything else including all member of the other set.

By summing over the eigenstates of $N_1$ and $N_2$,
\beq
Z = 1 + 2 e^{-\beta (\Omega_0 - \mu )}+ e^{-\beta (2(\Omega_0 - \mu )
+U))}
\eeq
where the factor of $2$ in the middle term reflects  the degeneracy of the one
fermion states. From the above,  we obtain by differentiation
\beqr
<N> &=& \lim_{\beta \rightarrow \infty} \frac{2(1 + e^{\beta (\mu -
\Omega_0 - U)})}{e^{-\beta (\mu - \Omega_0 )} + 2 + e^{\beta (\mu -
\Omega_0 - U)}}\\
&=& 0 \ \ \mu < \Omega_0 \\
&=& 1 \ \ \Omega_0 < \mu < \Omega_0 + U\\
&=& 2 \ \ \mu > \Omega_0 +U.
\eeqr
 Table III summarizes  results from this subsection.  In the next subsection
they will  be rederived using path integrals.

 \subsection{Fermion Coherent States}
 In this section we will be using {\em Grassmann numbers.} Here are the
rules for manipulating them:
 \begin{itemize}
 \item All Grassmann numbers anticommute with each other and with all
fermionic operators.
 \item As a result of the above, the square of any Grassmann number is zero
and the product of an even number of Grassmann numbers will commute with
anything. Likewise any Grassmann number will commute with an even
number of fermion operators such as $N=\Psi^{\dagger} \Psi$. When a
Grassmann number is  taken through a ket or bra containing an even (odd)
number of fermions it will not (will) change sign.
 \item Do not associate a numerical value to Grassmann numbers. There are
no big or small Grassmann numbers. All you will need are the above
definitions.
 \end{itemize}
 Consider  the state
 \beq
 |\psi > = |0> - \psi |1>.
\eeq
where $\psi$ is a Grassmann number.
This state, called a {\em fermion coherent state} is  an eigenstate of
  $\Psi$ with eigenvalue $\psi$:
   \beq
 \Psi |\psi> = \psi |\psi >\label{fermco}
 \eeq
 as is readily verified:
 \beqr
 \Psi |\psi> &=& \Psi |0> - \Psi \psi |1>\\
		  &=& 0 + \psi \Psi |1> \\
		   &=& \psi |0> \\
		    &=& \psi (|0> - \psi |1>)
		    \eeqr
		    where we have appealed to the fact that $\psi$ anticommutes
with $\Psi$ and that $\psi^2 =0$.
If we act on both sides of Eqn.(\ref{fermco}) with $\Psi$, the left vanishes
due to $\Psi^2 = 0$ and the right due to
$\psi^2 =0$.

		    It may be similarly verified that
		    \beq
		    <\overline{\psi} |\Psi^{\dagger} = <\overline{\psi} |
\overline{\psi}
		    \eeq
		    where
		    \beq
		    <\overline{\psi} | = <0| - <1|\overline{\psi}  = <0| +
\overline{\psi} <1|.
		    \eeq
 Please note two points. First, the coherent state vectors are not the usual
from a complex vector space  since they are linear combinations  with
Grassmann coefficients. Second,  $\overline{\psi}$ is not in any sense the
complex conjugate of $\psi$ and $<\overline{\psi} |$ is not the adjoint of
$|\psi >$. You should therefore be prepared to see a change of Grassmann
variables in which $\psi $ and  $\overline{\psi}$ undergo  totally unrelated
transformations.

 The inner product of two coherent states is
 \beqr
 <\overline{\psi} |\psi > &=& (<0| - <1|\overline{\psi} )(|0> - \psi |1>)\\
 &=& <0|0> + <1|\overline{\psi} \psi |1>\\
 &=& 1 + \overline{\psi} \psi \\
 &=& e^{\overline{\psi} {\psi} }.\label{cohprod}
 \eeqr
 Any function of a Grassmann variable can be expanded as follows:
 \beq
 F(\psi ) = F_0 + F_1 \psi
 \eeq
 there being no higher powers possible.

 We will now define integrals over Grassmann numbers. These have no
geometric significance and are formally defined. We just have to know how
to integrate $1$ and $\psi$ since that takes care of all possible functions.
Here is the list of integrals:
 \beqr
 \int \psi d\psi &=& 1 \\
 \int 1 d\psi &=& 0.
 \eeqr
 That's it! As you can see, selling tables of Grassmann integrals is no way to
make a living.
 There are no limits on these integrals. Integration is assumed to be a linear
operation. The differential $d\psi$ is also a Grassmann number. Thus $\int
d\psi \psi =-1$. The integrals for $\overline{\psi}$ or any other Grassmann
variable are identical.  A result we will use often is this:
 \beq
 \int \overline{\psi} \psi d\psi d\overline{\psi} = 1.
 \eeq
 Note that if
 the differentials or variables come in any other order there can be
a change of sign.
 For example we will also invoke the result
 \beq
 \int\overline{\psi} \psi d\overline{\psi} d\psi = -1.
 \eeq
 Let us now consider  some gaussian integrals. You are urged to show the
following:
 \beqr
 \int e^{-a\overline{\psi} {\psi}} d\overline{\psi} d\psi  &=& a
\label{gaussave}\\
 \int e^{-\overline{\psi} M\psi} [d\overline{\psi} d\psi] &=& det M
 \eeqr
 where in the second formula $M$ is a 2-by-2 matrix, $\psi$ is a column
vector with entries $\psi_1$ and $\psi_2$,  $\overline{\psi}$ a column vector
with entries $\overline{\psi}_1$ and $\overline{\psi}_2$ and $
[d\overline{\psi} d\psi] = d\overline{\psi}_1 d\psi_1 d\overline{\psi}_2
d\psi_2. $This result is true for matrices of any size.
 To prove these simply expand the exponential and do the integrals.

 Consider next the "averages" over the gaussian measure:
 \beq
 <\overline{\psi} \psi > = \frac{\int \overline{\psi} \psi e^{a\overline{\psi}
\psi} d\overline{\psi} d\psi}{\int e^{a\overline{\psi} \psi} d\overline{\psi}
d\psi} = \frac{1}{a} =-<\psi \overline{\psi} >.
 \eeq
 The proof is straightforward and left as an exercise.

  Consider now two sets of Grassmann variables (labeled 1 and 2). The reader
should verify that
  \beqr
  <\overline{\psi}_i \psi_j  > &=& \frac{\int \overline{\psi}_i \psi_j \psi_k
\psi_le^{a_1\overline{\psi}_1 \psi_1 + a_2\overline{\psi}_2 \psi_2}
d\overline{\psi}_1 d\psi_1 d\overline{\psi}_2 d\psi_2}{\int
e^{a_1\overline{\psi}_1 \psi_1 + a_2\overline{\psi}_2 \psi_2}
d\overline{\psi}_1 d\psi_1 d\overline{\psi}_2 d\psi_2}\\
 &=& \frac{\delta_{il}}{a_i}
  \equiv <\overline{i}l>.\label{fermiwick0}
  \eeqr

  We now have a Wick's theorem for fermions:
  \beqr
  <\overline{\psi}_i \overline{\psi}_j \psi_k \psi_l > &=& \frac{\int
\overline{\psi}_i \overline{\psi}_j \psi_k \psi_le^{a_1\overline{\psi}_1 \psi_1
+ a_2\overline{\psi}_2 \psi_2} d\overline{\psi}_1 d\psi_1 d\overline{\psi}_2
d\psi_2}{\int e^{a_1\overline{\psi}_1 \psi_1 + a_2\overline{\psi}_2 \psi_2}
d\overline{\psi}_1 d\psi_1 d\overline{\psi}_2 d\psi_2}\\
 &=& \frac{\delta_{il}}{a_i} \frac{\delta_{jk}}{a_j} -
\frac{\delta_{ik}}{a_i}\frac{\delta_{jl}}{a_i}\\
  &\equiv& <\overline{i}l><\overline{j}k>-
<\overline{i}k><\overline{j}l>.\label{fermiwick}
  \eeqr
  The reader not familiar with such objects is urged strongly to prove  this
simple case of Wick's theorem for  fermions. Note the strong similarities to
the bosonic case. Once again we find that the answer is zero unless each
Grassmann is accompanied by its partner.  The answer is once again a sum
over all possible parings. The only difference comes from the minus signs
which are determined as follows.   We first move each variable till it is next
to its partner. In the example above,  if $j=k$ and $i=l$, the middle two
Grassmanns are already next to each other and the ones at the ends can be
brought together without any minus signs since they are separated by a pair of
Grassmanns. This is why the first term is positive in Eqn.(\ref{fermiwick}).
On the other hand if $i=k$ and $j=l$, we must move  $j$ through $k$ to meet
its mate and this produces a minus sign.  When more than four variables are
averaged,    an obvious generalization holds: pair the fields in all possible
ways and put in a minus sign for every time a variable crosses another.
Although we did not see it here, the following thing can happen and does
happen in a calculation that comes later in this section: the variable and its
partner are next to each other, but in the wrong order, with $\psi$ to the left
of $\overline{\psi}$. In this case an extra minus sign is needed to rearrange
these.

Finally note that Jacobians behave counterintuitively for Grassmann variables.
Consider
\beq
\int a \psi d\psi =a.
\eeq
In terms of
\beqr
a\psi &=& \chi , \\
\int \chi J(\psi /\chi ) d\chi &=& a.
\eeqr
Assuming $J$ is a constant we pull it out of the integral (with no minus signs
since it involves an even number of Grassmann variables) , use the fact that
the integral of $\chi$ is unity to obtain
\beq
J(\psi /\chi ) = a
\eeq
while one might have expected the inverse.

   As an application of this resut, the reader may wish to rederive
Eqn.(\ref{gaussave})  by making the change of variables from $\psi$ to $\chi
= a \psi$. {\em Note that there is no need to transform $\sib$ at the same
time.}

We need two more results before we can write down the path integral. The
first is the resolution of the identity:
\beq
I = \int |\psi ><\overline{\psi} |
e^{-\overline{\psi} \psi}d\overline{\psi} d\psi
 \eeq
 In the following proof of this result we will use all the previously described
properties and drop terms that are not going to survive integration. (Recall
that only $\overline{\psi} \psi = - \psi \overline{\psi}$ has a non-zero
integral.)
 \beqr
 \int |\psi ><\overline{\psi}
 | e^{-\overline{\psi} \psi}d\overline{\psi} d\psi
&=& \int |\psi ><\overline{\psi}
| ( 1- \overline{\psi} \psi ) d\overline{\psi}
d\psi \\
 &=& \int (|0> - \psi |1>)(<0| -
 <1|\overline{\psi} ) (1- \overline{\psi} \psi )
d\overline{\psi} d\psi \\
 &=& \int ( |0><0| + \psi|1><1|\overline{\psi} ) (1- \overline{\psi} \psi )
d\overline{\psi} d\psi \\
 &=& |0><0|\int (-\overline{\psi} \psi )d\overline{\psi} d\psi + |1><1|\int
\psi\overline{\psi} d\overline{\psi} d\psi \\
 &=& I.
 \eeqr
 The final result we need is that for any bosonic operator (an operator made
of an even number of Fermi operators)
 \beq
 Tr \Omega = \int <-\overline{\psi} | \Omega | \psi > e^{-\overline{\psi}
\psi}d\overline{\psi} d\psi .
 \eeq
 The proof is very much like the one just given and is left to the reader.

 \subsection{The fermionic path integral}

 Consider the partition function for a single oscillator:
 \beqr
 Z &=& Tr e^{-\beta (\Omega_0 - \mu )\Psi^{\dagger} \Psi}\\
 &=&  \int <-\overline{\psi} |  e^{-\beta (\Omega_0 - \mu )\Psi^{\dagger}
\Psi}| \psi > e^{-\overline{\psi} \psi}d\overline{\psi} d\psi .
 \eeqr
 You cannot simply replace  $\Psi^{\dagger}$ and $\Psi$ by $-
\overline{\psi}$ and $\psi$ respectively in the exponential. This is because
when we expand out the exponential not all the $\Psi$'s will be acting to the
right on their eigenstates and neither will all $\Psi^{\dagger}$'s be acting to
the left on their eigenstates. (Remember that we are now dealing with
operators not Grassmann numbers. The exponential will have an infinite
number of terms in its expansion.) We need to convert the exponential to its
{\em normal ordered form} in which all the creation operators stand to the
left and all the destruction operators to the right. Luckily we can write down
the answer by inspection:
 \beq
 e^{-\beta (\Omega_0 - \mu )\Psi^{\dagger} \Psi} = 1 + (e^{-\beta (\Omega_0
- \mu )}-1)\Psi^{\dagger} \Psi
 \eeq
 whose correctness we can verify by considering the two possible values of
$\Psi^{\dagger} \Psi$. (Alternatively you can expand the exponential and use
the fact that $N^k = N$ for any nonzero $k$.) Now we may write
 \beqr
 Z &=&\int <-\overline{\psi} | 1 + (e^{-\beta (\Omega_0 - \mu )}-
1)\Psi^{\dagger} \Psi| \psi >
e^{-\overline{\psi} \psi}d\overline{\psi} d\psi  \\
 &=& \int <-\overline{\psi} | \psi> (1 + (e^{-\beta (\Omega_0 - \mu )}-1)(-
\overline{\psi} \psi ))e^{-\overline{\psi} \psi}d\overline{\psi} d\psi \\
 &=& \int (1 - (e^{-\beta (\Omega_0 - \mu )}-1)\overline{\psi} \psi ) e^{-
2\overline{\psi} \psi}d\overline{\psi} d\psi \\
 &=& 1+ e^{-\beta (\Omega_0 - \mu )}
 \eeqr
 as expected. While this is the right answer, this is not the path integral
approach. As for the latter the procedure is this. Consider
 \beq
 Z = Tr e^{-\beta H}
 \eeq
 where $H$ is a normal ordered operator $H(\Sid , \Psi )$. We write the
exponential as follows:
 \beqr
  e^{-\beta H} &=& \lim_{N \rightarrow \infty}(e^{-\frac{\beta}{N}
H})^{N}\\
  &=& \underbrace{(1-\varepsilon H)\ldots (1-\varepsilon H)}_{N \ times}\ \ \
\ \ \ \varepsilon = \beta /N
  \eeqr
 and introduce the resolution of the identity $N-1$ times:
 \beqr
 Z&=& \int <-\overline{\psi}_1|(1-\varepsilon H)|\psi_{N-1}>e^{-
\overline{\psi}_{N-1} \psi_{N-1}}<\overline{\psi}_{N-1}|(1-\varepsilon
H)|\psi_{N-2}>e^{-\overline{\psi}_{N-2} \psi_{N-2}}<\psi_{N-2}|
\nonumber \\
& & \ldots|\psi_2>e^{-\overline{\psi}_2 \psi_2}<\overline{\psi}_2|(1-
\varepsilon H)|\psi_1>e^{-\overline{\psi}_1 \psi_1}\prod_{i=1}^{N-
1}d\overline{\psi}_i d\psi_i .  \label{bo}
\eeqr
Now we may legitimately make the replacement
\beq
<\overline{\psi}_{i+1}|1 - \varepsilon H(\Sid , \Psi )|\psi_{i}> =
<\overline{\psi}_{i+1}|1 - \varepsilon H(\overline{\psi}_{i+1} , \psi_{i}
)|\psi_{i}> = e^{\overline{\psi}_{i+1}\psi_i}     e^{- \varepsilon
H(\overline{\psi}_{i+1} , \psi_{i} ) }
\eeq
where in the last step we are anticipating the limit of infinitesimal
$\varepsilon$. Let us now {\em define} an  additional pair of variables (not to
be integrated over)
\beqr
\overline{\psi}_{N} &=& -\overline{\psi}_1 \label{antisymm}\\
\psi_{N} & =& -\psi_1 .\label{antisymm1}
\eeqr
The first of these equations allows us to replace the left-most bra in
Eqn.(\ref{bo}),  $<-\overline{\psi}_1 |$, by $<\overline{\psi}_N |$. The
reason for introducing $\psi_N$ will follow soon.

Putting together all the factors (including the overlap of coherent states) we
end up with
\beqr
Z &=& \int \prod_{i=1}^{N-1}e^{\overline{\psi}_{i+1} \psi_{i}}e^{-
\varepsilon H(\overline{\psi}_{i+1} , \psi_{i} )}e^{-\overline{\psi}_{i}
\psi_{i}} d\overline{\psi}_i d\psi_i
\\
&=&  \int \prod_{i=1}^{N-1}e^{ \left[( \frac{(\overline{\psi}_{i+1} -
\overline{\psi}_{i})}{\varepsilon} \psi_{i} - H(\overline{\psi}_{i+1} ,
\psi_{i} )\right]\varepsilon } d\overline{\psi}_i d\psi_i \\
&\simeq & \int e^{\int_{0}^{\beta}\overline{\psi}(\tau ) ( -
\frac{\partial}{\partial \tau} -\Omega_0 + \mu )\psi (\tau ) d\tau}
[d\overline{\psi} d\psi]
\eeqr
where the last step need some explanation. With all the factors of
$\varepsilon $ in place we do seem to get the continuum expression in the last
formula. However the notion of replacing differences by derivatives is purely
symbolic for Grassmann variables. There is no sense in which
$\overline{\psi}_{i+1} - \overline{\psi}_{i}$ is small, in fact the objects
have
no numerical values. What this really means here is the following. In a while
we will trade $\psi (\tau )$ for $\psi (\omega )$ related by Fourier
transformation.
At that stage we will replace $ - \frac{\partial}{\partial \tau}$
by $i\omega$ while the exact answer is  $e^{i\omega}-1$. If we do not make
this replacement, the Grassmann  integral, when evaluated in terms of
ordinary numbers, will give exact results for anything one wants to calculate,
say the free energy. With this approximation, only quantities insensitive to
high frequencies will be given correctly. The free energy will come out wrong
but the
correlation functions will be correctly reproduced. (This is because the
latter are given by derivatives of the free energy and these derivatives make
the integrals sufficiently insensitive to high frequencies.) Notice also that
we
are replacing
$H(\overline{\psi}_{i+1} , \psi_{i} ) = H(\overline{\psi} (\tau +
\varepsilon ) ,
\psi (\tau ) )$ by $H(\overline{\psi} (\tau  ), \psi (\tau ) )$ in the
same spirit.

 Now turn to the Fourier expansions alluded to above. Let us write
 \beqr
 \overline{\psi}(\tau ) &=& \sum_{n} \frac{e^{i\omega_n \tau }}{\beta}
\overline{\psi} (\omega ) \\
 \psi(\tau ) &=& \sum_{n} \frac{e^{-i\omega_n \tau }}{\beta} \psi (\omega )
\eeqr
where the allowed frequencies are chosen to satisfy the antisymmetric
boundary conditions in Eqn.(\ref{antisymm} -\ref{antisymm1}). Thus
\beq
\omega_n = \frac{(2n +1)\pi}{\beta}\label{matsubara}
\eeq
where $n$ is an integer. Note that we have chosen the Fourier expansions as
if $\psi$ and $\overline{\psi}$ were complex conjugates, which  they are not.
This choice however makes the calculations easy.

The inverse   transformations are
\beqr
\psi (\omega ) &=& \int_{0}^{\beta} \psi (\tau ) e^{i\omega_n \tau }d\tau \\
\overline{\psi} (\omega ) &=& \int_{0}^{\beta} \overline{\psi} (\tau ) e^{-
i\omega_n \tau }d\tau,
 \eeqr
 where we use the orthogonality property
 \beq
 \int_{0}^{\beta} e^{i\omega_n \tau }e^{-i\omega_m \tau }d\tau =
\frac{e^{i(\omega_n - \omega_m ) \beta}-1}{i(\omega_n - \omega_m )} =
\beta \delta_{mn}.
 \eeq
Performing the Fourier transforms in the action and changing the functional
integration variables to $\psi (\omega )$ and $\overline{\psi} (\omega ) $ (the
Jacobian is unity)
and going to the limit $\beta \rightarrow \infty $, which
converts sums over discrete frequencies to integrals  over a continuous
$\omega$, we end up with
\beq
Z = \int e^{\intom \overline{\psi} (\omega ) ( i\omega - \Omega_0 + \mu )
\psi (\omega ) }[d\overline{\psi} (\omega )d\psi (\omega ) ]
\eeq
{\em Although $\beta$ has disappeared from the picture it will appear as $2
\pi \delta (0)$, which we know stands for the total time.} (Recall Fermi's
Golden Rule calculations.)
An example will follow shortly.

Let us first note that just as in the case of the scalar gaussian model, the
correlation function is related to the integral over just a single pair of
variables (Eqn.(\ref{gaussave})) and is given by:
\beqr
  <\overline{\psi} (\omega_1 ) \psi(\omega_2 ) > &=& \frac{ \int
\overline{\psi} (\omega_1 ) \psi(\omega_2 )e^{\intom \overline{\psi} (\omega
) ( i\omega - \Omega_0 + \mu ) \psi (\omega ) }[d\overline{\psi} (\omega
)d\psi (\omega ) ]}{\int d\overline{\psi} (\omega ) d\psi(\omega )
 e^{\intom \overline{\psi} (\omega ) ( i\omega - \Omega_0 + \mu ) \psi
(\omega ) } }\\
  &=& \frac{2\pi \delta (\omega_1 - \omega_2 )}{i\omega_1 - \Omega_0 +
\mu }
  \eeqr
  In particular
  \beq
  <\overline{\psi} (\omega ) \psi (\omega ) >= \frac{2\pi \delta (0)}{i\omega -
\Omega_0 + \mu } =\frac{\beta}{i\omega - \Omega_0 + \mu }
\eeq
Let us now calculate the mean occupation number $<N>$:
\beqr
<N> &=& \frac{1}{\beta Z} \frac{\partial Z}{\partial \mu}\\
&=& \frac{1}{\beta} \intom <\overline{\psi} (\omega ) \psi (\omega ) >\\
&=& \intom \frac{e^{i\omega o^{+}}}{i\omega - \Omega_0 + \mu } \\
&=& \theta (\mu - \Omega_0 )
\eeqr
as in the operator approach. Notice that we had to introduce the factor
$e^{i\omega o^{+}}$ into the $\omega$ integral. We understand this as
follows. If we had done the calculation using time $\tau$ instead of frequency
$\omega$, we would have calculated the average of $\Sid  \Psi $. This would
automatically
have turned into $\overline{\psi} (\tau + \varepsilon ) \psi (\tau
)$ when introduced into the path integral since the coherent state bra to the
left of the operator would have come from the next time slice compared to the
ket at the right. (Remember how $ H(\Sid , \Psi )$ turned into
$H(\overline{\psi} (i +1) \psi (i))$.) Notice that the integral over $\omega$
was not convergent,  varying  as $d\omega /\omega$. It was therefore
sensitive to the high frequencies and we had to intervene with the factor
$e^{i\omega o^{+}}$ . Later we will deal with  integrals that have two or
more powers of $\omega$ in the denominator and are hence  convergent.
We will not introduce  this factor in those cases.

Our final calculation will be the determination of $<N>$ for the toy Hubbard
model using path integrals.
The partition function is
\beqr
Z &=&  \int [d\overline{\psi}_1 d\psi_1 d\overline{\psi}_2
d\psi_2]e^{S_0}e^{U\int \overline{\psi}_1 \overline{\psi}_2 \psi_1 \psi_2}\
where \\
S_0 &=& \intom \sum_{i=1}^{2}\overline{\psi}_{i} (\omega )(i\omega -
\Omega_0 + \mu ) \psi_i (\omega )  ,\\
U\int \overline{\psi}_1 \overline{\psi}_2 \psi_1 \psi_2&=& \int_{-
\infty}^{\infty} \prod_{i=1}^{4}\frac{d\omega_i}{2\pi}
\overline{\psi}_1 (\omega_4 ) \overline{\psi}_2 (\omega_3 ) \psi_1
(\omega_2 ) \psi_2 (\omega_1 ) 2\pi \delta (\omega_4 + \omega_3 - \omega_2
- \omega_1 ).
\eeqr
Note that the hamiltonian already was in normal ordered form: each creation
operator was to the left of {\em its} destruction operator. This allowed us to
replace the operators by their eigenvalues when the coherent states were
introduced. For notational uniformity we have further arranged to have all
creation operators to the left of all destruction operators. This merely
introduces an extra minus sign here since operators corresponding to different
oscillators anticommute.

Prior to calculating $<N>$ let us calculate $<N_1>$. This is given by
\beq
N_1 = \intom <\overline{\psi}_1 (\omega ) \psi_1 (\omega )>  e^{i\omega
o^{+}}\label{N1}
\eeq
where $<\overline{\psi}_1 (\omega ) \psi_1 (\omega )> $ stands for the
correlation function with the full action and not just $S_0$. We may however
express it in terms of averages over the gaussian measure $S_0$ using the
same trick we used for bosons: we multiply and divide the exact $Z$ by
$Z_0$, the partition function with $U=0$ to obtain
\beq
<\overline{\psi}_1 (\omega ) \psi_1 (\omega )> = \frac{<\overline{\psi}_1
(\omega ) \psi_1 (\omega )e^{U\int \overline{\psi}_1 \overline{\psi}_2 \psi_1
\psi_2 }>_0}{<e^{U\int \overline{\psi}_1 \overline{\psi}_2 \psi_1 \psi_2
}>_0}.
\eeq
 In principle the reader has all the information needed to evaluate this
expression to any order in perturbation theory. The exponential is to be
expanded (in the numerator and denominator) to the desired order  and all
averages  done using Wick's theorem. We shall carry out the calculation to
order $U$ to show the details. The result to all orders will then simply be
stated and the sceptics are encouraged to check it to higher orders. The
integration measure and delta functions will be occasionally suppressed.

To first order we find
\beqr
<\overline{\psi}_1 (\omega ) \psi_1 (\omega )>=\frac{<\overline{\psi}_1
(\omega ) \psi_1 (\omega )>_0 + U\int \left< \overline{\psi}_1 (\omega )
\psi_1 (\omega ) \overline{\psi}_1 (\omega_4 )\overline{\psi}_2 (\omega_3
)\psi_1 (\omega_2 ) \psi_2 (\omega_1 ) \right>_0}{1 +U\int \left<
\overline{\psi}_1 (\omega_4 )\overline{\psi}_2 (\omega_3 )\psi_1 (\omega_2
) \psi_2 (\omega_1 ) \right>_0}
\eeqr
Both in the numerator and denominator we have just one pair of fields with
label $2$; these must obviously be paired, though it take a sign change to
bring them next to each other. The four fields with label $1$ can be paired in
two ways. One of the ways, in which the external fields
$\overline{\psi}_1 (\omega ) \psi_1 (\omega )$ get paired is precisely
cancelled by the denominator expanded out to order $U$. This corresponds to
the cancellation of disconnected diagrams. Reader new to this concept are
urged to verify this. What remains may be written as follows:
\beqr
<\overline{\psi}_1 (\omega ) \psi_1 (\omega )> &=& \frac{2\pi \delta
(0)}{i\omega -\Omega_0 + \mu } +
\frac{2\pi \delta (0)}{(i\omega -\Omega_0 + \mu )^2} U \int_{-
\infty}^{\infty}\frac{d\Omega_1}{2\pi} \frac{e^{i\omega_1
o^{+}}}{i\omega_1 -\Omega_0 + \mu } \label{orderu}\\
&=& \frac{2\pi \delta (0)}{i\omega -\Omega_0 - \mu - U<N_2>}
\eeqr
In going to the last step we have taken two terms of the power series in $U$
assumed that they represent a geometric series and summed the series. This
result is undoubtedly correct   to the order we are working in. We have also
replaced the integral over $\omega_1$ with $<N_2>$ which is also good to
this order.
It turns out that both these approximations are in fact exactly what
we would  get if we went to all orders, as will be explained shortly. Let us
accept this for the present and see what follows. Using the above correlation
function into the formula for $<N_1>$, Eqn.(\ref{N1}), we obtain
\beq
<N_1> = \theta [\mu - \Omega_0 - U<N_2> ].
\eeq
Obviously we can similarly derive another equation
\beq
<N_2> = \theta [\mu - \Omega_0 - U<N_1> ].
\eeq
Let us explore these equations for various cases. First if $\mu < \Omega_0$,
neither $\theta$ -function can be satisfied since $< \! N_i\! > \ge 0$. Thus we
get $<N>=0$ for this case as before. Likewise if $\mu > \Omega_0 + U$,
both $\theta$-functions will be satisfied satisfied since $<N_i> \le 1$. This
gives us $<N>=2$ as before. Finally consider $\Omega_0 < \mu < \Omega_0
+ U$.
Since each $<N_i>$ equals a $\theta$-function, it can equal only $0$ or $1$.
It is readily seen from these equations that the only two consistent  choices
are $<N_1> =1, <N_2> =0$ and vice versa, once again in agreement with the
operator solution.

Now for the higher terms in the expansion. These are best seen in
diagrammatic terms.       Consider Figure 4a. The dark line stands for the full
propagator and the thin line for the one computed in the gaussian measure. To
order $U$ we kept the two diagrams shown and these correspond to the
expressions in Eqn.(\ref{orderu}). The disconnected diagram that got
cancelled by the denominator is shown in Figure 4b.  If we go to higher
orders we can run into either iterates of the one loop or embellishments of it.
The embellishments convert the loop integral over the free propagator of
species $2$ to the integral over the full propagator which then reduces to
$<N_2>$. The iterations produce the remaining terms in the geometric series
that was presumed in going from Eqn.(\ref{orderu}) to the next one. This
leaves us with diagrams such as the "sunrise" diagram in Figure 4d. (Once
again the nomenclature is from field theory.) These diagrams and all the rest
vanish in this problem because the corresponding frequency integrals are
convergent  and have all the poles on the same half-plane, allowing us to
close the contour the other way.

The generalization of Grassmann integrals to many-body problems is
straightforward. The labels $1$ and $2$ from the toy Hubbard model can run
over, say the modes in the Brillouin zone. The action is once again obtained
by replacing the normal ordered hamiltonian by the corresponding function of
Grassmannian coherent state labels.  As for  the coupling functions,  just as
the coupling in the bosonic $u(4321)$ was symmetric under the exchange of
the first or last two labels among themselves, the fermionic couplings  will be
antisymmetric under such an exchange due to the anticommuting nature of the
Grassmann variables.

Finally a matter of notation. We will be switching from  upper case letters
$\Psi$ and $\Sid$ for fermion operators to lower case. It will be clear from
the context whether we are referring to the operators or the Grassmann
variables.

Table IV summarizes the results from the discussion on fermionic path
integrals.

\section{MOTIVATION AND WARMUP:   RG  IN  $d=1$}

We are now ready to turn to the main topic: application of the   RG  to
interacting nonrelativistic fermions. The best way to explain the method is to
deal with specific problems to which it applies. We begin with the problem of
charge density wave (CDW) formation in a system of spinless fermions at
half-filling. It has relatively simple kinematics and illustrates the   RG
approach very nicely.  In fact the  methods explained here were originally
developed (Shankar 1991) to deal with this problem in two dimensions.
Let us begin with a discussion of  the various terms used above in describing
the model.

We begin with  the justification for  the study of spinless fermions, which are
admittedly a  theorist's construction. As we progress with the paper, it will
become apparent that the   RG  is primarily concerned with the the  symmetry
properties of the
Fermi surface  of the noninteracting fermions. For example the
superconducting instability for arbitrarily small attraction is due to the
invariance of the Fermi surface  under time reversal: if $\vec{K}$ lies on the
surface, so does $-\vec{K}$. Likewise the  charge density wave (CDW)
instability on the square lattice,  towards  a ground state in which the charge
density is nonuniform and oscillates between the sublattices,   is due to
nesting: if $\vec{K}$ lies on the Fermi surface, so does $\vec{K} +
\vec{Q_N} $, where $\vec{Q_N} $ is a fixed nesting vector. Now these very
same properties of the Fermi surface  will also destabilize  a system of real
electrons.  The actual nature of the instability can  of course be different in
the two cases. For example the nested Fermi surface in $d=2$  will cause
electrons to go to an antiferromagnetic state in which the magnetization
(rather than charge density) oscillates in magnitude between the two
sublattices.  Likewise the time reversal invariant Fermi surface  will lead to
superconductivity but (Cooper)   pairing can  take place for any angular
momentum, while for spinless fermions only odd orbital angular momentum
states are allowed due to antisymmetry requirements.  To summarize, spin
really is an inessential complication if we are simply tying to understand how
the   RG  works. When comparing  theory to experiment, spin will of course
have to be included, but this  really will be straightforward.

Consider now the requirement of half-filling, which means that the system has
half the maximum number allowed by the exclusion principle.  For spinless
fermions this means one particle for every other site, while for electrons it
means one particle per site. Despite this, the two problems will have the
Fermi surface  in the noninteracting case. In both cases, the Fermi surface
will enclose  half the Brillouin zone and have the same shape,  decided by the
lattice parameters. However each filled momentum state will carry two
electrons (of opposite spin) but just one spinless fermion. In other words, the
condition of half-filling implies a different particle density in the two
cases,
but the same Fermi surface . \footnote{An aside for readers with a different
background, say particle physics, who are troubled by the following: why
bother with effects at such a special filling ? Is there any chance   that a
generic system will have filling factor of exactly half, as compared to say
.51?
Yes!  Consider a square lattice  with an atom at each site. Since each atom
contributes an {\em integer} number of electrons  to the conduction band, the
filling factor  (the ratio of the number of electrons to the maximum allowed
number of two per site) is bound to be a half-integer or integer.  For a more
complicated unit cell, or intercalated compounds, one can have other simple
fractions like $1/4$.}

We start here with the one dimensional version of the spinless fermion
problem  for two reasons. First, the one dimensional problem is very
interesting in itself and shows the power of the   RG  approach. Secondly, as
mentioned earlier, due to the fact the Fermi surface  in $d=1$ consists of just
two points, the problem resembles quantum field theory with a few coupling
constants (rather than a theory with coupling functions)  and affords  a
painless introduction to the use of the   RG  for fermions.  We will then be
better prepared for  the $d=2$ version in Section X.

\subsection{The $d=1$ model: definition and mean-field analysis}
Let us consider the following specific hamiltonian for a  spinless fermion
system on a $d=1$ lattice labeled by an integer $n$:
\beqr
H &=& H_0 + H_I \\
     &=&\!  -\frac{1}{2}
     \sum_j \sid (j+ 1) \psi (j) + h.c. + U_0 \sum _j (\sid
(j) \psi (j) -\! \frac{1}{2}  ) ( \sid (j + 1) \psi (j+1) -\! \frac{1}{2} )
\label{spinless}
     \eeqr
     where the fields obey
     \beq
     \{ \sid (j) , \psi (m) \} = \delta_{mj}
     \eeq
      with all other anticommutators vanishing.

 The first term represents hopping. The hopping amplitude has been
normalized to $1/2$.   The second term represents  nearest neighbor
repulsion of strength $U_0$. The role of the $1/2$-'s subtracted from the
charge densities $n_j$ ($ = \sid_j \psi_j$)  and $n_{j+1}$ is this. When we
open up the brackets, it is readily seen that they represent  a chemical
potential
\beq
\mu = U_0.
\eeq
This happens to be exactly the value need to maintain half-filling in the
presence of the repulsion $U_0$. To see this, make the change $\psi
\leftrightarrow \sid$
at all sites. This exchanges the site occupation number $n
= \sid \psi $ with $1-n$ or changes the sign of $n-1/2$. Thus both   brackets
in the interaction
term change sign under this, and  their product is unaffected.
As for the hopping term, it  changes sign under $\psi \leftrightarrow \sid$.
This can be compensated  by changing the sign of $\psi \ , \sid$ on just one
sublattice (which preserves the anticommutation rules and does not affect the
other term) .
Thus $H$ is invariant under exchanging particles with holes. This means the
ground state (if it is unique)  will satisfy $<n> = <1-n>$ which in turn means
$<n> = 1/2$. (If there is a degeneracy of the ground state, the result still
holds, but takes a little more work to establish.)

   Let us understand this model in the extreme limits $U_0 = 0$ and $U_0 =
\infty$.

   As for the  first case let us introduce momentum states via
    \beq
    \psi (j) = \int_{-\pi}^{\pi} \frac{dK}{2\pi}\psi (K) e^{iKj}
    \eeq
    and the inverse  transform
    \beq
\psi (K) = \sum_j e^{-iKj} \psi (j).
\eeq
Using $\sum_j = 2\pi \delta (0) $, we can verify that
\beq
\{ \psi (K) , \sid  (K') \} = 2 \pi \delta (K- K').
\eeq
 In terms of these operators
\beqr
H_0 &=& \intK \sid (K) \psi (K)  E( K) \\
E(K) &=& -\cos K .\label{freefer}
\eeqr
  The Fermi sea is obtained by filling all negative energy states, i.e., those
with $|K| \le K_F = \pi /2 $, which corresponds to half-filling.  The Fermi
surface  consists of just two points $|K| = \pm \pi /2$.
It is clear that the ground state is a perfect conductor since  we can move a
particle just below the Fermi surface  to just above it at arbitrarily small
energy cost.

Consider now the situation in the other extreme $U_0=\infty$. We now
ignore the hopping
term and focus on just the interaction. It is evident that the
lowest energy   states are those in which no particle has a neighbor: thus
either the A-sublattice consisting of  even sites is  occupied,  or the B-
sublattice, made up of  the odd sites, is occupied. This makes the product
$(n_j - 1/2)(n_{j+1} - 1/2)$ negative on every bond.  These two states, which
break the translational symmetry of the lattice are the  CDW states.  The
order parameter, which measures the difference between the mean occupation
of the odd and even sites is maximal (unity).   In the CDW state, the system is
an insulator. Any excitation of the ground state requires us to move the charge
and this will cost an energy of order $U_0$. (This is clearest as $U_0
\rightarrow \infty$.) One  expects that even for large but finite $U_0$, the
symmetry would still be broken, but with a smaller order parameter.

{\em Here is the question we want to answer: will the system develop the
CDW order and gap for arbitrarily small repulsion, or will it remain a
conductor up to some finite $U_0$?}

We will use the   RG  to answer it. But first let us see what  a very standard
tool, namely  {\em mean field theory},  can tell us.
In this approach one assumes a  CDW order parameter in the ground state
and asks  if the assumption is self-consistent. The self-consistency check is
approximate, as will be explained. Mean-field theory predicts that CDW will
set in for any repulsion however small. Here is a short description of the
calculation.

Let us begin with Eqn.(\ref{spinless}) and
 make the {\em ansatz}
\beq
<n_j> = \frac{1}{2} + \frac{1}{2}(-1)^j \Delta
\eeq
where $\Delta$ is the CDW order parameter. We will now see if the ground
state   energy of the system is lowered by a nonzero value of $\Delta.$ To this
end, we will find the ground state energy as a function of
$\Delta$ and minimize it and see if the minimum occurs at a nonzero
$\Delta$. However this last step will be done approximately since this is an
interacting many body system.  The approximation is the following. We start
with the interaction
\beq
H = -\frac{1}{2} \sum_j \sid (j+ 1) \psi (j) + h.c +U_0 \sum_j (n_j -
\frac{1}{2} )(n_{j+1}- \frac{1}{2})
\eeq
and make the substitution
\beqr
n_j &=& \frac{1}{2} + \frac{1}{2} (-1)^j \Delta + :n_j:\\
n_{j+1} &=& \frac{1}{2} + \frac{1}{2} (-1)^{j+1}\Delta  + :n_{j+1}:
\eeqr
where the normal ordered operator  $:n_j:$ has no expectation value in the
true ground state and represents the fluctuations in number density.
Upon making these substitutions and some rearrangements we find
\beqr
H &=& -\frac{1}{2} \sum_j \sid (j+ 1) \psi (j) + h.c \\ & & +U_0 [
\frac{1}{4}\sum_j \Delta^2 - \Delta \sum_j (-1)^j n_j ] + U_0 \sum_j
:n_j::n_{j+1}:
\eeqr
{\em In the mean field approximation we ignore the last term. } The rest of
the hamiltonian is quadratic and solved by Fourier   transformation  . Due to
the factor $(-1)^j$ which  multiplies $n_j$, states with momentum  $K$ and
$K' = K + \pi$ will mix.   The hamiltonian becomes
\beq
H = \int_{0}^{\pi} \frac{dK}{2\pi}  (\sid (K) , \sid (K') ) \left[
\begin{array}{ll}
E(K) & - U_0 \Delta \\
-U_0 \Delta & E(K')
\end{array} \right] \left[
\begin{array}{l}\psi (K) \\ \psi (K') \end{array} \right] + U_0 2\pi \delta (0)
\frac{\Delta^2}{4}
\eeq
Notice that we have halved the range of $K$ integration, but doubled the
number of variables at each $K$.    The two-by-two matrix, which is traceless
due to the relation
\beq
E(K')  = -\cos (K+\pi ) =  - E(K),
\eeq
is readily diagonalized. The one-particle energy levels come in equal and
opposite pairs and we fill  the negative energy states to obtain the following
ground state energy per unit volume:
\beq
\frac{E_0}{2\pi \delta (0)} = \frac{\Delta^2 U_0}{4} - \int_{0}^{\pi}
\frac{dK}{2\pi} \sqrt{E^2 (K)
+ \Delta^2 U^{2}_{0}}
\eeq
where the integral comes from the filled sea. Minimizing with respect to
$\Delta$ we obtain the relation
\beq
\Delta = \int_{0}^{\pi} \frac{dK}{\pi} \frac{U_0 \Delta }{\sqrt{E^2 (K)
+ \Delta^2 U^{2}_{0}}}.
\eeq
Assuming $\Delta \ne 0$, we cancel it on both sides. It is  clear that $U_0
<0$ is not acceptable since the two sides of the equation would then have
opposite signs. For positive $U_0$, a nontrivial solution requires that
\beq
 1 = U_0 \int_{0}^{\pi} \frac{dK}{\pi} \frac{ 1}{\sqrt{E^2 (K)
+ \Delta^2 U^{2}_{0}}}.
\eeq
which is called the {\em gap equation}. On the left hand is the number $1$
and on the right hand side, something of order $U_0$. It appears that we will
get a solution only above some minimum $U_0$. This is wrong. The
integrand
becomes very large at the Fermi points $|K|= K_F =\pi /2$, where $E(K)$
vanishes. Writing
\beqr
E(K) &=& k \\
k &=& |K| - K_F
\eeqr
we approximate the gap equation as  follows:
\beq
1 = U_0 \int_{-\Lambda }^{\Lambda } \frac{dk}{\pi} \frac{ 1}{\sqrt{k^2
+ \Delta^2 U^{2}_{0}}}\simeq  \frac{2U_0}{\pi} \ln \frac{
\Lambda}{\Delta U_0}
\eeq
where $\Lambda$, the upper cut-off on $|k|$,  is not very important. What is
important is that due to the logarithmic behavior of the integral near  the
origin in $k$, i.e., near the Fermi surface, there will always be a solution to
the gap equation given by
\beq
\Delta = \frac{\Lambda}{U_0} e^{-\pi / 2U_0 }.
\eeq
The logarithmic divergence is also reflected in the divergent susceptibility of
the noninteracting
system to a probe (or perturbation)  at momentum $\pi$. (At second order in
perturbation theory, the perturbation will link the ground state to states of
arbitrarily low energy in which a particle just below the right (left) Fermi
point is pushed to just above the left (right) Fermi point. The small energy
denominators, summed over such states will produce the logarithm.)

Mean-field theory also predicts that the same thing will happen in $d=2$. In
this case the nesting condition (readers unfamiliar with this concept should
consult Figure 17, Section X) ensures that the perturbation at $(\pi, \pi )$
will excite particles just below the Fermi surface  to states just above it on
the
"other side" {\em no  matter where the starting point is on the Fermi surface.}
Now for  any Fermi surface, if we take some  perturbation of any momentum
(which is not too large compared to the size of the surface ), there will
always
be some points  just below sea level, that will get knocked to points just
above sea level. But these points would have to come from some special
angular region of the Fermi surface which are connected by this momentum
(whose angular dimensions will decrease with the energy denominators)  and
the integral over the small energy denominators  will converge.   If we use a
coordinate $\theta$ on the surface  and a coordinate $\varepsilon$ normal to
it, the integrals will be off the form
\beq
 \int_{-\Lambda}^{\Lambda} \int_{\theta_1 (Q,\varepsilon )}^{\theta_2 (Q,
\varepsilon)} \frac {d\varepsilon d \theta }{\sqrt{\varepsilon^2 + \Delta^2}}.
\eeq
The cut-off $\Lambda$ focuses on the small energy denominator region. In
the absence of nesting, for a perturbation of some given momentum $Q$, the
range of $\theta$ will shrink with $\varepsilon$ and the integral will
converge. On the other hand, at the  nesting momentum, the angular integral
will over the entire Fermi surface (since every point on or near the Fermi
surface get scattered to another, as shown in Figure 17, Section X) and the
integral will have a  logarithmic divergence.

Returning to $d=1$, mean-field theory also predicts that the system will a
have non-zero   superconducting order parameter for the smallest attractive
coupling. In the corresponding calculation the instability will stem  from the
time-reversal symmetry of the problem: $E(K) = E(-K)$.

Unfortunately both these predictions are wrong. The error comes from the
neglected
quartic operator  $:n_j::n_{j+1}:$.  We know all this because the present
spinless hamiltonian  Eqn.(\ref{spinless}) can be solved exactly (Yang and
Yang 1976). \footnote{ The reader consulting this reference will find that
these authors solve a problem of quantum spins on a line. This {\em XXZ
chain} is related to the spinless fermions by a Jordan Wigner   transformation
. } The exact solution tells us that the system remains gapless for $U_0$ of
either sign until it exceeds a minimum value of order unity.
We will now develop the   RG  approach to this problem and obtain results in
harmony with this exact result.

\subsection{ The   RG  approach  for  $d=1$ spinless fermions}

Our goal is to explore
the stability of the noninteracting spinless fermions  to
weak interactions. We are not interested in the fate of just the model in
Eq.(\ref{spinless}) but in a whole family of models of which this will be
special case. Our strategy, stated earlier on, is the following.
First we   argue that at weak coupling, only modes near $\pm K_F$ will be
activated. Thus we will linearize the dispersion relation $E(K) = -\cos K$
near  these points and work with a cut-off $\Lambda$:
\beq
H_0 = \sum_i \intk \psi^{\dag}_{i} (k) \psi_i (k) k \label{freefercont}
\eeq
where
\beqr
k &=& |K| - K_F \\
i   &=& L, R \ \ \ (left \ \ or \ \ right).
\eeqr
Notice that $H_0$ is an integral  over fermionic oscillators which we studied
in Section III. The frequency $\Omega_0$  of the oscillator at momentum $k$
is simply $ k$.

 Next  we will write down a $T=0$ partition function for the noninteracting
fermions. This will be a Grassmann integral only over the degrees of freedom
within a cut-off $\Lambda$  of  the Fermi surface. We will then find an   RG
transformation   that lowers the cut-off but  leaves the free-field action,
$S_0$,  invariant.  With the   RG  well defined,  we will  look at the generic
perturbations of this fixed point and classify them as usual.
 If no relevant operators show up, we will still have a scale-invariant gapless
system. If, on
the other hand, there are generic relevant perturbations, we will
have to  see to which new fixed point the system  flows. (The new one could
also be gapless.) The stability analysis can be done perturbatively. {\em In
particular,  if a relevant perturbation takes us away from the original fixed
point, nothing at higher orders  can ever bring us back to this fixed point.
}The fate of the nearest neighbor model will then be decided by asking if it
had a relevant component in its interaction.

Let us then  begin with the partition function for our  system of fermions:
\beqr
Z_0 &=& \int \prod_{i=L,R}\prod _{|k|<\Lambda} d\psi_{i} (\omega k)
d\sib_{i} (\omega k)e^{S_0} \\
  S_0        &=& \sum_{i = L\ R} \intk \intom \sib_{i} (\omega  k) (i\omega -
k)\psi_{i} (\omega k)\label{freeaction}
  \eeqr

 This is just
 a product of functional integrals for the Fermi oscillators at each
momentum with $\Omega_0 (k)= k$.

  The first step in the   RG    transformation   is to integrate out all $\psi
(k\omega )$ and $\sib (k\omega )$ with
  \beq
  \Lambda / s \le |k| \le \Lambda  \label{newcut}
\eeq
and {\em all }$\omega$. Thus our phase space has the shape of a rectangle,
infinite
in the $\omega$ direction, finite in the $k$ direction. This shape will
be preserved under the   RG transformation. Since there is no real relativistic
invariance here, we will make no attempt to treat $\omega$ and $k$ on an
equal footing. Allowing $\omega$ to take all values  allows us to extract  an
effective
hamiltonian operator at any stage in the   RG  since  locality in time
is assured.

Since the
integral is gaussian, the result of integrating out fast modes  is just a
numerical prefactor which we throw out.
The surviving modes now have their momenta going from $-\Lambda  /s $ to
$\Lambda  /s$.  To make this action a fixed point  we define rescaled
variables:
\beqr
k' &=& sk \nonumber \\
\omega '  &=& s\omega \nonumber \\
\psi_{i} '(k'\omega ') &=& s^{-3/2} \psi_{i} (k\omega  )\label{rescale}
\eeqr
Ignoring a constant that comes from rewriting the measure in terms of the
new fields, we see that $S_0$ is invariant under the mode elimination and
rescaling operations.

We can now consider the effect of perturbations on this fixed point. Rather
than turn on the perturbation corresponding to the  nearest neighbor
interaction we will perform a more general analysis. The result for the
particular case will be subsumed by this analysis.
\subsection{Quadratic perturbations}
First consider perturbations which are quadratic in the fields. These must
necessarily be of the form
\beq
\delta S_2 = \sum_{i = L,\ R} \intk \intom \mu (k\omega  )\sib_{i} (\omega
k) \psi_{i} (\omega k)
\eeq
 assuming  symmetry between left and right fermi points.

 Since this action separates into slow and fast pieces, the effect of mode
elimination is simply to reduce $\Lambda$ to $\Lambda /s$ in the integral
above. Rescaling moments and fields, we find
 \beq
 \mu' (\omega' , k', i) = s^2 \mu (\omega , k ,i).
 \eeq
 We get this factor $s^2$  as a result of combining a factor $s^{-2}$ from
rewriting the  old  momenta and frequencies in terms of the  new and a factor
$s^3$ which comes from rewriting the old fields in terms of the new.

  Let us expand $\mu$ in a Taylor series
 \beq
 \mu (k, \omega) = \mu_{00} + \mu_{10} k + \mu_{01} i\omega + \cdots +
 \mu_{nm} k^{n} (i\omega )^{m} + \cdots \label{muexp}
 \eeq
 The constant piece is a relevant perturbation:
 \beq
 \mu_{00} \longrightarrow  s\mu_{00} .
 \eeq
  This relevant flow reflects the readjustment of the Fermi sea
  to a change in chemical potential. The correct way to deal with this term is
to include it in the free field action by filling  the Fermi sea to a point
that
takes $\mu_{00}$  into account. As for the next two terms, they clearly
modify terms that are already present in the action, can be absorbed into them
and correspond to marginal interactions. When we consider quartic
interactions, it will be seen that mode elimination will produce terms of the
above form even if they were not there to begin with just as $\phi^4$ theory.
The way to deal with them will be discussed
  in due course. As for higher order terms in  Eqn.(\ref{muexp}), they are
irrelevant under the   RG  mentioned above. This is however a statement that
is correct at the free-field fixed point. We shall have occasion to discuss a
term that is irrelevant at weak coupling but gets promoted to relevance as the
interaction strength grows.

  \subsection{Quartic perturbations: the    RG  at Tree Level}
We now turn on the quartic interaction whose most general form is
\beq
\delta S_4 = \frac{1}{2!2!}\int_{K\omega} \sib (4) \sib (3) \psi (2) \psi (1)
u(4, 3, 2 ,1)
\label{s4}
\eeq
where
\beqr
\sib (i) &=& \sib (K_i, \omega_i) \ \ etc., \\
\int_{K \omega} \! \! &=&\! \!  \left[ \prod_{i=1}^{4}\int_{-
\pi}^{\pi}\frac{dK_i}{2\pi}\int_{-
\infty}^{\infty}\frac{d\omega_i}{2\pi}\right] \left[ 2\pi
\overline{\delta} (K_1 + K_2 - K_3 - K_4) 2\pi \delta (\omega_1 + \omega_2
- \omega_3 - \omega_4 )\right]
\label{measure1}
\eeqr
and $\overline{\delta}$ enforces momentum conservation  mod $2\pi$, as is
appropriate to any lattice problem.  A process where lattice  momentum is
violated in multiples of $2\pi$ is called an {\em umklapp} process. The delta
function containing frequencies enforces time translation invariance.The
coupling function $u$ is  antisymmetric under the exchange of its first or last
two arguments among themselves since that is true of the Grassmann fields
that it multiplies. {\em Thus the coupling $u$  has all the symmetries of the
full vertex function $\Gamma$ with four external lines.}

To get a feeling for all these  ideas let us  consider the nearest-neighbor
repulsion from Eqn.(\ref{spinless}) and ask what $u$ it generates in the
action.
Let us first begin with the operator
\beq
H_I = U_0 \sum_{j} \sid_j \psi_j \sid_{j+1} \psi_{j+1} = - U_0 \sum_{j}
\sid_j  \sid_{j+1}  \psi_j  \psi_{j+1}
\eeq
and make a Fourier   transformation   to get
\beq
H_I = -U_0 \left[ \prod_{i=1}^{4}\int_{-\pi}^{\pi}\frac{dK_i}{2\pi} \right]
2\pi
\overline{\delta} (K_1 + K_2 - K_3 - K_4) \sid (K_4) \sid (K_3) \psi (K_2)
\psi (K_1) e^{i(K_1 - K_3)}.
\eeq
We now antisymmetrize $e^{i(K_1 -K_3)}$ with respect to $1
\leftrightarrow 2$ and $3 \leftrightarrow 4$ since the rest of the integrand is
antisymmetric under either of these operations.
This gives us the result
\beq
e^{i(K_1 -K_3)} \rightarrow \frac{1}{4}(e^{i(K_1 -K_3)} - e^{i(K_2 -
K_3)} - e^{i(K_1 -K_4 )} + e^{i(K_2 -K_4)})   \label{yokel}
\eeq
We next use the fact that due to the $\overline{\delta}$-function,
\beq
K_1 +K_2 = K_3 +K_4 + Q \ \ \ Q = 0 \ or 2\pi
\eeq
where $Q$ is the umklapp momentum. It cannot be any higher multiple of
$2\pi$ given that  $|K_i| \le \pi$. This means
\beq \sin Q/2 =0,\label{Q}.
\eeq
 This result and some  simple trigonometry applied to Eqn.(\ref{yokel}) lead
to   the following coupling:
\beq
\frac{u(4,3,2,1)}{2!2!} =  U_0 \sin (\frac{K_1 - K_2}{2} ) \sin (\frac{K_3 -
K_4}{2})\cos (\frac{K_1 +K_2 - K_3 -K_4}{2}) . \label{NN}
\eeq
 In arriving at this equation we have used Eqn.(\ref{Q}) and gone from the
interaction hamiltonian to the corresponding action in  the the path integral
the usual way. The integral of $-H_I$  (with Fermi operators replaced by
Grassmann numbers) from $\tau = 0$ to $\tau = \infty $ becomes the integral
over four frequencies with one overall delta function.  Notice that $u$ has no
frequency dependence.

Let us now return to the general interaction, Eqn.(\ref{s4} - \ref{measure1}),
and restrict the momenta to lie within $\Lambda$ of either Fermi point L or
R. Using a notation where L (left Fermi point) and R (right Fermi point)
become  discrete a label $i = l \ or \ R$  and 1-4 label   the frequencies and
momenta (measured from the appropriate Fermi points).   Eqns.(\ref{s4} -
\ref{measure1} ) become

\beq
\delta S_4 = \frac{1}{2!2!} \sum_{i_1 i_2 i_3 i_4= L,R}
\int_{K\omega}^{\Lambda} \sib_{i_4} (4) \sib_{i_3} (3) \psi_{i_2} (2)
\psi_{i_1} (1) u_{i_4 i_3 i_2 i_1}(4, 3, 2 ,1)
\eeq
where
\beqr
\int_{K \omega} ^{\Lambda}  &=&  \left[ \int_{-
\Lambda}^{\Lambda}\frac{dk_1\cdots dk_4}{(2\pi    )^4}\int_{-
\infty}^{\infty}\frac{d\omega_1\cdots d\omega_4}{(2\pi )^4}\right] \left[ 2\pi
\delta (\omega_1 + \omega_2 - \omega_3 - \omega_4 )\right]
\nonumber \\
						      & &\left[ 2\pi
\overline{\delta} ( \varepsilon_{i_1} (K_F + k_1 ) +\varepsilon_{i_2} (K_F
+k_2) -  \varepsilon_{i_3} ( K_F + k_3) - \varepsilon_{i_4} (K_F + k_4)
)\right]\label{measure2}
 \eeqr
and
\beq
\varepsilon_i = \pm 1 \ \ for \ \ R\ , \ L .
\eeq

Let us now implement the   RG    transformation    with this interaction. This
proceeds exactly as in $\phi^4$ theory.
Let us recall how it goes. If schematically
\beq
Z = \int d\phi_< d\phi_> e^{-\phi^{2}_{<} - \phi^{2}_{>}} e^{-u(\phi_< +
\phi_> )^4}
\eeq
is the partition
function and we are eliminating $\phi_>$, the effective $u$ for
$\phi_<$ has two origins.
First, we have a term $-u \phi^{4}_{<} $ which is there to begin with, called
the {\em tree level} term. Next, there are terms generated by the $\phi_>$
integration. These are computed in a cumulant expansion and are given by
Feynman diagrams whose internal momenta lie in the range being eliminated.
The loops that contribute to the flow of $u$ begin at order $u^2$.

Let us first
do the  order $u$ tree level calculation for the renormalization of
the quartic interaction. This gives us just Eqn.(\ref{measure2}) with
$\Lambda \rightarrow \Lambda /s$. If we now rewrite this in terms of new
momenta and fields, we get an interaction with the same kinematical limits as
before and we can meaningfully read off the coefficient of the quartic-Fermi
operators as the new coupling function. We find
\beq
u'_{i_4 i_3 i_2 i_1}(k_i', \omega_i'  ) =u_{i_4 i_3 i_2 i_1}(k_i'/s, \omega_i'
/s ) \label{scaling}
\eeq

The reader who carries out the intermediate manipulations will  notice  an
important fact:
$K_F$ never enters any of the $\delta$ functions: either all $K_F$'s cancel in
the nonumklapp cases, or get swallowed up in multiples of $2\pi$ (in inverse
lattice units) in the umklapp cases due to the periodicity of the
$\overline{\delta} $-function.  As a result the momentum $\delta$ functions
are free of $K_F$ and  scale very nicely under the   RG transformation:
\beqr
\overline{\delta}(k) &\rightarrow & \overline{\delta}(k'/s)\\
				  &=& s \overline{\delta}(k')
\eeqr
Turning now to Eqn.(\ref{scaling}), if we expand $u$ in a Taylor series in its
arguments and compare coefficients, we find readily  that the constant term
$u_0$  is marginal and the higher coefficients are irrelevant. Thus $u$
depends only on its discrete labels and we
can limit the problem to just a few coupling constants instead of the coupling
function we started with. Furthermore, all reduce to just one coupling:
\beq
u_0 = u_{LRLR}         = u_{RLRL}
	  = -u_{RLLR}
	    = -u_{LRRL} .
	    \eeq
	    Other couplings corresponding to $LL \rightarrow RR$ are wiped
out by the Pauli principle since they have no momentum dependence and can't
have the desired antisymmetry.

	    As a concrete example consider the $u$ that comes from the nearest-
neighbor  interaction Eqn.(\ref{NN}) reproduced here for convenience:
\beq
\frac{u(4,3,2,1)}{2!2!} =  U_0 \sin (\frac{K_1 - K_2}{2} ) \sin (\frac{K_3 -
K_4}{2})\cos (\frac{K_1 +K_2 - K_3 -K_4}{2})  \nonumber
\eeq
and ask what sorts of couplings are contained in it.

	     If 1 and 2 are both from  R, we find the following factor in the
coupling
	    \beqr
	    \sin (\frac{K_1 - K_2}{2}) &=& \sin (\frac{k_1 - k_2 }{2} )
\nonumber \\
						    &\simeq & (\frac{k_1 - k_2
)}{2})
						       \eeqr

						   which leads
						   to  the requisite
antisymmetry but makes the coupling irrelevant (due to the $k$'s). There will
be one more power of $k$ from $3$ and $4$ which must also come from near
just one Fermi point so as to conserve momentum modulo $2\pi$. For
example the umklapp process, in which $RR \leftrightarrow LL$, has a
coupling
\beq
u_{NN}(umklapp) \simeq (k_1 - k_2 ) ( k_3 - k_4)
\eeq
and is strongly irrelevant at the free-filed fixed point.

							On the other hand  if 1 and 2
come from opposite sides,
\beq
\sin (\frac{K_1 - K_2}{2}) \simeq \sin (\pi /2 +O(k))
\eeq
 and likewise for $3$ and $4$, and we have a marginal interaction $u_0$ with
no $k$'s in the coupling.

 The tree level analysis readily extends to couplings with six or more fields.
All these are irrelevant, even if we limit ourselves to constant ($\omega$ and
k independent) couplings.

 To determine the ultimate fate of the coupling $u_0$, marginal at tree level,
we must turn to the one loop   RG  effects.

\subsection{  RG  at one loop: The Luttinger Liquid}

Let us  begin with the action with the quartic interaction and do a mode
elimination. To order $u$, this leads to an induced quadratic term represented
by the tadpole graph in Figure 5. We  set $\omega = k=0$ for the external
legs and have chosen them to lie at L, the left Fermi point. The integral given
by the diagram produces a momentum independent  term of the form $\delta
\mu \sib_L \psi_L$. But we began with no such term. Thus we do not have a
fixed point in this case. Instead we must begin with some term $\delta
\mu^{*}  \sib_L \psi_L$
such that upon renormalization it reproduces itself. We find it by demanding
that
\beq
\delta \mu^* = s \left[  \delta \mu^* - u_0^* \int_{-
\infty}^{\infty}\frac{d\omega}{2\pi}
\int_{\Lambda /s < |k| < \Lambda} \frac{dk}{2\pi} e^{i\omega 0^+}
\frac{1}{i\omega - k}\right]
\eeq
where we have used the zeroth order propagator and the fact that to this order
any $u_0 = u_0^*$. The exponential convergence factor is the one always
introduced  to get the right answer for, say, the ground state particle density
using $<\sib \psi >$. Doing the
$\omega $ integral, we get
\beqr
\delta \mu^* &= &s \left[ \delta \mu^* - u_0^* \int_{\Lambda /s < |k| <
\Lambda} \frac{dk}{2\pi} \theta (-k) \right] \\
		      &=& s \left[\delta \mu^* -
		      \frac{\Lambda u_0^*}{2\pi} ( 1-
1/s) \right] .
\eeqr
It is evident that the fixed point is given by
\beq
\delta \mu^* = \frac{\Lambda u_0^*}{2\pi}.\label{deltamu}
\eeq

Alternatively, we could just as well begin with the following relation for the
renormalized coupling
\beq
\delta \mu^{'} = s \left[  \delta \mu - u_0^* \int_{-
\infty}^{\infty}\frac{d\omega}{2\pi}
\int_{\Lambda /s < |k| < \Lambda} \frac{dk}{2\pi} e^{i\omega 0^+}
\frac{1}{i\omega - k}\right]
\eeq
which implies the flow
\beq
\frac{d\mu}{dt} =\mu -\frac{u_{0}^{*}}{2\pi}\label{mueqn}
\eeq
assuming we  choose to measure $\mu$ in units of $\Lambda$. The fixed
point of this
equation reproduces Eqn.(\ref{deltamu}).

We can find $\delta \mu^*$ in yet another way with no reference to the   RG .
If we calculate the inverse
propagator in the cut-off  theory to order u, we will
find
\beq
G^{-1} = i\omega - k - \frac{\Lambda u_0 }{2\pi}
\eeq
indicating that the Fermi point is no longer given by $k = 0$. To reinstate the
old $K_F$ as interactions are turned on, we must move the chemical potential
away from zero and to the value $\delta \mu = \frac{\Lambda u_0}{2\pi}$.
Thus the correct action that gives us the desired $K_F$, for this coupling, to
this order, is then schematically given by
\beq
S = \sib ( i \omega - k )\psi  + \frac{\Lambda u_0}{2\pi}\sib \psi +
\frac{u_{0}}{2!2!} \sib \sib \psi \psi .
\eeq
An  RG    transformation  on this action would not generate the tadpole graph
contribution.

{\em A very important point which will appear again is this: we must fine
tune the chemical potential as a function of u, not to maintain criticality (as
one does in $\phi^4$ where the bare mass is varied with the interaction to
keep the system massless)
but to retain the same  particle density.} (To be precise, we are keeping fixed
$K_F$, the momentum at which the one-particle Greens function has its
singularity.  This amounts to keeping the density fixed ( Luttinger(1960).)
If we kept $\mu$ at the old value of zero, the system would flow away from
the fixed point with $K_F = \pi /2$, not to a state with a gap, but to another
gapless one   with a smaller value of $K_F$. This simply corresponds to the
fact if the total energy of the lowest energy  particle that can be added to
the
system, namely $\mu $, is to equal 0, the kinetic energy at the Fermi surface
must be
slightly negative so that the repulsive potential energy with the others
in the sea brings the total to zero.

Now, we do not have to work with fixed density; we could take the given
$\mu$ and accept whatever $K_F$ it leads to.  However, we will frequently
work at fixed density for two reasons.   First, there is a simple experimental
way to fix the density, namely send in the desired number of particles and
seal the system, i.e., work with the canonical and not grand canonical system.
(This simple and viable  procedure looks rather complicated in the grand
canonical language.)  Secondly  we will keep $K_F$ fixed in many two and
three  dimensional cases to make contact with the  pioneering  work of
Landau,  done at fixed density.

Let us now turn our attention to the order $u_0^2$ graphs that renormalize
$u_0$. These are shown in Fig. 6. The increment in $u_0$, hereafter simply
called $u$, is given by the sum of the ZS (zero-sound),  ZS' and BCS graphs.
The analytical formula for the increment in $u$ is
\beqr
du ( 4321 ) & =& \int u (6351) u(4526) G(5) G(6) \delta (3+6-1-5) d5 d6
\nonumber \\
  & &      -\int u (6451) u(3526) G(5) G(6) \delta (6+4 - 1-5) d5 d6 \nonumber
\\
		     & &
		     - \frac{1}{2} \int u (6521) u (4365) G(5) G(6) \delta (5
+ 6 - 1  -2 ) d5 d6 \label{flow}
		     \eeqr
where $1$ to $4$ stand for all the attributes of the (slow) external lines, $5$
and $6$ stand for all the attributes of the two (fast) internal lines: momenta
(restricted to be within the region being eliminated), and   frequencies;  G
are
the propagators and the $\delta $ functions are for ensuring the conservation
of  momenta and frequencies and $\int d5d6$ stands for sums and integrals
over the attributes $5$ and $6$. (In the figure the momenta $1$ to $6$ have
been assigned some special values (such as $5=K$ in Fig.6a) that are
appropriate to the problem at hand. The formula is very general as it stands
and describes other situations as well.)
The couplings $ u$
are functions of all these  attributes, with all the requisite
antisymmetry properties. (The order in which the legs are labeled in $u$ is
important due to all the minus signs. The above equations have been written
to hold with the indicated order of arguments. In their present form they are
ready to be used by a reader who wants to include spin.)

This is the master formula we will invoke often. It holds even in higher
dimensions, if we suitably modify the integration region for the momenta.

To derive this formula , we do exactly what we did in Section II with bosons.
We split the modes into slow and fast ones and do the fast integral,  using the
cumulant expansion to collect the terms that feed back to the quartic coupling.
The calculation uses the fermionic Wick's theorem and exactly the same three
diagrams (ZS,ZS' and BCS) that we saw in the scalar field example appear.
The major  difference compared to the scalar case is in the extra minus signs
in the fermionic Wick's theorem. Of course the propagators now have
different forms and the range on loop variables is different.  Readers familiar
with Feynman diagrams may obtain this formula by drawing  all the diagrams
to this order in the usual Feynman graph expansion, but allowing the loop
momenta to range only over the modes being eliminated. In the present case,
these are given by the four thick lines  labeled a,b,c and d in Fig. 7 where
each  line  stands for a region of width $d\Lambda $ located at the cut-off ,
ie., a distance $\Lambda$ from the Fermi points. The external momenta are
chosen to be $(4321) = (LRLR)$, at the Fermi surface. All the external $k$'s
and $\omega$'s are set equal to zero since the marginal coupling $u$ has no
dependence on these. This has two consequences. First, the loop frequencies
in the ZS and ZS' graphs are equal,while those in the BCS graph are equal
and opposite. Second,  the momentum transfers at the left vertex are $Q =
K_1 - K_3 = 0$ in the ZS graph,
$Q' = K_1 - K_4 = \pi $ in the ZS' graph, while the total momentum   in the
BCS graph is $P = K_1 + K_2 = 0$. Therefore if one  loop momentum
$5=K$ lies  in any of the four shells in Fig.7,
so does the other loop momentum $6$which equals $K$, $K + \pi$ or  $-K$
in the ZS, ZS' and BCS graphs respectively. Thus we may safely eliminate the
momentum conserving $\delta $ function in Eqn.(\ref{flow}) using  $\int d6$.
This fact, coupled with
\beqr
 E(-K) &=& E(K) \\
E(K' =  K\pm \pi ) &=& -E(K)
\eeqr
leads to
\beqr
du(LRLR) \! \! &=& \! \!  \int_{-
\infty}^{\infty}\int_{d\Lambda}\frac{d\omega dK}{4\pi^2} \frac{u(KRKR)
u(LKLK)}{(i\omega - E(K))(i\omega - E(K))}
- \int_{-\infty}^{\infty}\int_{d\Lambda}\frac{d\omega dK}{4\pi^2}
\frac{u(K'LKR) u(RKLK')}{(i\omega - E(K))(i\omega + E(K))}\nonumber \\
		&  &
-\frac{1}{2} \int_{-\infty}^{\infty}\int_{d\Lambda}\frac{d\omega
dK}{4\pi^2} \frac{u(-KKLR) u(LR-KK)}{(i\omega - E(K))(-i\omega -
E(K))} \label{oneloop} \\
&\equiv & ZS + ZS' + BCS
 \eeqr
 where $\int_{d\Lambda}$ means the momentum must lie in one of the four
slices in Fig.7.

 The reader is reminded once again that the names ZS, ZS or BCS refer only
to the topologies
of the graphs. To underscore this point,especially for readers
who have seen a similar integral in zero sound calculations, we will now
discuss the ZS graph. In the present problem the loop momentum $K$ lies
within a sliver $d\Lambda$ of the cut-off. Both propagators have poles at the
point $\omega = -iE(k = \pm \Lambda)$. No matter which half-plane this lies
in, we can close the contour the other way and the $\omega$ integral
vanishes. This would be the case even if a small external momentum transfer
($Q = K_3 - K_1 << \Lambda$) takes place at the left vertex  since both
poles would still be on the same side. This is very different from what
happens in zero sound calculations where    the loop momenta roamed freely
within the
cut-off, and in particular, go to the Fermi surface.  In that case, the
integral becomes very sensitive to how the external momentum transfer $Q =
K_3 - K_1 $ and  frequency transfer $\Omega = \omega_3 - \omega_1$  are
taken to zero since  any nonzero $Q$, however small, will split the poles and
make them lie on different  half planes for $k < Q $ and the integral will be
nonzero.  It is readily seen that
 \beq
 \int_{-\infty}^{\infty}\int_{-\Lambda}^{\Lambda} \frac{d\omega
dk}{4\pi^2} \frac{1}{(i\omega - k)(i\omega  -k - Q +i\Omega ) } = \int_{-
\Lambda}^{\Lambda} \frac{dk}{2\pi} \frac{i}{\Omega +iq} (\theta (k) -
\theta (k+q))
 \eeq
 where the
 step function $\theta (k)$ is is simply related to the Fermi function:
$f(k) = 1- \theta (k)$.    If we keep $\Omega \ne 0$ and send $Q$ to zero we
get zero. On the other hand of we set $\Omega = 0$ and let $Q$ approach
zero we get (minus) the derivative of the (Fermi)  $\theta $ function, i.e., a
$\delta$-function at the Fermi surface . Thus reader used to zero-sound
physics should not be disturbed by the fact that the ZS graph makes  no
contribution  since the connotation here is entirely different.

\footnote{Before we move on the ZS' graph, let us notice another related fact.
Suppose we choose to find the one loop $\beta$-function using the field
theory method. Then we will  calculate the  the four-point function $\Gamma
$in a cut-off theory and demand that it be cut-off independent. The same three
graphs will  appear in the expression for  $\Gamma$,  but the loop integrals
will in fact
go all the way up to the cut-off.  Consequently   the ZS graph will
make a contribution that is very sensitive to how the small external momenta
and frequencies are chosen. However since this contribution, when nonzero,
comes from integrating a $\delta$-function at the Fermi surface, it will not
have any sensitivity to the cut-off and will make no contribution to the
derivative with respect to the cut-off  i.e., to the $\beta$-function.  The
situation is parallel to what we saw in massless $\phi^4$ theory in Section II.
There the  expression for $\Gamma$ had an infrared divergence when $r_0 =
0$, but
 that did not affect the derivative with respect  to the upper cut-off
$\Lambda$.  On the other hand in the modern approach one never saw any
singular behavior even in the intermediate steps. }

Now  for the ZS' graph, Fig.6b, Eqn.(\ref{oneloop}). We see that $K$ must
lie near $L$ since $1=R$ and there is no RR scattering. As far as the
coupling at the left vertex is concerned, we may set $K =  L$ since the
marginal coupling has  no $k$ dependence. Thus $K + \pi = R$ and the
vertex becomes $u(RLLR) = -u$. So does  the coupling at the other vertex.
Doing the $\omega$ integral (which is now nonzero since the poles are
always on opposite half-planes) we obtain, upon using the fact that there are
two shells (a and b in Fig.7) near L and that $|E(K)| = |k| = |\Lambda |$,
\beqr
ZS' &=& u^2 \int_{d\Lambda \in L} \frac{dK}{4\pi |E(K)|} \nonumber \\
 &=& \frac{u^2}{2\pi} \frac{d|\Lambda |}{\Lambda}
 \label{ZS'}
 \eeqr
The reader may wish to check that the ZS' graph will make the same
contribution to the $\beta$-function in the field theory approach.

 The BCS graph (Eqn.(\ref{oneloop}), Fig.6c) gives a nonzero contribution
since the propagators have opposite frequencies, opposite momenta, but equal
energies due to time-reversal invariance $E(K)= E(-K)$.  We notice that the
factor of $\frac{1}{2}$ is offset by the fact that K can now lie in any of the
four regions a,b,c, or d. We obtain
 a contribution of the same magnitude but opposite sign as ZS' so that
 \beqr
 du &=& (\frac{u^2}{2\pi} - \frac{u^2}{2\pi} ) \underbrace{\frac{d|\Lambda
|}{\Lambda}}_{dt} \\
  \frac{du}{dt}      &=& \beta (u) = 0.
  \eeqr

  Thus we find that $u$ is {\em still} marginal.  The flow to one loop
  for $\mu$ and $u$ is
  \beqr
  \frac{d\mu}{dt}& =& \mu - \frac{u}{2\pi}\\
  \frac{du}{dt}&=& 0.
  \eeqr
  There is a line of fixed points:
  \beqr
  \mu & =&  \frac{u^{*}}{2\pi}\\
      u^{*} & & arbitrary.
      \eeqr
  Notice that $\beta$ vanishes due to a  cancellation between two diagrams,
each of which by itself would have led to the CDW or BCS instability.  When
one does a mean-field calculation for CDW, one focuses on just the ZS'
diagram and ignores the BCS diagram. This amounts to taking
  \beq
  \frac{du}{dt} = \frac{u^2}{2\pi}
  \eeq
  which, if correct, would imply that any positive $u$
  grows under renormalization. If this growth continues we expect a CDW.
  On the other hand,  if just the BCS diagram is
  kept we will  conclude a run-off for negative couplings leading to a state
with $<\psi_R \psi_L > \neq 0$.

		    What the $\beta$ function does  is to treat these
		    competing instabilities simultaneously  and predict  a
		    scale-invariant theory.

		     Is
		     this the correct prediction
		     for the spinless model, which as
we saw, had the marginal interaction in its interaction? Yes,  the exact
solution of Yang and Yang (1976) tells us there is no no gap till $u$ is of
order unity. If the RG  analysis
were extended to higher loops we would keep getting
		    $\beta = 0$ to all orders. This follows
		     from the
		    the Ward identity
		    in the cut-off continuum model
		    (De Castro and Metzner 1991)
		    which reflects the fact that in
this  model, the number of fermions of type $L$ and $R$ are separately
conserved. How do we ever reproduce
		    the eventual CDW instability known to exist in the exact
		    solution? The answer is as follows. As we move along the
		    line of fixed points, labeled by $u$,  the dimension of
		    various operators will change from the free-field values.
Ultimately the umklapp coupling , ($RR \leftrightarrow LL$) , which was
suppressed by a factor $(k_1 - k_2 )(k_3 - k_4)$ , will become marginal and
then relevant. If we were not at half-filling such a term  would  be ruled out
by momentum conservation and the scale invariant state, called a Luttinger
liquid, (Luttinger 1961, Haldane 1981),
will persist for all $u$.  While this liquid provides us with an example of
where the   RG  does better than mean-filed theory, it is rather special and
seems to occur
in $d=1$ systems where the two Fermi points satisfy the conditions for
{\em both} CDW and BCS instabilities.  In higher dimensions we will find
that any instability due to a divergent susceptibility is not  precisely
cancelled by another.

As an aside, note that in the ZS' and BCS diagrams, the integrand is a
function of just $\omega^2 + k^2 $ so that we have rotational (Euclidean)
invariance. In this case we can, if we wish, work with a disk of radius
$\Lambda$ in $\omega - k$ space rather than the rectangle of width
$\Lambda$
and infinite height. You may check that if we integrate out a shell of
thickness $d\Lambda$ in the $\omega - k$ space, we get the same
contribution
to the $\beta$-function.

The main results from this section are summarized in Table V.
\section{THE    RG  IN  $d>1$: ROTATIONALLY INVARIANT CASE AT
TREE LEVEL}
We now proceed to apply exactly the same approach to spinless fermions
in $d>1$.   The nontrivial  geometry of the Fermi surface  will play a
profound
role and the application of the   RG  leads to some phenomena not seen in
the usual applications to critical phenomena, which is of course what makes
it interesting. We start with the simplest case of a circular Fermi surface  in
$d=2$. The extension of the analysis to spherical surfaces in $d=3$ is very
direct
and will be explained. Only in Section X will we take up the nested Fermi
surface
that leads to CDW formation.
\subsection { Tree Level in $d=2$ }

Let us begin with a
square lattice containing
spinless fermions at very low filling. In this case we
can approximate
the free-particle dispersion relation  as follows:
\beqr
E &=&  -\cos K_x - \cos K_y  \\
    &\simeq &-2+K^2/2
    \eeqr
    Since the problem now has rotational invariance, it is isomorphic to the
problem of electrons in free space with a dispersion relation
    \beq
    E = \frac{K^2}{2m}
    \eeq
      {\em We will therefore study the latter since this allows us to make
contact with Landau's work on it.} Let us  introduce  a chemical potential so
that the ground state is a Fermi circle of radius $K_F = \sqrt{2m\mu}$. Next,
we linearize the dispersion relation
near the Fermi surface   :
\beqr
\varepsilon(K) &=& E(K) - \mu \\
		       &=& \frac{K^2 - K^{2}_{F}}{2m} \\
			  &=&  \frac{kK_F}{m} +  O(k^2)\ \ (k=|K| - K_F)\\
			  &\equiv & vk
			  \eeqr
where $v$ is the Fermi velocity.
			     The free-field action now becomes
			     \beq
			     S_0 = \intom \iT \intk
			     \left[ \sib  (\omega \theta k)
(i\omega - vk) \psi(\omega \theta k) \right]
\eeq
To obtain this, we must  replace the measure  Kdk by $K_F dk$, as the
difference will prove irrelevant under  the   RG  and  absorb a factor
$\sqrt{K_F}$ into each of the two Fermi-fields.
Mode elimination proceeds just as in $d=1$: we eliminate all
modes obeying  $\Lambda /s \le |k|\le \Lambda$ for all $\omega$ and
$\theta$.
The same scaling of $ k$
and $\omega$ and the fields as in Eqn.(\ref{rescale}) leaves $S_0$
invariant.   The only  difference is that the internal index $i$
which took just two values (L = left and R = right) is now
replaced by a continuous
parameter $\theta$. {\em The $d=2$ theory thus looks like an
integral over one dimensional theories, one for each direction,
each with infinitesimal weight. }

Next we dispense with rotationally invariant quadratic interactions as in
$d=1$: either they modify the chemical potential,  rescale existing terms, or
are irrelevant.
 (Also irrelevant is  the difference between K and $K_F$ in the measure.) Let
us then move on to the really interesting case of the quartic interaction. This
has the general form
\beq
\delta S_4 = \frac{1}{2!2!}\int_{K\omega \theta }
\sib (4) \sib (3) \psi (2) \psi
(1) u(4, 3, 2, 1)
\label{s42}
\eeq
where
\beqr
\sib (i) &= &\sib (K_i, \omega_i, \theta_i) \ \  etc., \\
\int_{K \omega \theta } &= &\left[
\prod_{i=1}^{3}\int_{0}^{2\pi}\frac{d\theta_{i}}{2\pi}\int_{-
\Lambda}^{\Lambda}\frac{dk_i}{2\pi}\int_{-
\infty}^{\infty}\frac{d\omega_i}{2\pi}\right] \theta (\Lambda - |k_4|)
\label{theta9}\\
k_4 &=&
|\vec{K_4} | - K_F.\label{s422}
\eeqr
Much of the new physics stems from this measure for quartic interactions. Let
us understand it in some detail focusing on the factor  $\theta (\Lambda -
|k_4|)$ which plays a crucial role.

We start with a quartic  interaction invariant under space-time translations
and Fourier transform it,  getting  an integral over four $\omega$'s subject to
a
$\delta$-function constraint and four momenta $\vec{K}_i$ subject to a
momentum conserving $\delta$-function. Let us now eliminate one of the four
sets of variables, say the one numbered $4$,  by integrating them against the
$\delta$-functions. The $\omega$-integral is easy: since all $\omega$'s are
allowed, the condition $\omega_4 = \omega_1 + \omega_2 - \omega_3$ is
always satisfied for any choice of the first three frequencies. The same would
be true for the momenta if all momenta were allowed. But they are not, they
are required to lie within the annulus  of thickness $2\Lambda$ around the
Fermi circle.
Consequently, if one freely chooses the first three momenta from the annulus,
the fourth could have a length  as large as  $3K_F$. The role of $\theta
(\Lambda - |k_4|)$ in Eqn.(\ref{theta9}) is to prevent exactly this.

Now such a $\theta$-function will arise in the $\phi^4$ theory also if we
eliminate $\vec{k}_4$
by integrating it against the momentum conserving $\delta$-function. Its
effect is however quite different there. For one thing, even if we ignore it,
nothing very serious happens
since the first three $k$'s lie in the tiny  ball of
size $\Lambda$ and $k_4$ can never stray too far, being bounded by
$3\Lambda$. In particular, it will be controlled by $\Lambda$ and decrease
with it. In the present case, even if the first three momenta lie on the Fermi
surface, the fourth can be off
by an amount of order $K_F$ rather than $\Lambda$. Secondly, even if keep
the $\theta$ function in the $\phi^4$ case, its response to renormalization  is
very simple.  Under the action of
$\Lambda \rightarrow \Lambda /s $,
\beqr
\theta (\Lambda - |  k_4|)
&\rightarrow &  \theta (\Lambda /s - |\vec{  k_4}|)\\
					&=&
					\theta (\Lambda - s|\vec{  k_4}|)\\
					&=& \theta
					(\Lambda - s|\vec{  k_1} + \vec{
k_2} - \vec{  k_3}|)\\
					&=& \theta (\Lambda - |\vec{  k'}_4|)
					  \label{thetaf}
					\eeqr
 {\em Thus the $\theta$ function of the old variables goes into exactly the
 same function of the new variables.} Since the rest of the integration
 measure goes into itself upon rescaling from   $k$ to   $k'$ (and absorbing
factors of $s$ into the new fields), we get the usual
 result
 \beq
 u'(  k') = u(  k'/s).
 \eeq
Upon doing a Taylor series in its arguments, we get the familiar result
that the constant part
$u_0$ is marginal and the other Taylor coefficients are irrelevant in $d=4$.

Let us try to do the same here starting with Eqns.(\ref{s42} - \ref{s422}).
We first reduce
the range of each $k_i$ by a factor $s$. Then we rescale all momenta to bring
the range back to the old value. We must finally see if the $\theta$ function
responds as it did above in the case of the $\phi^4$ theory. If it does, we
could
conclude that
$$u'(k', \omega' ,\theta ) = u(k'/s,  \omega' /s,  \theta). $$
{\em But it does not!} The problem is that $k_4$ is not a function of just
the other three little k's but also of $K_F$:
\beq
k_4 = |(K_F + k_1)  \vec{\Omega}_1 + (K_F + k_2)  \vec{\Omega}_2 -
(K_F + k_3)  \vec{\Omega}_3 | - K_F
\eeq
where $\vec{\Omega}_{i}$ is a unit vector in the direction of $\vec{K_i}$:
\beq
\vec{\Omega}_i = \vec{i} \cos \theta_i + \vec{j} \sin \theta_i
\eeq
where $\theta_i$ is the orientation of the unit vector along momentum
$\vec{K}_i$. (In $d=2$, we will use $\vec{\Omega}_i$ and $\theta_i$
interchangeably. )

It is now easy to check that if we  carry out the manipulation that led
to Eqn.(\ref{thetaf}) we will find:
\beq
\theta (\Lambda - |k_4 (k_1, k_2, k_3, K_F)|) \longrightarrow \theta
(\Lambda - |k'_4 (k'_1, k'_2, k'_3, sK_F)|).
\label{theta2}
\eeq
{\em Thus the $\theta$ function  after the  RG    transformation   is not the
same function of the new variables as the $\theta$ function before the
  RG    transformation   was of the old variables due to the fact that $K_F
\rightarrow sK_F$. }
As mentioned earlier, we cannot ignore the $\theta$ function, since unlike in
$\phi^4$ theory,
it is possible for very large $k_4$ (order $K_F$) to arise, even if the
first three are of order $\Lambda$.
We have a real problem implementing the   RG  program:  how are we to say
what the new coupling is if the integration measure does not come
back to its old form?

Before describing the solution this impasse, let us restate the problem in more
geometric terms.
Imagine that we have renormalized with a large $s$ and are down to a shell
of very small thickness, i.e., $\Lambda /K_F $ is very tiny. Thus all the
momenta are essentially on the Fermi surface and the only freedom is in their
direction, $\vec{\Omega}_i$ or $\theta_i$.
 {\em Now the point is that we cannot choose three of these angles freely, but
only two, if all vectors are to lie on the Fermi circle. }
For example, if we choose angles $\theta_1$ and $\theta_2$, the sum of the
corresponding vectors lies along the bisector of the angle between them. The
only way this initial state momentum can equal the final state momentum
$\vec{K}_3 + \vec{K}_4$ is for the final angles to equal the initial angles:
\beqr
Case \ I: \theta_3 &=& \theta_1 \\
	       \theta_4 &=& \theta_2
	       \eeqr
	       or
\beqr
Case \ II: \theta_3 &=& \theta_2 \\
	       \theta_4 &=& \theta_1 .
	       \eeqr
In the case of identical spinless fermions Cases I and II are physically
equivalent.

There is only one exception: if the initial angles are exactly opposed to each
other, leading to a total momentum $\vec{P} =0$, the final momenta are free
to point in any direction as long as they oppose each other:
\beqr
Case \ III: \theta_2 &=& -\theta_1 \\
		\theta_4 & = & -\theta_3.
		\eeqr
{\em In summary either $3$ and $4$ are slaved to (be equal to ) $1$ and $2$
or    $2$ and $4$ are slaved to (be opposite to ) $1$ and $3$.}

Let us now back off from the limit  $\Lambda /K_F=0$ and discuss the
problem when $\Lambda /K_F$ is small but not zero.  Figure 8 depicts the
situation.  First let us  ask for all pairs of  momenta that
\begin{itemize}
\item Lie within the annulus, and
\item Add up to some $\vec{P}$.
\end{itemize}
The construction in Fig.8 gives all of them. First we draw two annuli with
centers separated by
$\vec{P}$. They intersect in two regions (called
I and II in the figure) of size
of order  $\Lambda$ in each direction.         If we start at the center of the
left
annulus and draw a vector to any point in I or II, and then a vector from that
point to the  center of the
right annulus; we get two vectors that meet the twin
requirements listed above. For example, the initial vector $\vec{K}_1$ and
$\vec{K}_2$ correspond to choosing this point from region II. Since
$\vec{K}_3 + \vec{K}_4 = \vec{P}$, the latter pair must also stem from this
construction. The figure shows them linked to region I. It is clear that the
direction of the final vectors is within $\Lambda / K_F$ of the initial
vectors,
with
$\theta_3 \simeq \theta_2$ and $\theta_4 \simeq \theta_1$ for this choice.
(Had we chosen the final vectors to come from region II also, the other paring
would have occurred. )The figure makes it clear that if choose the orientations
of the first three vectors without paying attention to this restriction, i.e.,
choosing a point outside regions I and II (but inside the annulus) for
terminating $\vec{K}_3$, $\vec{K}_4$ can  end up being  much shorter or
longer than $K_F$.

The problem of the changing $\theta$-function may be stated as follows in
terms of this figure. Let us begin with some cut-off and a choice of four
angles that is allowed at that cutoff as per this construction. There is some
coupling $u$ for this choice of angles. If we reduce the cut-off, the allowed
choices of angles shrinks and a coupling that  was previously allowed may no
longer allowed. Since the angles $\theta$ play the role of internal (isospin-
like) degrees of freedom, we have a situation in which the range of internal
labels is changing under mode elimination. This makes it impossible to make
meaningful comparison of the couplings before and after since certain
processes are no longer allowed. (Imagine an $SU(4)$ theory renormalizing
to an $SU(3)$ theory.)

 {\em
Since we do not want the momentum or internal labels to change their
allowed  range of values (if we are  going to follow the standard   RG
procedure of comparing the couplings before and after to see how they are
flowing),
we will take the view that all $\theta$'s are allowed, but that
some
$u$'s are abruptly renormalizing to zero as the cut-off is reduced. }

 The situation is a lot clearer if we use a smooth cut-off for $k_4$:
 \beq
 \theta (\Lambda - |k_4| ) \longrightarrow e^{-|k_4|/\Lambda}
\label{soft}
\eeq
for now no process is disallowed, but only exponentially suppressed as we
renormalize. We may interpret that as the exponential decay  of the
corresponding  $u$ under renormalization.
\footnote{Nothing is gained by using a soft cut-off for the rest.} Now recall
\beq
k_4 = |K_F \underbrace{(\val{\Omega}{1} + \val{\Omega}{2} -
\val{\Omega}{3})}_{\val{\Delta}{}} +  k_1 \val{\Omega}{1} +k_2
\val{\Omega}{2} - k_3 \val{\Omega}{3}| - K_F
\label{modk}
\eeq
where $\vec{\Omega_{i}}$ is a unit vector in the direction of
$\vec{K_{i}}$.
In what follows we shall  keep just the $\val{\Delta}{}$ piece and
ignore the $O(k)$ terms.  This is because the only time the latter
are comparable to the former   is when both are of order $\Lambda$,
in which case $k_4 \simeq K_F $, and this regions is exponentially
suppressed by the smooth cut-off Eqn.(\ref{soft})  anyway.

Under the   RG    transformation  at tree level,
\begin{eqnarray}
\! \! \! \! \! \prod_{i= 1}^{3}
\intki\iTi \intomi e^{-(K_F / \Lambda)| |\Delta| -
1|} u(k\omega \theta)
\sib \sib \psi
\psi &\longrightarrow & \nonumber  \\  \! \! \! \! \! \! \! \!
\prod_{1}^{3}\intkpi
\iTi \intompi e^{-(sK_F / \Lambda)| |\Delta| - 1|} u(k'/s
\ \omega' /s     \ \theta)
\sib \sib \psi \psi  .
\end{eqnarray}
Let us write
\beq
e^{-(sK_F / \Lambda)| |\Delta| - 1|} = e^{-(K_F / \Lambda)| |\Delta| - 1|}
e^{-((s-1)K_F / \Lambda)| |\Delta| - 1|}
\eeq
so that the measures before and after have the same factor $e^{-(K_F /
\Lambda)| |\Delta| - 1|} $. Now that the measures are identical before and
after the   RG    transformation  we can  compare apples to apples and
identify  the new quartic coupling:
\beq
u'(k'\omega ' \theta )
= e^{-((s-1)K_F / \Lambda)| |\Delta| - 1|} u(k'/s \  \omega'
/s \  \theta ).
\label{nuu}
\eeq
We may conclude that the only couplings  that survive the  RG
transformation  without any  decay correspond to the cases where
\beqr
|\val{\Delta}{} | &=& |\val{\Omega}{1} + \val{\Omega}{2} -
\val{\Omega}{3} | = 1
\label{survivors}
\eeqr
In $d=2$ this equation has only three solutions:
\beqr
Case\ I:\ \val{\Omega}{3} &=& \val{\Omega}{1}  \ \ (Hence\ \
\val{\Omega}{2} = \val{\Omega}{4}) \label{case1} \\
Case\ II:\ \val{\Omega}{3} &=& \val{\Omega}{2}  \ \ (Hence\ \
\val{\Omega}{1} = \val{\Omega}{4}) \label{case2} \\
Case\ III:\ \val{\Omega}{1} &= &- \val{\Omega}{2}  \ \ (Hence\ \
\val{\Omega}{3} = -\val{\Omega}{4}) \label{case3}
\eeqr
       This result which was anticipated earlier, can be deduced algebraically
or
seen geometrically
       by considering Fig.8
	in the limit of  zero shell thickness
since this gives the allowed region in the hard cut-off scheme after
infinite amount of renormalization i.e., at the fixed point.

Notice that  we have an extra conservation
law at the fixed point $\Lambda / K_F =0$: not only is total momentum
conserved, the set of individual
momenta is  also conserved. The only exception is when the incoming
momenta are equal and
opposite. Then they add up to zero and the final momenta can be any two
opposite points
on the Fermi circle. In all cases, two of the momenta determine the other two
rather than three of them determining the fourth.

For couplings that do obey this condition, Eqn.(\ref{nuu}) becomes
\beq
u'(k'\omega ' \theta )_{|\Delta | = 1}
=  u(k'/s \  \omega' /s\  \theta )_{|\Delta| =
1}.
\label{nuu2}
\eeq

 Performing  a Taylor expansion in $k$ and $\omega$
 \footnote{A common concern frequently expressed by those familiar with
many-body physics of these fermions is this: is there not some
ambiguity on how the limit $\omega \rightarrow 0$
is to be taken as $k_i \rightarrow 0$ ? If so, how  does $u$  even
have a Taylor expansion at the origin of $\omega k $ space ? The answer
is that $u$ is just the bare coupling that goes into the cut-off theory.
It is not the full four-point function $\Gamma$ which is calculated in the
cut-off theory  by summing all Feynman diagrams with $u$ as the coupling,
$\Lambda$ as the cut-off,  and propagators whose momenta can go right up
the Fermi surface .
Indeed the $\Gamma$ so computed has all the above mentioned
singularities. In contrast $u$
is obtained by taking some analytic
(short range) interaction in the full momentum space and eliminating the
modes {\em outside} the cut-off. This procedure  cannot produce any
nonanalyticity. The situation is just as in $\phi^4$ theory where the
bare couplings in the action are analytic functions of momenta while
the full Green's functions are plagued with infrared singularities coming
from the soft modes.}
 and comparing  coefficients of
separate powers, we   conclude that the leading term, with no
dependence on either variables is marginal, while all
the rest are irrelevant.
We shall refer to this term as $u$ hereafter.

 We see that {\em the tree level fixed point is
characterized by two independent  functions} and not a handful of coupling.
They are
\beqr
u\left[ \theta_4(\theta_1 \theta_2 \theta_3 ) =
\theta_2 ;\theta_3 = \theta_1 ;
\theta_2 ; \theta_1 \right]
&=& F(\theta_1 ; \theta_2) = F(\theta_1 - \theta_2
\equiv \theta_{12} ) \ \ (rot \ \ inv ), \label{F}\\
u\left[ \theta_4(\theta_1 \theta_2 \theta_3 )
= \theta_1 ;\theta_3 = \theta_2 ;
\theta_2 ; \theta_1 \right] &=& - F(\theta_{12}
) \ \ Pauli  \label{-F} \ , \ \\
u\left[ \theta_4(\theta_1 \theta_2 \theta_3 )
= -\theta_3 ;\theta_3 ; -\theta_1 ;
\theta_1 ; \right] &=& V(\theta_1 ;
\theta_3) = V(\theta_1 - \theta_3 \equiv
\theta_{13} ) \ \ (rot \ \ inv ).\ \label{V}
\eeqr
 Note that the manifestation of the Pauli principle on F and $V$ is somewhat
subtle: F will not be antisymmetric under $1 \leftrightarrow 2$ since the way
it is defined above, we cannot exchange 1 and 2 without exchanging 3 and 4
at the same time.  On the other hand, since
 3 and 4 can be exchanged without touching 1 and 2 in the definition of $V$,
$V$ must go to $-V$ when $\theta_{13} \rightarrow \theta_{13} + \pi $.

 A concrete example is useful here. Let us begin with the nearest neighbor
 interaction of spinless fermions in $d=2$, transcribed into momentum space,
\beq
\frac{u(\vec{K}_4,\cdots \vec{K}_1)}{2!2!} = U_0 \left[ \sin (\frac{ K_{1x}
- K_{2x}}{2})  \sin (\frac{ K_{3x} - K_{4x}}{2})\cos(\frac{K_{1x} +
K_{2x} - K_{3x} - K_{4x}}{2}) + x \leftrightarrow y \right]
\label{NN2}
\eeq
and expand it to second order in $K$, (as we did in the kinetic energy)
obtaining a rotationally invariant interaction:
 \beq
 u \simeq (\val{K}{1} - \val{K}{2} ) \cdot (\val{K}{3}- \val{K}{4})
 \eeq
 If we evaluate this with all $|K|$ at $K_F$ and use Eqns.(\ref{F}, \ref{-F} ,
\ref{V} )
 we obtain
 \beqr
 F(\theta_{12} ) &\simeq & U_0 | \val{K}{1} - \val{K}{2}|^2\simeq  U_0 (
1 - \cos \theta_{12} )\\
 V(\theta_{13} ) &\simeq & U_0 \val{K}{1}  \cdot\val{K}{3} \simeq U_0
\cos \theta_{13}
 \eeqr
 Observe the effect of particle exchange on these functions.

Let us understand the significance of the coupling function $F$. If we
calculate
the four-point vertex $\Gamma$, it will be given to lowest order by $u$.
Since the
only $u$'s that survive require that the final directions $\theta_3$ and
$\theta_4$
equal the initial ones $\theta_1$ and $\theta_2$ up to a permutation, only
forward
scattering exists at the fixed point. \footnote{This is true at lowest order
when $\Gamma = u$. But higher orders in the forward scattering coupling
can only give forward scattering.}
The only exception occurs  when the incoming particles
have opposite directions: then they can scatter to another pair with equal and
opposite
momenta, the corresponding amplitude being given of course by $V$.

When we study noncircular Fermi surfaces, we will find yet another coupling
function $W$
that
survives if there is nesting. This  will correspond to processes in which
the  momentum transfer equals the nesting vector  $\vec{Q}_N$.

 \subsection{Tree Level analysis in $d=3$}
 Let us repeat  the preceding analysis in $d=3$ where the Fermi surface  is
 parametrized by two angles $\theta$ and $\phi$. All integrals
 over $\theta$ get replaced by integrals over solid angle. The process of mode
 elimination and rescaling of fields and momenta $k$ proceeds exactly
 as in $d=1$ or $d=2$ since the coordinates on the Fermi surface   play
 the role of an internal variables (like isospin) and are
  unaffected by the RG  transformation. We will then end up with
Eqn.(\ref{nuu}). The equation for
  the couplings
  that survive is still the same as Eqn.(\ref{survivors}), but now
  the unit vectors $\vec{\Omega}_i$
  can point anywhere in three dimensions.
This in turn
  means
   that
 the condition $|\val{\Delta}{}| = 1 $ not only has the
 solutions given in Eqns.(\ref{case1}, \ref{case2} , \ref{case3} )
 but a continuum of others.  First consider
 Eqns.(\ref{case1} - \ref{case2}) which tell us that
 $\val{\Omega}{3}$ and $\val{\Omega}{4} $  must coincide with
 $\val{\Omega}{1}$ and $\val{\Omega}{2} $  up to a permutation.
 In $d=3$, the former can rotate about their sum. In other words, Figure 8
 with zero shell thickness must now be viewed as depicting not the
 intersection of two circles, but two spheres. Thus the vectors 3 and
 4 do not have to coincide  with 1 and 2
 (up to a permutation) but can rotate around a cone with opening angle equal
 to that between 1 and 2. The planes containing $1$ and $2$ can have an
angle $\phi_{12,34}$ with the plane containing $3$ and $4$. (The  cases I
and II considered in $d=2$ correspond to  $ \phi_{12;34}= 0 $ or $\pi$.)

Let us review this.  The incoming particle momenta $1$ and $2$
lie on the Fermi sphere. Their sum lies in the  plane they define  and bisects
the angle between them.
The final particles also on the sphere, can give the same sum by lying
anywhere on the
cone generated by rotating the incoming pair around their sum.

  Although
the individual momenta are no longer conserved, we have the additional
requirement that {\em the angle between the
final pair is the same as the angle between the initial pair:}
\beq
\vec{\Omega}_1 \cdot \vec{\Omega}_2 = \vec{\Omega}_3 \cdot
\vec{\Omega}_4 .
\eeq

Once again the only exception is
when the incoming momenta add up to zero.
In this case the final momenta are free to point in any direction as long as
they are mutually opposite.
 {\em Thus couplings corresponding to non-forward
 scattering (in which the initial and final directions are not the same) do not
vanish under the   RG transformation, but survive as marginal couplings. }
 The fixed point is then characterized by a function
 \beqr
  F &= &F(z_{12}, \phi_{12;34})\\
      &\equiv & F(z, \phi)\label{3df}
      \eeqr
      of two variables: $$z_{12} = \val{\Omega}{1}  \cdot\val{\Omega}{2}
$$
 and $\phi_{12;34}$ which is the angle between the planes containing 12 and
34 respectively.

 In addition to this we still have the $V$ function coming from
Eqn.(\ref{case3}) with
 $V = V(\val{\Omega}{1}  \cdot\val{\Omega}{3} )$.

 If we go back to nearest neighbor  coupling evaluated in this case we find:
 \beqr
 F &=& U_0  (1-z_{12}) \cos{\phi} \label{DP} \\
 V &=&U_0 \  z_{13}
 \eeqr
 Notice that since 3 and 4 are not slaved to 1 and 2 anymore, we can
exchange just the latter. This causes
 $\phi \rightarrow \phi + \pi $, which in turn changes the sign of F.
As for $V$ , it changes sign under $z_{13} \rightarrow -z_{13}$.

 Table VI and VII contain a list of fixed points, couplings,  and their flows
including ones to be discussed later in the paper.
\section{  RG  IN $d>1:$ ROTATIONALLY INVARIANT CASE AT ONE
LOOP}
We have found that the couplings $F$ and $V$  are marginal at tree level.
The next thing to do is to buckle down, as we did in $d=1$,  and go to the
one loop graphs and see if they tilt the marginal ones towards relevance or
irrelevance or preserve marginality.
We first do the analysis in $d=2$ and then discuss briefly the changes
encountered upon going to $d=3$.
\subsection{  RG  for the two-point function}
Just as in $d=1$, mode elimination produces at one loop (tadpole graph as in
Fig.5)  a nonzero change in the quadratic term in the form of a chemical
potential. To retain the old Fermi surface, we must find an input $\delta
\mu^*$ which will reproduce itself under the   RG transformation. Carrying
out the same analysis as in $d=1$ we find
\beqr
\delta \mu^* ( \omega \ k \ \theta \ )  &= &\int \frac{d\omega' \ dk'\ d\theta'
}{(2\pi )^3} \frac{F (\theta - \theta' )}{i\omega - v^* k} \\
&=& \int_{-\Lambda}^{0}\frac{dk'}{2\pi} \int_{0}^{2\pi}
\frac{d\theta'}{2\pi}
F (\theta - \theta' ) \label{2dself}\\
&=& \delta \mu^{*}
\eeqr
where the last equation signifies that $\delta \mu^{*}$ is a constant
independent   of $\omega \ ,k \ $ or $\theta$.

    We must begin with this   quadratic term if it is to reproduce itself under
mode elimination.
We may also see this  as the counter term one must
add as the interaction is turned on to maintain the same Fermi surface.
 The same holds in
 $d=3$:
\beqr
\delta \mu^* ( \omega \ k \ \val{\Omega}{} \ )  &=& \int_{-
\Lambda}^{0}\frac{dk'}{2\pi} \int \frac{d\Omega'}{4\pi^2}
F(z=\val{\Omega'}{}  \cdot  \val{\Omega}{},\phi =0 )\\
&=&  \delta \mu^{*}
\label{3dself}
\eeqr

Note that only the forward scattering $F$ (
with $\phi =0$ )  enters this equation.

   \subsection{The one loop $\beta $-function for $F$}
Now we turn to the real issue: the renormalization of the quartic couplings.
Since  some of these couplings (called $F$ and $V$) proved to be marginal at
tree level, we must go, as we did in $d=1$, to one loop to see if they flow.
The analysis will be done using the modern approach,
though a passing remark may be
occasionally made about the field theory approach.The same  diagrams that
appeared in $d=1$ will appear, the integrals will be of the  same form as
Eqn.(\ref{oneloop}),  but the range of loop  momentum integration will reflect
the higher dimensionality.
Eqn.(\ref{flow}) gives in this case (suppressing vector symbols)
\beqr
du(4321) &=& \int_{-\infty}^{\infty} \int_{d\Lambda} \frac{d\omega
dK}{4\pi^2} \iT \frac{ u(K + Q, 3,K,1) u(4,K , 2, K +    Q)}{(i\omega -
E(K))(i\omega - E(K+Q))}\nonumber \\
& &- \int_{-\infty}^{\infty} \int_{d\Lambda} \frac{d\omega dK}{4\pi^2}\iT
\frac{u(K+Q', 4,K,1) u(3,K,2,K+Q')}{(i\omega - E(K))(i\omega - E(K+Q'))}
\nonumber\\
& & -\frac{1}{2}\int_{-\infty}^{\infty} \int_{d\Lambda} \frac{d\omega
dK}{4\pi^2} \iT \frac{u(P-K,K,2,1) u(4,3,P-K,K)}{(i\omega - E(K))(-
i\omega - E(P-K))}\\
&\equiv & ZS+ ZS'+ BCS \label{oneloop2d}
\eeqr
where $Q = K_3 - K_1$, $Q' = K_4 - K_1$ and $P = K_1 + K_2$. The
subscript $d\Lambda$ on $\int_{d\Lambda}$ reminds us that both loop
momenta must be in the shells being eliminated. Although $K$ is restricted to
this region explicitly, the restriction on the other momentum, $K+Q$ in the
ZS graph; $K+Q'$ in the ZS' graph  and $P-K$ in the BCS graph is implicit.
The Feynman diagrams are shown in Fig.9.

On the left hand side of the above equation, we choose the variables $4321$
such that the corresponding coupling survives renormalization at tree level,
which means it must be an $F$ or a $V$.

 Let us begin with the renormalization of $F$. We set all external legs at zero
frequency and on the Fermi surface  ($k = \omega = 0$) since the dependence
on these variables is irrelevant.  As for the angles, we chose
 $\val{\Omega}{1} = \val{\Omega}{3}$. Consider the ZS diagram in Fig.9
given by the first  integral  in Eqn.(\ref{oneloop2d}). Since $Q=0$ here, we
know that if $K$ lies in the shell being eliminated, so does $K+Q$ for any
direction  of $K$.  In other words, $\theta$ runs over the full range. However,
this diagram vanishes for the
 same reason it did in the $d=1$ case: both poles in the $\omega$ plane are
on the same side. Even if we put in a small momentum transfer $Q <<
\Lambda$ at the left vertex, it will not change anything. This is because the
loop momenta are near $k = \pm \Lambda$ and it takes a minimum
momentum transfer of order $\Lambda$ to knock something from
 below $K_F$ to something above it and vice versa. This is in accord with the
general statement that the bare coupling has no singular limit at small
$\omega$ or
 $Q$.

 \footnote{What if we choose to follow the field theory approach? The
situation
 is exactly as in $d=1$. We must
 evaluate the one
 loop graphs contributing to $\Gamma$ which is  sensitive to whether we take
the
 the  $Q << \omega $ or vice versa.
 The  limit which gives a nonzero $\Gamma $ , ($\omega << Q$), makes a
contribution of the form
 $
 \int d\omega \rightarrow \frac{f(\val{K} {} ) -  f( \val{K}{} + \val{Q}{}
)}{E(\val{K}{} ) - E( \val{K}{} + \val{Q}{} )} $
 where $f$ is the Fermi function. As $Q \rightarrow 0$, this becomes a
$\delta$ function at $K_F$  and makes a contribution that is $\Lambda$-
independent
 and hence irrelevant to the $\beta$-function.}

 Now consider the ZS' diagram. {\em  Due to the momentum transfer
$\vec{Q'}$
 of order $K_F$ at the left vertex, not only is  the magnitude of the loop
 momentum  restricted to lie within the  shell being eliminated, its angle is
 also restricted to a range of order $d\Lambda / K_F$. } This is clarified by
 Fig.10.     The dark circles now
 represent the thin slices being integrated. The intersection regions,
 of order $d\Lambda^2$, give us the allowed loop momenta for the $\beta$
 function calculation: they lie in the shell and have the right momentum
 transfer $K_4 - K_1$. Of the eight intersection, only the
 four marked ones satisfy the condition of being on opposite
 sides of $K_F$ so that the $\omega$ integral survives.
 Since the $\omega$ integral gives a denominator of order $\Lambda$,
 the contribution to  $du $ is order $\frac{d\Lambda }{\Lambda}
 \frac{d\Lambda}{K_F}$
 so that the $\beta $-function vanishes as we take the limit
 $dt = \frac{d|\Lambda |}{\Lambda} \rightarrow 0$.

 It is also clear that if $\vec{Q'} = 0$ and  $\vec{Q}$
is not, (which obtains if $\vec{\Omega}_4 = \vec{\Omega}_1$) we can
repeat the above argument with $ZS \leftrightarrow ZS'$.

 You may check that in  the field theory approach we would get a $\beta$-
function that went as $ \frac{\Lambda}{K_F}$ Since we are ultimately going
to send $ \frac{\Lambda}{K_F} \rightarrow 0$, this will not matter in that
limit. But note that the two schemes do not agree in detail except for the
marginal flows.

 Finally for the same kinematical reason, the BCS diagram does not
renormalize $F$ at one loop. Consider  Fig.8 with $K_3$ and $K_4$
replaced by the two momenta in the BCS loop, $K$ and $P-K$. In each
annulus keep just two  shells of thickness $d\Lambda$ at the cut-off
corresponding to the modes to be eliminated. The requirement that $K$ and
$P-K$ lie in these shells and also add up to $P$ forces them into intersection
regions of order $d\Lambda^2$.  This means the diagram is just as ineffective
as the ZS' diagram in causing a flow. Thus any $F$ is a fixed point to this
order. Notice that unlike the $d=1$  $\beta$-function which vanished due to a
cancellation of two terms, the present one vanishes trivially.
 \subsection{The one loop $\beta$-function for $V$ }
 Let us now look at the evolution of $V$ .  We choose the external
 momenta
 equal and opposite and on the FErmi surface. The ZS and ZS' diagrams do
not contribute to
 any marginal flow for the same reason BCS and ZS' did not contribute
 to the flow of $F$: since $Q$ and $Q'$ are of order $K_F$, they are
kinematically  suppressed by an extra factor $ \frac{\Lambda}{K_F}$
 in the field theory approach and by $\frac{d\Lambda}{K_F}$ in the
 modern approach. But the BCS diagram  produces a flow in either approach.
We
 follow the modern approach. The flow is due to the following factors:

 \begin{itemize}
 \item  The loop angle can run freely over its full range
 because no matter  what value K takes in the shell being eliminated, the other
momentum
 $P-K = -K $ automatically lies in the shell.
			The two energies $E(K)$ and
$E(-K)$ are equal while the two $\omega$'s are equal and opposite.
 \item The $\omega$ and $k$ integrals behave as in $d=1$ and produce  a
 factor $dt =  \frac{|d\Lambda |}{\Lambda}$.
 \end{itemize}
 We find
 \beq
 \frac{dV(\theta_{1} - \theta_3) }{dt} = -\frac{1}{8\pi^2} \iT  V(\theta_{1} -
\theta )
 V(\theta - \theta_{3})
 \eeq
 This is an interesting example of the $\beta$-function for the coupling {\em
function}. Fortunately we can simplify the picture by going to angular
momentum eigenfunctions:
 \beq
 V_l = \iT e^{il\theta } V(\theta )
 \eeq
  and obtain an infinite number of flow equations, one for each angular
momentum $l$:
  \beq
  \frac{dV_l}{dt} = - \frac{V_l^2}{4\pi} \label{bcsflow}.
  \eeq
The flow tells us that the couplings $V_l$ , marginal at tree level, are
marginally relevant if  negative, and marginally irrelevant if  positive. The
former is just the BCS instability. As for the repulsive case, if we integrate
the flow
 we get
 \beq
 V_l (t ) = \frac{V_l (0) }{1 + t \left[ V_l (0) / 4\pi \right]}
 \eeq
 which is just the observation of Morel-Anderson (1962) that inter-electron
repulsion is logarithmically reduced if we develop an effective theory for the
modes close to $K_F$.

 If this analysis were to be repeated in $d=3$, the only difference would be
that
 the BCS  $\beta$-function for $V(z ) $ would be decoupled using the
Legendre polynomials $P_l(z) $ with $l$ odd. The decoupled equation would
have the same form as Eqn.(\ref{bcsflow}) with the same implications.

\subsection{Fixed point structure at one loop}
  Let us take stock: the tree level fixed point is characterized by
  two marginal functions $F$ and $V$ .  The function $F$ is marginal at one
loop also,
  while $V$ is  marginally relevant in an infinite number of ways,
  one for each
  angular momentum $l $, if attractive, and marginally irrelevant if
  repulsive. It appears that even if a single $V_l$ is negative, we run off
  to some other massive fixed point with a BCS gap. What if all the $V_l$'s
are
  positive? It appears that these couplings will renormalize to zero
  logarithmically and we will end up with a fixed point characterized by $F$.
  This turns out be incorrect, atleast  in principle, due to a reason
  first pointed out
  by Kohn and Luttinger (1965). This is tied to some irrelevant operators
which cannot be
  ignored.  Here is the point.  An irrelevant operator by definition is
something
that renormalizes to zero, not something you can set equal to zero at the
outset without any consequences. Before  it renormalizes to zero, it can
modify the flow of the relevant couplings. Recall the case of $u_0$, the scalar
coupling in $d=4$. Although it was irrelevant, adding it to the gaussian fixed
point  generated a mass term $r_0$ which then took off. A very similar thing
happens here: an irrelevant term produces a small  negative  BCS coupling ,
which then takes off. This subtle  issue is discussed in  the next subsection.

 We discuss here another irrelevant term which does not destabilize the fixed
point, but
 modifies our description of it.

  Consider the sunrise diagram, Fig.11. In  mode elimination this diagram
  comes from taking two quartic terms and seeing how they
  feed back on the quadratic term.Though it is
  also of order $u^2$, it has  two loops and may be ignored in the one loop
discussion we are having. But if one evaluates it, one finds it is irrelevant
due
to phase space restrictions in the limit of small $\Lambda /K_F$.
\footnote{The details of the evaluation  will not be provided here. The
interested reader is asked to do the phase space analysis.}  However, before
we reach this limit, it can produce interesting effects that the   one loop
analysis did not include.

    Let us  write its contribution  as
  $-\Sigma (k \omega )\sib \psi$ where $\Sigma$ is called the {\em self
energy}. ( It has  no $\theta $ dependence here
  due to rotational invariance.)   If we Taylor expand $\Sigma$  as follows:
  \beq
 \Sigma (k \omega ) = \Sigma (00) + i \omega \underbrace{( 1 - Z^{-1}
)}_{\frac{\partial \Sigma}{\partial i\omega}} + k \frac{\partial
\Sigma}{\partial k} + \ irrelevant \ \
 pieces
 \eeq
 we see that
 \begin{itemize}
 \item $\Sigma (0 0) $ is to be eliminated using a counter-term $\delta \mu $
of order $u^2$.This is to ensure that $K_F$ is unchanged.
 \item $( 1 -Z^{-1}  )$ changes the coefficient of $i\omega $ to $Z^{-1}. $
 \item $\frac{\partial \Sigma}{\partial k} $ changes the coefficient of k from
$K_F / m$ to $(1 + \frac{m}{K_F} \frac{\partial \Sigma}{\partial k} )K_F
/m. $
 \end{itemize}
 At the free -field fixed point we rescaled the field to keep the coefficient
of
both the quadratic terms fixed.  Now we have seen that there is a manifold of
fixed points parametrized by nonzero $u$.  In this case  $i\omega $ and $k$
may receive different contributions from eliminated modes and there is no
rescaling which will keep both terms fixed.   So we will keep the coefficient
of $i\omega $ fixed at unity , i.e., define
 \beq
 \psi' = s^{-3/2} Z^{-1/2} \psi
 \eeq
 This has two effects. The coefficient of $k$ changes as follows:
 \beqr
 \frac{K_F}{m} &\rightarrow & Z ( 1 + \frac{m}{K_F} \frac{\partial
\Sigma}{\partial k} )\frac{ K_F}{m} \nonumber \\
			&\equiv & \frac{K_F}{m^*}
			\eeqr
 which defines the effective mass $m^*$.

 Next, the new quartic coupling is given by
 \beq
 u' = ( u + \delta u )Z^2
 \eeq
 where $\delta  u $ is the contribution we have already discussed. Question:
does this modify the $\beta$ function we calculated? Answer: not to order
$u^2$ since $Z$ deviates from unity at  order $u^2$ and  this  produce
changes of order $u^3 $ in the equation above.

 How about the fact that $m^{*}$ is now moving as we renormalize ? Upon
looking at the kinematics of   the sunrise diagram one can tell that as
$\Lambda$ goes to zero, its contributions will vanish. Thus although $m$ will
evolve to $m^*$ in the early stages of renormalization, a fixed point
characterized by some  $m^*$ will emerge asymptotically.

 How can we have a fixed point in a theory where there is a nonzero
dimensionful parameter $K_F$? The answer is that the fixed point theory
described above has   no knowledge of $K_F$: the ZS graph gets all its
contribution from the delta function at the Fermi surface, while the other two,
which know about $K_F$, are suppressed by the factor $\Lambda /K_F$ and
vanish at the fixed point.This situation is not changed by going to higher
loops, as will be shown in the next section on the $1/N$ expansion.

 \subsection{The Kohn-Luttinger Effect}
  The flow of $V_l$ was such that attraction led to instability, while
repulsion
meant downward renormalization to 0. There is no doubt the former is true,
but the latter is incorrect in principle.
  Now the fault is not with the solution to Eqn.(\ref{bcsflow}),
  but with the equation itself. In deriving it, we ignored the
  contribution from the ZS and ZS' channel graphs  on the grounds
  that they were finite   and down by powers of $\Lambda / K_F$,
  which made them unimportant in the limit $\Lambda / K_F \rightarrow 0.$
  But there is a surprise waiting for us if we go ahead and compute their
  contribution to the flow. As shown in  the Appendix , the modified flow in
  $d=3$ is (upon setting all positive numbers independent of $l$ and $t$ to
  unity):
  \beq
  \frac{dV_l}{dt} = - \frac{V_l^2}{4\pi}  - \frac{V^2 (\pi )
\lambda^{7/4}}{l^{15/2} \left[ \lambda^{7/4} + l^{-7/2}
\right]^2}\label{klbcsflow}.
  \eeq
where $V(\pi )$ is the backward scattering amplitude in the BCS channel and
$\lambda  = \Lambda /K_F$. Notice that as $\Lambda /K_F \rightarrow 0$,
the second term vanishes as
$(\Lambda /K_F)^{7/4}$ which is why we ignored it. (The strange power
comes in because the intersection region scales in a special way when the
momentum transfer is $\simeq 2K_F$, which is the region that dominates this
piece; note the backward scattering amplitude in the answer.) Why  do we
care about this piece?

Let us imagine that we are just beginning our   RG .
The input potential is the projection on the angular momentum $l$
channel of some  short range potential.  It follows that $V_{l} (t=0)
\simeq e^{-l} $ as $l \rightarrow \infty $ in order that the sum of such
coefficient times the $P_l (z)$ and all the derivatives of the sum
with respect to
$z$ converge to given analytic function $V(z)$.
By contrast, the second term, at fixed $\lambda$ goes as $l^{-15/2}$ as
$l \rightarrow \infty .$
It is clear that as soon as the flow begins, the exponentially small
initial coupling $V_{l}(0)$
will very quickly be driven to negative values by the second term.
Thereafter both terms will drive the instability.

This is the   RG  version of the famous Kohn-Luttinger argument,
which is discussed at some length in the  Appendix . The argument
implies that at $T=0$, the fixed point we studied   always faces the
BCS instability.  However one is still interested in  this fixed point
characterized by F.  The reason is that $T_c$ for the Kohn-Luttinger
superconductor could be very low, or the $l$ value for pairs, absurdly
high. Thus if we imagine a tiny nonzero temperature above this $T_c$,
the instability will disappear. (For a recent survey of the Kohn-Luttinger
effect see Baranov {\em et al}  1992).) Recall from our
analysis in Section III that a temperature $T = 1/\beta$
leads to  an imaginary  time coordinate of range $0 \le \tau \le \beta$.
In other words a quantum system infinite in space and at inverse temperature
$\beta$ is mapped by the path integral method to a  system in $d+1$
dimensions which is infinite in the spatial direction and of width
$\beta$ in the (imaginary) time direction. As we renormalize, the thickness in
the time direction
(in the sliding units) will get reduced by $s$ just like the correlation
length,
and just unlike the momenta. Thus we will flow to smaller $\beta$ or
larger $T$, the fixed point being
$T = \infty$. But if the cross-over is  very slow,  then in the interim
the fixed point described by $F$ will control the physics. Interestingly
enough there are many real world systems described in exactly that
fashion. More on this in the section on Landau's Fermi Liquid Theory
Table VI contains a summary of results from the one loop analysis.

\section{$\ \ \ $THE 1/N PICTURE, LEAP TO ALL LOOPS}
So far we have followed the   RG  program to one loop in some detail.
We must now see what happens at all loops. In general this would be a
formidable
problem.  Luckily in the present problem a great simplification arises
 in the limit $\frac{\Lambda}{K_F}\rightarrow 0$.
The presence of this small parameter allows us to relate  the  sum
over all loops
to the one loop result.

Now a similar thing happens
in theories where
$N$, the number of species of fields,
becomes very large. In this case, in the limit
$1/N \rightarrow 0$, it is possible to sum over all loops and the answer is
expressed in terms of                                the one
loop graphs. It is also possible to correct this answers in powers of   $1/N$.
This is called the $1/N$ expansion.

We will begin with a review of the $1/N$ approximation. It will then be
shown that
$\Lambda /K_F$ plays the role of $1/N$. We begin as usual in $d=2$.

\subsection{ $1/N $ in d=2 }

Consider a  $\phi^4$ theory with  action
\beq
S_0 = -\sum_{i=1}^{N} \phi_{i}^{*} \frac{k^2}{2} \phi_i - \frac{1}{N}
\sum_{i=1}^{N} \sum_{j=1}^{N} \phi_{i}^{*}  \phi_i V_{ij} \phi_{j}^{*}
\phi_j .
\label{1overnaction}
\eeq
Lots of integrals are suppressed and only the internal index is highlighted
since what we are about to say is independent of dimension. All we need to
note is that
 there are $N$ species of fields (or particles) and they have a quartic
interaction $V$. The interaction has a factor $1/N$ in front of it to ensure
that
we have a nontrivial limit as $N \rightarrow \infty$.
 Note that in the interaction vertex, if an $i$ and a $j$ come in, the same
indices also exit as
 shown in Fig.12.

Let us look at a four-point function to one loop, as shown in Fig.12. Among
the one loop graphs, only the first is of the same order as the tree level
graph:
this is because the extra factor of $1/N$ coming from the extra vertex is
compensated by a sum over the loop index which is free to take all values.
This in turn was because the index $i$ that came in went out at once, leaving
the loop
index $l$ free to roam over all values.   By  contrast, the external indices
have
insinuated themselves into the loops in the other two diagrams (ZS' and BCS)
and there is no sum over the indices there. These graphs are then of order
$1/N^2$ and hence suppressed by a factor $1/N$ relative to the tree graph.  It
is clear that the  sum over iterated ZS loops ("the bubble sum")  gives the
leading behavior (in $1/N$) of the four-point function.

 Note that the $\beta$-function of this theory is completely given by the one
loop answer. This is because iterates of the ZS loop ("the bubble sum")
merely produce the n-th power of $\ln \Lambda$ when iterated $n$ times
while higher order terms in $\beta$ come from subdominant logarithms. This
conclusion is also evident from Weinberg's (1993) discussion o the graphical
content of the effective action.

Let us look at our theory now and consider a four-point function $\Gamma
(\theta_2 , \theta_1 , \theta_2 , \theta_1 )$ in obvious notation with all
external
$\omega = k = 0$. Return to Fig.12, this time using integrals over $\theta$'s
in
place of a sum over  discrete indices. In the ZS graph, where the incoming
"index" $\theta _1$ immediately exits, there is no momentum transfer at the
left (and hence right) vertex.
Thus the loop angle runs over the full range $0 -
2\pi$. In the ZS' and BCS diagrams on the other hand, there is a large
momentum transfer $Q'$ or large total momentum $P$ (Figures 8 or 10) tell
us that   loop angles must lie within $\Lambda  / K_F$ of the external angles.
In other words, the external angles have insinuated themselves into the loops
and frozen  the loop sum. Since the ratio of the ZS' and BCS  graphs to the
ZS  graphs is $\Lambda /K_F$, we expect that this ratio will play the role of
$1/N$.

It should not be too surprising  that we have a $1/N$ description available for
us here: the noninteracting $d=2$ theory was written as an integral over the
internal index $\theta$ which labeled pseudo two-dimensional theories, (with
a phase space $dkd\omega$), one for each direction.
But what should be the $N$ assigned to this integral (rather than sum) over
$\theta$?
Is it infinity, since there are infinite directions or is it of order unity
since each
has infinitesimal measure? Let us sharpen the analogy with $1/N$ analysis to
answer this question.
As a first step let us write the free-field action with all factors of $K_F$
intact:
\beq
S_0 =
\intk \iT K_F \intom \left[ \sib (i\omega - v^* k) \psi \right]
\label{i/naction}
\eeq
Now chop the angular integration into regions of width $\Delta \theta =
2\Lambda /K_F $
so that the annulus breaks up into $N = 2\pi K_F /2\Lambda $ cells labeled
by an index $i$. The momentum of a point within a cell $i$ is
\beqr
\vec{K} &=& K_F \ \val{\Omega}{i} + k_{i}\  \val{\Omega}{i} + k_{\perp
i}\  \val{t}{i}  \\
  &\equiv & K_F\  \val{\Omega}{i} + \val{k}{i}
  \eeqr
  where $\val{\Omega}{i}$
  is the unit  radial vector at  the center of cell $i$,
$\val{t}{i}$ is a unit tangent, $k_i$ and $k_{\perp i}$ are radial and angular
displacements from the cell-center. We refer to the first piece of order $K_F$
as the large momentum and the other as the small one.
  The measure per cell is
  \beqr
  \intk \int_{-\Lambda /K_F}^{\Lambda /K_F}K_F \frac{d\theta }{2\pi} &=
&\intk \int_{-\Lambda }^{\Lambda } \frac{dk_{\perp} }{2\pi} \equiv
   \int \frac{d^2 k}{(2\pi )^2}.
  \eeqr
  In this notation
  \beq
  S_0 = \sum_{i=1}^N \int \sib_{i} (\val{k}{i} \omega_i ) \left[ i \omega -
v^* k_i \right]
  \psi_i (\val{k}{i} \ \omega_i ) \frac{d^2k_{i}d\omega_{i} }{(2\pi )^3} .
  \eeq
  We now write down an interaction term
  \beqr
  S_F &= & -\frac{1}{K_F} \sum_{i, j = 1}^{N} \int  \sib_{j} (\val{k}{4}
\omega_4 )\psi_j (\val{k}{2} \ \omega_2 ) F_{ij} \sib_{i} (\val{k}{3}
\omega_3 )\psi_i (\val{k}{1} \ \omega_1  ) \\
F_{ij} &=& F(\val{\Omega}{i}  \cdot\val{\Omega}{j} ) \\
\int &=& \int \prod^{4}_{1}\frac{d^2k_i d\omega_i}{(2\pi )^3} 2\pi \delta
(\omega_1 +\omega_2 - \omega_3 - \omega_4 ) (2\pi )^2 \delta^{(2)} (
\val{k}{1} + \val{k}{2} - \val{k}{3} - \val{k}{4} )\label{SF}
\eeqr
Notice that the interaction conserves momentum: the fact that the index $i$
appears once in a $\psi$ and once in $\sib$ means the large momentum  is
conserved. The small one is also conserved because of the explicit
$\delta^{(2)} $ function.  If we express all frequencies and momenta in units
of $\Lambda$, it will be found that
the only place where $\Lambda$ appears will be front of the interaction term
in the form $\Lambda /K_F $ exactly playing the role of $1/N$.  The proof is
left as an exercise.

Now, this interaction is not exactly the one we have dealt with so far; it
forbids certain processes that were previously allowed. Consider a
process where the initial particles have momenta $\val{K}{1}\ \val{K}{2} $
from cells $i_1 \ i_2$ respectively adding to a total $\val{P}{}$. If we draw
Figure 8, we will indeed find that $\val{K_3}\ \val{K}{4} $ lie with the same
intersection regions as $\val{K}{1}\ \val{K}{2} $ , {\em but this need not
mean they are in the same cells.} In other words, the intersection region can
straddle more than one cell, if we imagine these cells permanently etched on
the annulus. The old interaction would allow
the full intersection region, while the new one would only allow the part of it
in which the cell  indices match in pairs. But notice that the difference
between the two interactions shows up only in situations which are
kinematically suppressed by a power of $\Lambda /K_F$. On the other hand,
if we concentrate on just the sum over iterated ZS loops, with $\omega , Q <<
\Lambda$,  it can be verified that the two give identical answers.

The careful reader will find one problem with the $1/N$ analogy. In the usual
examples, $u$ in Eqn.(\ref{1overnaction}) is held fixed as $N$ varies. Here
$N$ is related to the cut-off $\Lambda$. As we lower the latter  to increase
the former, we must follow the flow of $F(\Lambda ) = F(1/N)$. However,
we have seen that there is no flow within the $F$ couplings  in the asymptotic
region. Thus $F_{ij}$ is essentially constant independent of $N$ for large
$N$.

 In view of the above, we make the employ the following  two-stage attack on
the fermion problem:
\begin{itemize}
\item Reduce the given problem in the full K-space  to a small $\Lambda $
(large N)  theory using the   RG .
\item Solve the resultant theory using the smallness of $1/N$ .
\end{itemize}

Consider now another type of coupling corresponding to $V$ . In schematic
form and the same notation used above,
\beq
S_V = \frac{\Lambda }{K_F} \sum_{ij}\int \sib_{-j} (4) \ \sib_{j} (3)  \
V_{ij}  \
\psi_{-i} (2) \psi_{i} (1)
\eeq
This interaction leads to a bubble sum (iterates of the one loop graph) in the
BCS channel. Here the coupling $V$ grows as $N$ increases since there is a
flow now. If we want to increase $N$ keeping $V(N)$ fixed, we must start
with weaker and weaker $V's$. If we do this, we can handle the BCS
problem also by summing over bubble diagrams.

Let us hereafter assume the BCS amplitudes $V$ are absent in the spirit
discussed earlier. Then we are left with iterated ZS loops and F's.
The physics of this is Fermi liquid theory, to be discussed in the next
section.

\subsection{$\ \ $ $1/N $ in $d = 3$}
We have seen that   RG  allows nonforward amplitudes to survive in $d=3$
as marginal interactions. If we divide the spherical Fermi surface  into
patches
of size $\Lambda^2$, and label them by an index $i$, this will run over
roughly $(K_F / \Lambda )^2$ values. The interaction $F_{ij}$ will not be of
the form Eqn.(\ref{SF}) since non-forward scattering amplitudes are allowed.
Let us however divide the possible interactions into a set involving just
forward scattering and the rest.
If we consider a forward scattering four-point function $\Gamma_{ij}^{ij} $
(with cell index conserved) it will be given as
the bare vertex plus a sum of iterated ZS diagrams all involving  only forward
scattering amplitudes.
The insertion of a large angle scattering anywhere will produce a factor
$\Lambda /K_F$. (The suppression factor is {\em not} $(\Lambda /K_F )^2$
because kinematics only restricts the the angle between 1 and 2 to be that
between 3 and 4, the plane containing the latter is still free rotate by the
angle
we called $\phi_{12;34}$. )If we consider instead $\Gamma_{ij}^{kl} $,
which is nonforward, we will find that it is given by just the tree level
coupling we put in the action.  All loop corrections   will be down by powers
of $\Lambda /K_F$. If we consider any response function to a soft (low wave
number) probe, these amplitudes will not enter the physics. If we compute the
lifetimes of particles,  these amplitudes will play a role, but phase space
will
again introduce powers of  $\Lambda /K_F$. Thus the nonforward amplitudes
live in shadow world, perhaps large in magnitude but small in their effects.

As for the BCS amplitudes, we can find the stable state with a gap, using the
$1/N$ expansion to limit ourselves to summing bubble graphs. Since
nonforward amplitudes never enter the computation, this explains why it is
permissible to  use the reduced BCS hamiltonian in which only scattering
within pair states is kept. (These amplitudes are just our $V$ 's of course.)

Note that as in any $1/N$ theory, the $\beta$-function at one loop (which we
have calculated) is all there is in the large $N$ or small $\Lambda$ limit.
\subsection{$\ \ \ $Two-point functions at large $N$}

Let us close by asking what happens to two-point functions at large $N$. Let
us look at some of the graphs contributing to $G^{-1}$ shown in Fig.13 if we
use the one loop fixed point action Eqn.(\ref{3dself}).  We see that all
iterates
of the tadpole are exactly canceled by the one loop counterterm or fixed point
chemical potential $\delta \mu^*$: whenever we can draw one more loop, we
can also use the counter term which exactly kills it. The sunrise diagram on
the other hand, brings in new corrections. But this diagram is suppressed by
$1/N$ and ignored in the limit we are interested in. {\em This means that in
this theory, we know the exact self-energy of a particle for the given
interaction F. }This will play a big role in  Fermi liquid theory to which we
will turn our attention shortly.  (One way to understand the above result is
the
following. When the shell thickness goes to zero, the limiting theory acquires
particle-hole symmetry. We are quite accustomed to this symmetry
determining the chemical potential, even on a lattice.  This will continue to
be
true even for nonspherical Fermi surfaces. )

The picture we developed in the large $N$ limit agrees with rigorous
calculations of Feldman {\em et al} (1990,1991,1992), as explained to me by
E.Trubowitz. According to these authors:
\begin{itemize}
\item The system always goes to a BCS state at $T=0$.
\item If the BCS diagrams are eliminated the rest define a convergent series
with a finite radius of convergence.
\end{itemize}

Further details may be found in the references given.
\section{$\ \ $LANDAU'S FERMI LIQUID THEORY}
Nearly four decades ago Landau (1956, 1957, 1959) attacked the problem of
interacting fermions at very low temperatures $T << K_F$. Assuming the
system evolved continuously from the noninteracting limit, he developed a
phenomenological theory which proved very successful, for example in the
study of Helium -3
(Pines and Nozieres (1963), Leggett (1975), Vollhardt and W\"{o}lfle
(1990), Baym and Pethick (1991), Lifshitz and Pitayevskii 1980.) {\em The
picture he arrived at, called  Fermi liquid theory , may be described in the
terminology of this paper  as the  the fixed point described by $F$. } For
many readers of Landau's work there was an element of mystery surrounding
some of the manipulations. This had to be so, since he substituted forty years
of subsequent developments   (the RG  in particular) with his remarkable
intuition.

Following his work, a diagrammatic derivation of  Fermi liquid theory  was
provided by Abrikosov {\em et al}  and is described in their book (Abrikosov
{\em al} (1963)). While the details were rather tedious, they did a lot to
clarify where everything came from. I believe the approach developed here,
using the   RG ,  provides an even simpler route to  Fermi liquid theory at
least for those with a certain background.

So let us pretend we do not know what  Fermi liquid theory is and ask how
we would arrive at it. (This paper specializes in $T=0$, and one expects the
results to work also for the crossover region at low $T$ with minor changes.)
Using the   RG  developed here, we would find that after repeated
renormalization, we would have mapped the initial problem to one with
$\Lambda / K_F \rightarrow 0 $.  Setting  $V=0$  eliminates the BCS
instability, i.e.,  implements Landau's requirement that there be  be no phase
transitions, leaving us  with a fixed point theory parametrized by the marginal
couplings $m^* , F(z\phi)\ $. The physics of this fixed point is Landau's
theory. The excitations of this  effective theory are the  {\em
quasi-particles}
of  Landau, in contrast to the "bare" particles created by the fields we began
with prior to mode elimination. The fact that the quasi particles have infinite
lifetimes  was established  by Landau using   phase arguments.  In the present
analysis, the lifetime terms, which appear  as   $O(\omega^2)$ terms in the
self-energy, are   irrelevant under the RG transformation. (At the free field
fixed point, they fall down by $s^{-1}$ under a factor $s$ reduction of the
cut-off.)

Is there any interesting physics  in this limit $\lk$ ? If the cut-off is going
to
zero in laboratory units, are there any Feynman diagrams at all or, do we just
read off all the physical scattering amplitudes from the vertices in the
action,
with no loop corrections, there being  nothing left to run in the loops? In
other
words, is the full vertex function $\Gamma$ the same as the (duly
antisymmetrized) couplings $F$ in the action?

The answer depends on whether we are looking at $\phi =0$ or $\phi \ne 0$,
ie.e at processes in which the final pair lies or does not lie in the same
plane
as the initial pair. As for the latter it is indeed true that tree level
amplitudes in
the action would  not be dressed by any loops due to the kinematics in $d>1$.
For example at one loop, the loop momentum would be  restricted in
magnitude and angle to size $\Lambda $ and $\Lambda /K_F$ respectively.
This, coupled with the fact that these diagrams have no singularities,  would
make them negligible and we would have $\Gamma = F$.

For  forward scattering however, the iterated loops in the    ZS channel, with
only  forward scattering couplings (F)  appearing in all the vertices, would
have  no restriction on the loop angle, which is why the graphs survived in the
large $N$ picture. However the {\em magnitude} of loop momentum $k$
would still be bounded by $\Lambda$. Why would  these graphs survive in
the limit of vanishing cut-off?
The answer is that  the integrand has a $\delta$-
function singularity at the Fermi surface  (derivative of the Fermi function),
if
the external frequency transfer is zero,  rendering  the integral  insensitive
to
$\Lambda$ as long it is non-zero.

Any attempt to introduce a nonforward amplitudes $F(\phi \ne 0)$ into the
iterated ZS loops  would bring in a suppression factor, call it $1/N$ or
$\Lambda /K_F$. Thus the forward and nonforward  amplitudes do not mix.
If we focus our attention on computing responses to soft probes ($(\omega \
Q ) << \Lambda$), we can ignore the nonforward $F$'s. {\em The resultant
theory is just Landau's  Fermi liquid theory . The function $F( z, \phi = 0)$
is
called the Landau  parameter $F(z)$. }

Landau's  $F(z)$ can be introduced in another equivalent  way. Let us begin
with our fixed point theory (hereafter in $d=3$)
\beq
S = \int \sib \left[ i \omega - v^* k \right] \psi \frac{dk d\Omega
d\omega}{(2\pi )^4}
+ \delta \mu^* \int \sib \psi + \frac{1}{2!2!} \int u \sib \sib \psi \psi
\label{flfp}
\eeq
where $u$ can contain nonforward amplitudes as well. We have seen that if
\beq
\delta \mu^* ( \omega \ k \ \vec{\Omega} \ )  = \int_{-
\Lambda}^{0}\frac{dk'}{2\pi} \int \frac{d\Omega'}{4\pi^2} F
(\val{\Omega'}{}  \cdot \val{\Omega}{} )
\eeq
then there are no self-energy corrections in the limit we are in. Imagine now a
state with a macroscopic number of (quasi) particles added to the ground
state  so that all states up to momentum $k = r(\val{\Omega}{} )$ are
occupied in the direction $\val{\Omega}{}$.
We can make such a state the ground state by modifying the hamiltonian to :
\beq
H' = H - \int \int \frac{dk d\Omega}{(2\pi )^3}\psi^{\dag} v^* r(
\val{\Omega}{} ) \psi.
\label{modh}
\eeq
The action now becomes
\beq
S = \int \sib \left[ i \omega - v^* (k - r( \val{\Omega}{} )) \right] \psi
\frac{dk
d\Omega d\omega}{(2\pi )^4}
+ \delta \mu^* \int \sib \psi + \frac{1}{2!2!} \int u \sib \sib \psi \psi  .
\label{flfpmod}
\eeq
Let us now calculate the energy $\varepsilon' (k \ \vec{\Omega} )$
(associated with H') of a particle in a state labeled by $(\vec{\Omega} k )$.
It
is found from $G^{-1}$, which we can calculate exactly in the large $N$
limit since we just need to evaluate the tadpole. This  gives:
\beq
\varepsilon' (k \ \vec{\Omega} ) = v^* ( k - r( \val{\Omega}{})) - \int
\frac{d\omega' dk' d\Omega' }{(2\pi )^4}
\frac{F (\val{\Omega}{}  \cdot  \val{\Omega'}{} )}{i \omega' - v^* (k' - r(
\val{\Omega'}{} ))}
- \delta \mu^* .
\eeq
Doing the $\omega$ integral gives us $\theta ( - k' + r (\val{\Omega'}{} ))$
whereas the integral in $\delta \mu^*$ gives $\theta (-k)$. As a result, the
energy $\varepsilon $ of a particle, now measured with respect to H (which
differs by $v^* r $ from $\varepsilon'$) is :
\beq
\varepsilon (k \vec{\Omega} ) = v^* k +  \int \frac{d\Omega'}{(2\pi )^3} F
(\val{\Omega}{}  \cdot  \val{\Omega'}{} )r(\val{\Omega'}{})
\label{qpe}
\eeq

It is evident that  the integral involving F represents the interaction between
the (quasi) particle
in question and the rest. Consider now a state with $\delta
n ( k\ \vec{\Omega}{} )$  quasi-particles at the point $(k\ \vec{\Omega} ) $,
with $\delta n = -1$ if it is a hole.  In terms of $r(\vec{\Omega} )$,
\beq
\delta n(k\vec{\Omega} ) = \theta (k) \theta ( r(\vec{\Omega} ) - k) - \theta
(-
k) \theta (k - r(\vec{\Omega} )).
\eeq

 The energy of such a state, with reference to the ground state follows from
Eqn.(\ref{qpe}):

 \beq
E\left[ \delta n \right] = \sum_{k \ \val{\Omega}{}} v^* k \delta n(k\
\vec{\Omega} ) + \frac{1}{V} \sum_{k \val{\Omega}{}
}\sum_{k'\val{\Omega'}{}} \delta n(k\ \val{\Omega}{} )  \overline{F}
(\val{\Omega}{}  \cdot  \val{\Omega'}{} )\delta n(k\ \val{\Omega'}{} )
+ O(\delta n^3 )\label{energyexp}
\eeq
where V is the volume, kept finite so we can do a sum rather than integral
over momenta,  and $\overline{F}$ is proportional to $F$, which is how
Landau introduced his $F$.

Some texts (Mahan 1981) devote some time to  why Landau went on  to keep
the quadratic term in $\delta n$, and if he did, why he did not go to higher
orders. The   RG  approach provides its own version of the answer. {\em
Both terms (coming from $\sib (i\omega - v^{*} k ) \psi$ and $\sib \sib \psi
\psi$)  are marginal, whereas higher powers of $\psi$ would bring in
irrelevant operators.} In particular the term with four powers of $\psi$
competes with a term with just two, since the latter has an extra k or
$\omega$ multiplying it and these renormalize downwards under the   RG
transformation.

Of course we did not have to wait for this analysis to understand why Landau
did what he did. If we go from Eqn.(\ref{qpe} ), which gives the energy of an
excitation, to the energy of all of the excitations we find:
$$\sum \rightarrow \frac{V}{(2\pi )^3} \int K_{F}^{2} dk d\Omega $$
\beqr
\frac{E\left[ r(\vec{\Omega}) \right]}{V} &=& \frac{K_{F}^{2}}{(2\pi )^3}
\left[ v^* \int_{0}^{r(\vec{\Omega} )} k \ dk \ d\Omega + \int \int
\frac{d\Omega}{(2\pi )^3}\frac{d\Omega'}{(2\pi )^3}
r( \vec{\Omega'} ) \lf r(\vec{\Omega} ) \right]\nonumber \\
\! \! \! &=& \frac{K_{F}^{2}}{(2\pi )^2 }\left[ v^* \int r^{2}(\vec{\Omega
})\frac{d\Omega}{4\pi} + \frac{1}{4\pi^2} \int \int
\frac{d\Omega}{4\pi}\frac{d\Omega'}{4\pi}r(\vec{\Omega'} ) \lf
r(\vec{\Omega})\right] .
\label{elastic}
\eeqr
We may view the above expression is representing the elastic energy of a
membrane or rubber band representing the Fermi surface  with
$r(\vec{\Omega} )$ as the deformation parameter. (This equation may be
found in page 54 of Nozieres and Pines.) It is now clear that both terms are of
the same order in the deformation. Haldane (in his lectures, yet to be
published) emphasizes this aspect of Landau theory. He writes down the
effective H as a {\em quadratic} function of some current densities obeying
an algebra familiar in Conformal Field Theory. It is clear that we are
discussing a solvable theory. The fact that only forwards scattering
interactions ($F(z,\phi =0)$)  enter this theory is what makes it solvable and
also endows it with additional symmetries. A very concrete application  of
Haldane's approach maybe found in the work of Houghton and Marston
(1992).

More recently Castro Neto and Fradkin (1993) used a coherent state path
integral to sum over the configurations of the Fermi surface.

       This concludes the link between  the present formalism and   Fermi
liquid
theory .  Once we have the concept of the  Fermi liquid theory there is no
need for the   RG . However, for the sake of those   RG  minded readers who
have followed all these arguments, but  are not familiar with  Fermi liquid
theory ,   three sample problems will be discussed, not only to provide some
instant gratification but also because each of them tells us something very
instructive.

\subsection{$\ \ \ $Landau Theory for the masses}
Our fixed point theory parametrized by $v^*$ and $u^* \simeq F$  has
evolved from some bare theory with mass $m$, coupling $U$ and so on. In
the  RG  approach, one does not usually attempt to reconstruct the bare
parameters in terms of the final fixed point quantities due to the unavoidable
loss of information that accompanies mode elimination. There is however an
ingenious argument due to Landau which does exactly that by relating $m$ to
$m^*$ and $F$. The reason not all information about  bare quantities is
irreversibly lost is due, as always, to a symmetry, Galilean invariance being
the operative one here. Let $U$ be a unitary operator that acts on a state $|
\psi >$ and gives it an infinitesimal  boost with velocity $$ \delta \vec{v} =
\delta \vec{p} /m . $$ Under this active   transformation  , the energy of the
eigenstate $|\psi >$  changes as follows:
\beq
\label{active}
\delta E = <U\psi | H |U \psi > - <\psi | H | \psi >.
\eeq
{\em Now Galilean invariance is the statement that the boost affects only the
kinetic energy of the particles and not the interaction energy.} Thus the
response of $H$ to this   transformation   is
the same as in free-field theory and given by :
\beq
\label{gi}
U^{\dagger}\ H \ u = H + {\em \vec{P}}  \cdot\frac{\delta \vec{p}}{m} +
\cdots
\eeq
where ${\em {\vec{P}}}$ is the total momentum operator.
Thus to first order
\beq
<U\psi | H |U \psi > - <\psi | H | \psi > = <\psi |{\em \vec{P}}
\cdot\frac{\delta \vec{p}}{m}|\psi>.
\label{passact}
\eeq
Let $|\psi >$  be a state  containing  one extra particle at the Fermi surface
Fermi surface   . Since the ground state has zero momentum, this is a state of
momentum $K_F$ in the chosen direction.
  Let the boost be in the same direction. The energy change  according to the
right hand side of of the above equation is
  \beq
  \delta E = K_F\frac{\delta p}{m}.
\eeq
As for the left hand side, the active   transformation    has three effects
which
change the energy to {\em first order} in the boost:
\begin{itemize}
\item The quasi-particle momentum goes up by $\delta p$ and by the very
definition of effective mass or Fermi velocity, its energy goes up by
$K_F\frac{\delta p}{m^*}.$
\item The sea gets bodily shifted by $\delta \vec{p}$. This does not affect the
sea kinetic energy since the total momentum of the sea was zero.
\item  The interaction of the quasi-particle with the shifted sea changes its
energy as per Eqn.(\ref{qpe}) with $r(\theta ) = \delta p \cos \theta $, where
the angle $\theta $ is measured relative to the boost.
\end{itemize}

Adding all the pieces and equating the result to what was given in the
previous equation, we get the famous relation:
\beq
K_F\frac{\delta p}{m} = K_F\frac{\delta p}{m^*}  + \delta p \int F(z) z
\frac{dz}{4\pi^2}
\label{mass}
\eeq
where $z = \cos \theta $. In terms of the dimensionless function
\beqr
\Phi &=& \frac{m^*}{2\pi^2 K_F} F = \frac{F}{2 \pi^2 v^*},\label{Phi}
\eeqr
 and the expansion in terms of Legendre polynomials
 \beq
 \Phi (z) = \sum_{l} \Phi_{l} P_{l} (z)
\eeq
we obtain
\beq
\frac{m^*}{m} = 1 + \frac{1}{3} \Phi_1 .
\eeq

\subsection{$\ \ \ $Zero Sound}
Zero sound  refers to natural oscillations in a Fermi liquid  resulting from
interactions between the particles. We find it just as we would find the
natural
frequencies of an oscillator: by looking for poles in certain response
functions.  This means the density-density response function or
compressibility in our problem.

 Let us imagine an external probe $\phi (Q\omega )$ which couples to the
density $\rho (Q\omega )$ of the fermions  producing  density fluctuations.
The compressibility $\chi $ is given by
\beq
(2\pi )^4
\delta^{(4)} (0) \chi(Q\omega ) = - <\rho (Q \omega ) , \rho (-Q \ -
\omega) >
\label{chi}
\eeq
We are interested in the limit $Q, \omega \ <<\ \Lambda $.  We will compute
the correlation function by using the diagrammatic rules we have employed so
far. The first few diagrams are given by Fig.13. Because of the fact that the
radial part of the $\delta $-function ($\delta (k - k') $) that we pull out of
our
graphs differs from
the traditional one   ($\delta (k - k' ) / K_{F}^{2}) $ used
in Eqn.(\ref{chi}), we find:
\beqr
\frac{\chi(Q\omega )}{\left[ -K_{F}^{2} \right]} &=&  \int
\frac{dk_1 d\Omega_1 d\omega_1}{(2\pi )^4} \left[ \underbrace{\frac{1}{
i\omega - v^* k_1}
\frac{1}{i\omega_1 + i\omega - v^* k_1 - v^*Q   z_{1Q}    }
}_{I(\omega_1 k_1 )      }\right]  \nonumber \\
& & + \left[ \int \frac{dk_1 d\Omega_1 d\omega_1}{(2\pi )^4} I(\omega_1
k_1 )  \cdot
\int \frac{dk_2 d\Omega_2 d\omega_2}{(2\pi )^4} I(\omega_2 k_2 )
F(\val{\Omega}{1}  \cdot\val{\Omega}{2} ) \right] + \cdots  \\
&=& \int \frac{Q z_{1Q} dz_{1Q}}{i\omega - v^* Q
z_{1Q}}\frac{1}{4\pi^2} +
\int \frac{Q z_{1Q} dz_{1Q}}{i\omega - v^* Qz_{1Q}}\frac{1}{4\pi^2}
\int \frac{Q z_{2Q} dz_{2Q}}{i\omega - v^* Q
z_{2Q}}\frac{1}{4\pi^2}F(\val{\Omega}{1}  \cdot\val{\Omega}{2} )  +
\cdots
\eeqr
where $z_{iQ}$ is the cosine of the angle between $\vec{Q}$ and the
direction of the loop momentum $K_{i}$.

Now, we would like to study the simplest problem of this kind and therefore
like to choose a constant F. Unfortunately, F, although  not antisymmetric
under the exchange of 1 and 2, still vanishes when the initial angles coincide.
(See the nearest neighbor  example wherein $F \simeq (1 - \cos
\theta_{12})$.)   We will
compromise and for once  introduce spin. We will assume only up and down
particles scatter, with a constant $F_0$. If we now look at Figure 14, we see
that the first loop gets an extra factor of 2 due to the spin sum. The second
also gets only
a 2 since spins at the vertex must be opposite. This restriction continues down
the chain. Now we find that the series is geometric. The sum gives
\beqr
\chi (Q \omega )  &=& \frac{-2K_{F}^{2}I_0}{1 - F_0 I_0} \\
I_0 &=&  \int \frac{Q zdz}{\omega - v^* Qz}\frac{1}{4\pi^2}
\eeqr
where we have chosen to look at real (rather than Matsubara)   frequency
$\omega$ since we are looking for real propagating excitations.

 Notice  how we get an answer that is very sensitive to whether $Q  / \omega
\rightarrow 0$  or vice versa.  In the former case $I_{0}$ vanishes whereas in
the latter case it equals $- \frac{1}{2\pi^2 v^{*}}$. This is the kind of
sensitivity that plagues the four point function $\Gamma$ also, as alluded to
earlier.

Clearly a pole occurs occurs in $\chi$ when
\beqr
\frac{1}{F_0} &=& \frac{1}{4\pi^2 v^*}\int \frac{ zdz}{s  -  z}\nonumber \\
&=& \frac{1}{2\pi^2 v^*} \left[ \frac{s}{2} \ln \frac{s +1}{s -1} - 1 \right]
\eeqr
where
\beq
s = \frac{\omega}{Qv^*}
\eeq
In terms of the dimensionless $\Phi_{l}$ introduced in Eqn.(\ref{Phi}):
\beq
\frac{1}{\Phi_0} = \left[ \frac{s}{2} \ln \frac{s +1}{s -1} - 1 \right] .
\eeq
For the solution
to exist, we require $s > 1$, i.e., the velocity of propagation
$\omega / Q$ must  exceed the Fermi velocity $v^*$.

We will not discuss the extensive physics of this phenomenon. (For example,
Mermin (1967) has showed that we will always have a zero-sound mode, for
any reasonable F. )The main point was to show that the narrow cut-off theory
has a lot of life in it.

\subsection{$ \ \ \ $Static Compresssibility}
To find the equilibrium compressibility, we simply set $\omega \equiv 0\ , Q
\rightarrow 0$ in the preceding calculation. This means that
\beq
\frac{Q z}{ \omega - v^* Qz} \rightarrow - \frac{1}{v^*}.
\eeq
The iterated integrals then simplify to the point that we no longer have to
introduce spin or assume F is a constant. Going back to the spinless case, we
find:
\beq
\chi = \frac{m^* K_F / 2\pi^2}{1 + \Phi_0}
\label{static}
\eeq
Had we computed $\chi$ in free-field theory we would have found
\beq
\chi_0 = \frac{mK_F}{2\pi^2}.
\eeq
 Thus
 \beqr
\frac{\chi}{\chi_0} &=&  \frac{m^* / m}{1 + \Phi_0} = \frac{1 + \Phi_1
/3}{1 + \Phi_0}.\label{renormchi}
\eeqr
Now, no one will
dispute that this is indeed the ratio of $<\sib \psi \sib  \psi
>$ correlation
functions in the fixed point theory to the free-field theory. But
in Landau theory one equates this to the ratio of compressibilities. This is
not
so obviously correct and I thank N.Read for forcing me to clarify this point.
The problem is this. Let us begin with the full path integral over the bare
fields prior to the   RG    transformation  with a coupling of the bare charge
density to some external field A:
\beqr
Z(A)  &= &\int d\psi_0 d\sib_0 e^{ S(A)} \label{za}\\
S(A) &=& \int \sib_{0} (i\omega - E(K) ) {\psi_{0}} + \frac{1}{2!2!} \int
\sib_{0} \sib_{0} {\psi_{0}}{ \psi_{0} U} +  A \int \sib_{0} {\psi_{0}} .
\eeqr
Now,  it is certainly true that
\beqr
\frac{\partial^2 \ln Z}{\partial A^2} &=&  <\sib_{0}  \psi_{0} \sib_{0}
\psi_{0} > = \chi
\eeqr
Suppose we now perform  the   RG transformation. In the process we rescale
the field: $\psi_0 Z^{-1/2} = \psi $. (I apologize for using Z to denote two
different things. Hereafter we will only see the above definition, as the wave-
function renormalization factor.) This means
$$A \sib_0 \psi_0 \rightarrow A Z \sib \psi .$$
Now the partition function is preserved by mode elimination, and we can take
its second logarithmic derivative with respect to $A$ after the   RG
transformation
to find $\chi$. But this will give $Z^2 <\sib \sib \psi \psi > $
whereas we computed the operator without the Z's in what we called $\chi$
of the fixed point theory.

So this is the mystery. The resolution lies in the fact that besides the
rescaling, a term of the form $A\sib \psi$ is generated (in addition to what
was already there) as we carry out  the   RG transformation,
 and this  precisely cancels the effect just discussed. This is an example of a
Ward identity based on charge conservation. (See Abrikosov {\em et al}
(1963) for a discussion.)
 Here is a {\em glimpse} of how it works with no numerical factors. Let us
look at two graphs that cancel. (All loop momenta lie in the eliminated region
and correspond to fast modes.) Take the sunrise diagram (Figure 15a) whose
$i\omega$ derivative at $\omega = 0$ contributes to $Z$,  the field rescaling
factor:
 \beqr
 i\omega &\rightarrow & i\omega ( 1 - \frac{\partial \Sigma}{i\partial
\omega}) \equiv i\omega Z^{-1}.
 \eeqr
 Imagine routing the external momentum through  the upper line. Taking the
$i\omega$ derivative clearly squares that propagator.This is shown in Figure
15b, with the cross denoting the place where  the second propagator joins the
first.

 Consider now the other phenomenon: generation of new terms. In the mode
elimination scheme  the coupling between $A$ and $\sib \psi$ can take place
via the fast modes  as shown in Figure 15c.  Notice that this diagram
coincides with that in Figure 15b in the limit when the probe brings in zero
momentum and frequency. {\em Consequently the field rescaling effects (due
to the  the self-energy diagram)  precisely cancel induced terms effects (due
to the vertex correction diagram). Although we took  just a pair of diagrams,
the result is exact. It reflects the fact that
 even though the quasi-particle can break up into many particles, (so that its
chance of being a single particle is reduced), the field can couple to the
fragments now, and the total charge of the fragments (which is all the field
couples to in the limit of zero frequency and wavelength) is that of the quasi-
particle.} (The careful reader will ask: what about the  c-number term of the
form $A^2$
in the action that comes from integrating fast modes? These contributions
from the fast modes drop out as the external momentum and frequency
vanish,  which is the limit we are interested in.)

 Thus a lot of Landau theory acquires its  power due to the fact that not only
are many  quantities (like $<\sib  \psi \sib  \psi > $) computable in the fixed
point theory, they directly correspond, with no intervening, unknown
prefactors, to physical observables (like compressibility) due to Ward
identities.

 \subsection{$\ \ \  $Notes for the experts}
  Here are some notes for readers who are familiar with the details of one or
other ideas invoked earlier.
 \begin{itemize}
 \item
 In the diagrammatic treatment of  Fermi liquid theory , one organizes the
graphs as follows (in the notation of Abrikosov et al.(1963)
First, one looks at the theory in the limit where the external transfers $Q,
\omega \rightarrow 0$. The full four-point function is called $\Gamma^k$ and
corresponds to the limit $\omega / Q \rightarrow 0$. It is given as a sum of
diagrams where the bare vertex is called $\Gamma^{\omega}$, which
corresponds to the limit $Q/ \omega \rightarrow 0$,  and is irreducible with
respect to a pair of  particle hole lines which are singular at the Fermi
surface
. It is assumed that the bare vertex is analytic in its  arguments,  and the
trouble makers, the   particle-hole lines which produce all the singularities
in
the small $(Q, \omega )$ limit,  are explicitly displayed.

 In the   RG  approach, the bare vertex $u$ contains all the {\em safe} modes
which include particle-hole lines with at least one of them outside the
cut-off.
{\em The only lines shown explicitly in the cut-off theory are particle -hole
lines both within the cut-off, these being the modes yet to be integrated. }
 \item Couplings corresponding to non-forward scattering, called $F(z, \phi
\ne 0)$ in this paper, are very important for the study of lifetime effects and
transport properties.  A nice discussion of this may be found, for example in
Mahan's book (1981).  Note in particular the discussion of the work of Dy
and Pethick in page 947. These authors asked how the forward scattering
Landau parameters F(z) may be extended to non-forward scattering and
argued for a certain $\phi$ dependence based on symmetry under exchange.
The additional factor of  $\cos \phi $ they came up with is exactly what we
find in Eqn. (\ref{DP}) of this paper.
 \item Even though F is a marginal coupling, there are no anomalous
dimensions for the operators, in contrast to the fixed line in $d=1$ along
which  the fermion field and other operators have continuously varying
dimensions. It is worth finding out if the interaction is a redundant operator
in
the sense of   RG .
  \item Landau theory appears very much like a classical self-consistent
theory. We understand this as coming from the large $N$ saddle point,
which, like any saddle point, represents a form of classical limit.
 \item Some readers familiar with $1/N$ expansions may ask how the $F$'s
manage to change physical quantities like $\chi$ by factors  of order unity,
when one always thinks of interactions as producing changes of order $1/N$.
(In other words, why are the dimensionless numbers $\Phi$, appearing in,
say,  Eqn.(\ref{renormchi}), producing corrections of  order unity? ) The
central feature  of $1/N$ expansions is that loops of a certain kind are order
unity since the loop sum pays for the extra factors of the interaction. If we
compute a scattering  amplitude, which exists only due to interactions, we get
a term of order $1/N$ since that is the strength of any one coupling. This is a
sum of a tree level or bare term and iterated loops, all of the same order. In
other words the quantum corrections in the iterated loops  are of the same
order as the tree level term. But if we compute something like the density-
density correlation function (not often done in field theory),  this is given
by
the polarization bubble which  is nonzero in free-field theory and hence of
order unity in  the $1/N$ series . Loop corrections to it (ZS graphs) are of
the
same order as well.
 \end{itemize}

 \section{ NON-CIRCULAR FERMI SURFACES: GENERIC}
We now discuss a Fermi surface  which has no special symmetries other than
time-reversal invariance: $E(\vec{K} ) = E(-\vec{K} )$.  Once again we
focus on $d=2$, discussing in passing the extension to $d=3$. A surface that
meets these conditions is an ellipse and is depicted in Figure 16.

The first step is to set up the  RG    transformation  for the noninteracting
problem. Since $|K|$ is no longer a measure of energy, we must draw
contours of constant {\em energy} $\varepsilon$ (measured from the Fermi
surface) and retain a band of width $\Lambda$ in either side of the Fermi
surface. Thus our starting point is the action:
\beq
S_0 = \int_{0}^{2\pi}\int_{-\Lambda}^{\Lambda} \int_{-\infty}^{\infty}
\sib (i\omega - \varepsilon ) \psi \frac{J(\theta \varepsilon ) d\theta
d\varepsilon d\omega}{(2\pi )^3}
\label{noncirc}
\eeq
where $\theta$ parametrizes the Fermi surface, and J is the Jacobian for
$(K_x K_y ) \rightarrow (\theta \varepsilon )$.  We will expand
\beq
J(\varepsilon \theta ) = J(\theta ) + \varepsilon J_1 (\theta ) + \cdots
\label{jacob}
\eeq
around the Fermi surface  and keep just the first term; the rest will prove
irrelevant.

The  RG    transformation  is exactly as before, with $\varepsilon$ in place of
$k$. It is clear that higher order terms in {J} renormalize to zero with
respect
to this   RG transformation.
The interaction term is

\beq
\delta S_4 = \frac{1}{2!2!}\int \sib (4) \sib (3) \psi (2) \psi (1) u(4, 3, 2
,1)\label{es4}
\eeq
where
\beq
\int \equiv \left[ \prod_{i = 1}^{3} \frac{J(\theta_{i} ) d\theta_{i}
d\varepsilon_{i} d\omega_{i}}{(2\pi )^3}\right]  e^{  - |\varepsilon_{4}|/
\Lambda}  .
\eeq
\subsection{   Tree Level analysis.}
The analysis proceeds exactly as in the rotationally invariant case. As
$\Lambda$, (the cut-off in energy now) is reduced to zero (in fixed laboratory
units) by mode elimination, we find once again that the set of initial momenta
$1$ and $2$, must coincide with the final set $3$ and $4$.
To see this, we must simply  replace figures  with intersecting circles with
say, intersecting ellipses (Fig.16) or whatever may be the shape of the Fermi
surface. In smooth cut-off only such couplings will be spared exponentially
downward renormalization.

So once again the marginal couplings at tree level obey
$\theta_3$
equal  to $\theta_1$ or $\theta_2$ unless $\theta_1 = -\theta_2 $ in
which case $\theta_3 = -\theta_4 $. (The figure shows the second possibility).
The tree level amplitudes are labeled as before:
\beqr
u(\theta_2 \theta_1 \theta_2 \theta_1 ) &=& -u(\theta_1 \theta_2 \theta_2
\theta_1) = F(\theta_2 \ \theta_1 ) \\
u(-\theta_3 \theta_3 -\theta_1 \theta_1 ) &=& V(\theta_3  \ \theta_1 )
\eeqr
{\em but are no longer functions of the differences of their arguments. }

  \subsection{   Tadpole graph.}
  We now see a new feature with the tadpole. Let us understand this without
any reference to the   RG . Suppose we begin with the action Eqn.(\ref{es4}),
and go to one loop. The tadpole graph, Fig.5, makes the following
contribution  to the self-energy
  \beqr
  \Sigma  (\theta \omega \varepsilon ) & = & \Lambda \int F(\theta \ \theta' )
\frac{J(\theta' ) d\theta'}{(2\pi )^2} \equiv \varepsilon_0 (\theta ).
  \label{2dself1}
  \eeqr
Thus
\beq
 G^{-1} = (i\omega - \varepsilon - \varepsilon_0 (\theta ) )
 \eeq
and the Fermi surface  has moved to $\varepsilon = -\varepsilon_0 (\theta ) $.
{\em This generally involves a change in shape.}

Now even in the rotationally invariant problem the Fermi surface  moves;
turning on interactions changes the   Fermi surface  radius from $K_{F}^{0}
= \sqrt{2m\mu}$ to $K_F (\mu ) $ such that
\beq
\frac{K_{F}^{2} (\mu )}{2m} + \Sigma (u) = \mu . \label{newrad}
\eeq
In that case
we can add a counter term $\delta \mu^* = \varepsilon_0 (\theta )
= \varepsilon_{0}$ to the action to restore the old Fermi surface  radius.
Recall that  the action (schematic)
\beq
S = \int \sib (i\omega - \varepsilon ) \psi + \varepsilon_0 \int \sib \psi
 + \frac{1}{2!2!} \int u \sib \sib \psi \psi \label{nrifp}
 \eeq
 was invariant under the   RG  at one loop and order u.

Now, it was pointed out that  adding a counter term to maintain the radius of
the Fermi surface  was not fine-tuning since it corresponded to maintaining a
fixed density, which is experimentally viable.  It was also pointed out that if
we did not add the counter term, the system would not acquire a gap, as in
$\phi^4$, theory, but would simply move to the new radius defined in
Eqn.(\ref{newrad}).

In the non-rotationally invariant problem  clinging to the old Fermi surface
is
certainly a case of fine tuning. There is no experimental way (such as sealing
off the system) to preserve the detailed shape of the Fermi surface .  We
could add a {\em constant} $\delta \mu $ to keep the density (i.e., volume
enclosed by the Fermi surface ) constant, but we will not;  we will let the
system find the true Fermi surface  for the given $\mu$ and $u$.  We can use
the   RG  to determine the final Fermi surface  as follows. {\em We rewrite
the initial action by adding and subtracting a presently unknown term $
\varepsilon_0 (\theta )$:}
\beqr
S &=& \int \sib (i\omega
- (\varepsilon + \varepsilon_0 (\theta ) ) ) \psi + \int
\varepsilon_0 (\theta )  \sib \psi
 +\frac{1}{2!2!} \int u \sib \sib \psi \psi \nonumber \\
 &\equiv &  \int \sib (i\omega - \overline{\varepsilon })\psi  +\int
\varepsilon_0 (\theta ) \sib \psi
 + \frac{1}{2!2!}\int u \sib \sib \psi \psi \label{renormpert}
\eeqr
and demand that this be a fixed point.  To one loop and order $u$ this means:
\beq
 \varepsilon^*_0 (\theta )
 = s \left[ \varepsilon^*_0 (\theta ) - \int_{-\Lambda
/s}^{\Lambda}\int_{0}^{2\pi} \int_{-\infty}^{\infty}
\frac{d\overline{\varepsilon'}
d\theta' d\omega' J(\theta' ) F(\theta \ \theta'
)}{(2\pi )^3 (i\omega - \overline{\varepsilon'})} \right]
 \label{bird}
 \eeq
which leads to
 \beq
 \varepsilon^*_0 (\theta )
 = \Lambda \int \frac{d\theta' J(\theta' ) F(\theta' \
\theta )}{(2\pi )^2}.
  \eeq
 To obtain this we have to do the $\omega $ integral in Eqn.(\ref{bird}). Note
that in that
equation everything in the integral is evaluated at order $u^0$ due
to an explicit $F$ in it. This means that the propagator in the integral has an
angle-independent $\overline{\varepsilon' } $.

 Since we now have a fixed point it must be true that we have found the
correct Fermi surface   . From the knowledge of $\varepsilon^*$ we can
reconstruct the new Fermi surface   . In principle one could go order by order
in this "renormalized perturbation theory". However in the large $N$ limit
which appears here also, the one loop answer gives the full self-energy
correction or change in Fermi surface   .

 \subsection{   One loop at order $u^2$}
 If we do the mode elimination as in the rotationally invariant case, we will
find once again that               The ZS and ZS' graphs once again do not
contribute to the $\beta$-function for $F$ the same reason as in the
rotationally invariant case: either there is not enough momentum transfer to
knock the internal line at $-\Lambda $ to $\Lambda$ or, there is a kinematical
suppression factor $d\Lambda /K_F$.

 If we do mode elimination as in the rotationally invaraint case, we find once
again that the ZS and ZS' graphs do not contribute to the flow of $F$ for the
same reason: either there is not enough
 momentum transfer to knock an internal line at $-\Lambda$ to $\Lambda$, or
there is a kinematical suppression factor  $d\Lambda /K_F$.
 The  flow  in the BCS channel is unaffected by non-rotational invariance. As
long as  $E(K) = E(-K)$ the BCS diagram, given by the third term in
Eqn.(\ref{oneloop}),  will make a contribution and we will obtain:
 \beq
 \frac{dV(\theta_{1} ; \theta_3) }{dt} = -\frac{1}{8\pi^2} \iT  V(\theta_{1} ;
\theta )
 V(\theta ; \theta_{3}) J(\theta )
 \eeq
though  we can no longer use rotational invariance to decouple this equation
using angular momentum eigenfunctions. It is however possible to do a
double Fourier expansion. This deserves further study analytically and
numerically.

Weinberg (1993) has recently derived such a flow equation for
superconductors whose Fermi surfaces obey just time-reversal invariance
using the notion of effective actions from quantum field theory. Besides the
flow,
his paper has a careful derivation and analysis of the effective action for
superconductors. All the important properties of the superconductor may be
derived from its effective action and the notion of broken gauge invariance
(Weinberg 1986s).

If we ignore the BCS interaction, we expect the Fermi liquid, $1/N$ etc., to
work as before except for lack of rotational invariance. The F function will
now be a function of two variables. We expect to find zero sound. We do not
expect any simple relation between $m$ and $m^* $ due to lack of Galilean
invariance.

Finally if we consider a Fermi surface  without time-reversal  symmetry, we
can get rid of the
BCS amplitudes even at $T=0$. (However,  in drawing the analog of Fig.8,
we must  draw the time-reversal inverted version of the second Fermi surface
(displaced by $\vec{ P}$ ) since the previous construction assumed that if
$\vec{K}$ is an allowed vector, so is $-\vec{K}$.
\section{\ \  NON-CIRCULAR FERMI SURFACES: NESTED}
We are finally going to discuss  spinless fermions on a square lattice at half-
filling.
A specific model for this problem is the one with nearest-neighbor
interaction;
\beqr
H &=& H_0 + H_I \\
     &=& -\frac{1}{2}
     \sum_{<jj'>} \sid (j) \psi (j') + h.c.+ U_0 \sum _{<jj'>}
(\sid (j) \psi (j) -\frac{1}{2})
( \sid (j') \psi (j') -\frac{1}{2}) \label{spinless2},
     \eeqr
where $j$ labels sites on a square lattice and the subscript $<jj'>$ on the
sums means $j'$ is restricted to be the nearest neighbor of  $j$   in the
direction of
increasing coordinates. (Thus if  $j$ is  the origin $(0,0)$ , $j'$ is
restricted to be  $(1,0)$ or $(0,1)$.)  The chemical potential (found by
opening up the brackets in the interaction term) is
\beq
\mu = 2U_0
\eeq
where the
the factor $2$ comes from the number of nearest neighbors.

At $U_0 =0$, the half-filled system will once again be a perfect conductor as
can be seen by going to momentum states. Likewise at $U_0 =\infty$ there
will be a charge density wave with more charge on one sublattice, say the one
whose $x$ and $y$ coordinates (which are integers in lattice units) add up to
an even number.

Once again we will focus not so much on the fate of this one model but rather
on a class of models described by the same  free-field fixed point and its
perturbations. Let us therefore find the fixed point describing  the
noninteracting problem.

Let us take as the free-fermion dispersion relation:
\beq
E = - \cos K_x - r \cos K_y.\label{balloon}
\eeq
{\em   which corresponds to the problem with unequal hopping in the two
directions.} The reason for the choice $r \ne 1$  will follow shortly. Notice
that we expect to see a CDW state at large repulsion in the nearest neighbor
model even if $r \ne 1$ (since in this limit the hopping term is not the
deciding factor,
the interaction term is).   Notice also that for any $r$,   $E$
still has the symmetry:
  \beq
  E(\vec{K} +
  \vec{Q_N}  ) = - E(\vec{K} )\ \ \ \ \vec{Q_N}  \equiv ( \pi \ ,
\pi )
  \eeq
  which is all we will need.
  The chemical potential that
  gives rise to half-filling at $U=0$ is $\mu = 0$
which means that
    $E = \varepsilon$ and that the latter also changes sign when we add the
nesting vector:
  \beq
  \varepsilon
  ( \vec{K} + \vec{Q_N}  ) = - \varepsilon (\vec{K} )\label{nest}
  \eeq
  The Fermi surface  for $r>1$ is sketched in Figure 17.
  The dark   line shows the Fermi surface which now has two branches
$\alpha = \pm 1$. Each point  on the Fermi surface  goes to another  point on
the Fermi surface  upon adding $\vec{Q_N} $. This means that if we shift the
figure by $\vec{Q_N} $, the shifted one (in the repeated zone scheme) will fit
perfectly with the original like something out of Escher's drawings.
  This means that if the momentum transfer is $\vec{Q}_N$, the analog of
Fig.10  will show  the complete {\em overlap} of the two displaced surfaces
rather than their   intersection. Also shown  contours of energy  $\varepsilon
= \pm \Lambda$. This where where the modes to be eliminated under
infinitesimal renormalization lie. Points at $\pm \Lambda$  are knocked to
$\mp \Lambda$ upon  transfer of $\vec{Q_N} $. This in turn ensures that if
one loop momentum lies in a shell being eliminated, the other (in the graph
with momentum transfer $\vec{Q}_N$ also does. This will lead to a flow of
the  coupling in which final momenta differ from the initial ones by
$\vec{Q}_N$.

  For  nesting to take place, we need half-filling and the symmetry in
Eqn.(\ref{nest}). The latter comes if we assume that hopping is always from
one sub-lattice
to the other in a bipartite lattice. Since hopping is usually just
nearest neighbor  to an excellent approximation, studying the effects of
nesting once again do not constitute fine-tuning.

  Let us write down the action for the noninteracting problem. We will use as
the final coordinates $\varepsilon $ and $\theta  \equiv  K_x$ together with a
discrete index $\alpha = \pm 1$ which tells us which of the two branches we
are on. Thus
  \beq
  S_0 = \sum_{\alpha}
  \intom \int_{-\pi}^{\pi} \frac{d\theta}{2\pi} \int_{-
\Lambda}^{\Lambda}
 \frac{d\varepsilon}{2\pi}
 J(\varepsilon \theta ) \sib_{\alpha} (i\omega -
\varepsilon ) \psi_{\alpha}
 \label{freeballoon}
 \eeq
 where
 \beq
 J(\varepsilon \theta )
 = \frac{1}{\sqrt{r^2 - (\varepsilon + \cos \theta )^2 }}.
 \eeq
It is clear why we introduced $r \ne  1$: if it equaled unity, $J$ would be
plagued  with (van Hove)  singularities on the Fermi surface  and the
expansion in  $\varepsilon$ would be impossible.
Since nesting, and not van Hove singularities  is what we are interested in
here, we will study $r >  1$.
In this case the Fermi surface  value of the Jacobian is
\beq
J_{0}(\theta ) =  \frac{1}{\sqrt{r^2 -   \cos^{2} \theta  }}.
 \eeq
Henceforth the subscript on $J$ will be dropped.

 Mode elimination of the action in Eqn.(\ref{freeballoon}) and the rescaling
of fields and $\varepsilon$ go as before to render the action the fixed point.

 As for the quartic
 term, there are now {\em three sets of couplings}  that are
marginal at tree level. Besides $F$ and $V$, we also have
 \beqr
  u \left[ \theta_2 +
  \pi , -\alpha_2  ; \theta_1 + \pi , -\alpha_1\  ; \theta_2
\alpha_2 ;
  \theta_1 \alpha_1 \right]
  &=& - u \left[ \theta_1 + \pi , -\alpha_1  ; \theta_2 +
\pi , -\alpha_2 ; \theta_2 \alpha_2  ;
  \theta_1  \alpha_1 \right] \nonumber \\
  & &
  \equiv W\left[ \theta_2\ \alpha_2 \ ;
  \theta_1\ \alpha_1 \right] \label{w}
  \eeqr
  which corresponds to processes wherein the momentum transfer between 1
and 3 or 2 and 3 equals $\vec{Q_N} $.  In this case, because of the nesting
property of the Fermi surface   ,  we  are assured that the fourth momentum
will lie
on the Fermi surface  if the first three do, and this, you recall, is the
condition for
the coupling to survive the tree level   RG transformation.   Note
also that $W$ describes umklapp scattering if both particles start out on the
same branch and hop to the opposite one.

As  for the one-loop $\beta$-function, the $F$'s do not flow and $V$'s have
the usual flow from the BCS diagram
 \beq
du \left[ -K_3 \ K_3
\ -K_1 \ K_1 \right] =- \frac{1}{2}\intom \int_{shell}
\frac{d^2 K}{(2\pi )^2}
\frac{u\left[ -K_3\ K_3\ -K \ K \right] u\left[ -K \ K \ -
K_1\ K_1 \right] }{(i\omega -
\varepsilon (K) )(-i\omega - \varepsilon (-K)) }
\label{nestbcs}
\eeq
where "shell" refers to the shell being eliminated around both branches. The
two contribute equally to give, in terms of $V$:
\beq
\frac{dV (\theta_3\ \alpha_3\ ; \ \theta_1\ \alpha_1 )}{dt} =
-\frac{1}{2}\sum_{\alpha}\int \frac{J(\theta )d\theta}{(2\pi )^2}V\left[
\theta_3\ \alpha_3\ ;\
\theta\ \alpha \right] V\left[  \theta \ \alpha \ ;\ \theta_1 \
\alpha_1 \ \right]
\label{bcsballoon}
\eeq

Since we do not have rotational invariance, we cannot separate this using the
angular momentum variables. We can however use the double Fourier
transform
and reduce it to discrete coefficients which will be coupled in their
evolution.

Let us now look at the flow of
$u\left[ {K'}_2 {K'}_1 K_2 K_1 \right]$ where ${K'}_{i} = K_{i} +
Q_N$:due to the ZS diagram:
\beq
du \left[ K'_2 \ K'_1 \ K_2 \ K_1 \right] = \intom \int_{shell} \frac{d^2
K}{(2\pi )^2}\frac{u\left[ K'_2\ K\ K_2
\ K' \right] u\left[ K' \ K'_1 \ K \ K_1
\right] }{(i\omega - \varepsilon (K) )(i\omega - \varepsilon (K')) }
\eeq
where "shell" means both $\varepsilon (K)$ and $\varepsilon (K')$ lie in the
thin shells of width $d\Lambda$ near $\pm \Lambda$.
\footnote{We have defined  $W$ with $K_3 = K^{'}_{1}$ and for this
choice only the ZS diagram contributes to flow.  Had we reversed the role of
$K_3$ and $K_4$ in the definition of $W$, an extra minus sign would have
entered it definition, and the ZS' (rather than  ZS ) diagram would have
contributed
to its flow. The latter would also have had a minus sign relative to
the ZS diagram. } Due to the nesting property two interesting things happen
leading to a flow:

\begin{itemize}
\item If $\varepsilon
(K)$ lies in the shell, so does $\varepsilon ({K'} ) = -
\varepsilon (K) $.
\item the $\omega$ integral never vanishes since the poles always lie on
opposite half-planes.
\end{itemize}
Doing the $\omega$ integral, we get, in terms of $W$
\beq
\frac{dW\left[ \alpha_2 \theta_2\ \alpha_1 \theta_1 \right]}{dt} = - \int
\sum_{\alpha} W\left[
\alpha_2 \theta_2 \ \alpha'\ \theta' \right] W\left[ \theta
\alpha \ \theta_1 \alpha_1 \right] \frac{J(\theta )d\theta }{(2\pi )^2}
\label{wdot}
\eeq
where $(\alpha'\ \theta' )$ refer to ${K'}$.

To get a feel
for this problem, let us evaluate the nearest neighbor  interaction
on the present Fermi surface  to obtain:
\beq
\label{weqn}
W\left[ \theta_2 \alpha_2 \ \theta_1 \alpha_1 \right] =
 -U_0 \left[ \sin^2 (\frac{\theta_1 - \theta_2}{2}) + \frac{1}{2} ( 1 -
\frac{\cos \theta_1
 \cos \theta_2}{r^2} - \frac{\alpha_1 \alpha_2 }{r^2} \sqrt{
 r^2 - \cos^2 \theta_1 }\sqrt{r^2 - \cos^2 \theta_2} ) \right]
 \eeq
 and
 \beq
V\left[ \theta_3 \alpha_3 \
\theta_1 \alpha_1 \right] = U_0 \left[ \sin \theta_1
\sin \theta_3 + \frac{\alpha_1 \alpha_3}{r^2}\sqrt{r^2 - \cos^2 \theta_1}
\sqrt{r^2 - \cos^2 \theta_3} \right]
 \label{vqn}
 \eeq

It is readily verified that $W$ always has the same sign for all values of its
arguments; opposite to that of $U_0$. Specializing to the repulsive case, it is
clear from the above equation that $\frac{d|W|}{dt} > 0$. Thus we have
proven the instability for this initial condition.
(This is a weak coupling argument. To describe the nearest neighbor  problem
in the small $\Lambda$ theory it is not enough to simply restrict the full
interaction to within the cut-off, we must take into account induced terms and
renormalization due to elimination of modes. These are however higher order
effects.) As for some other interaction, if it has any overlap with this
direction
we have found, it will be unstable. Note that unlike in the rotationally
invariant problem, we have not explicitly displayed an infinite number of
unstable directions (one for each $l$).

Unlike in the rotationally invariant case, the present problem has many open
questions. Here is list some of the more important ones.
\begin{itemize}
\item
The study of  flow equations Eqns.(\ref{bcsballoon} - \ref{wdot}) is a very
important follow-through to the work presented here.  Calculations are being
performed with G. Murthy  in which
the couplings from  the nearest neighbor model (Eqn.(\ref{weqn}-\ref{vqn}) )
are taken as initial conditions.  In studying the flows, it was  important to
remember that there are some couplings which  flow for more than one
reason. An example is when the initial momenta are $\vec{K} $ and $-
\vec{K} $  and the final ones are  $\vec{K} '$ and $-\vec{K} '$ where
$\vec{K}' = \vec{K} + \vec{Q_N}$: this coupling  is   equal to a $V$ and  a
$W$. The general idea is to run the flow till the cut-off is small and then
solve
the theory by summing over diagrams that survive in the limit of vanishing
$\Lambda$ (our $1/N$.)

\item Although the $1/N$ formalism was discussed in connection with Fermi
liquids with  a rotationally invariant Fermi surface, it applies to all
problems
discussed here: one can always reduce the cut-off (keeping track of the
evolving couplings) and then use the smallness of $\Lambda$ to do a sum
over diagrams that lead in the $1/N$ expansion. For rotationally noninvariant
problems, this can however be complicated by the constant motion of the
Fermi surface as we renormalize, even if we keep its volume constant by
modifying  the  chemical potential. In other words, as the modes are
eliminated, the new Fermi surface and new one-particle energies must be used
in  defining  contours of constant energy and choosing  modes for the next
round of elimination. At the one loop level considered here, this was a
nonissue.

\item It is  important to consider the problem just below half-filling: here we
expect that there will be  an initial growth of interactions as we lower the
cut-
off, which will then freeze once the cut-off is comparable to the deviation of
the Fermi surface  from nesting.  One must  see if in this case an attractive
interaction is generated in the meantime in the BCS channel. If so, this
coupling will continue to grow, since it does not rely on nesting to do so.
This will be yet another  test of the notion that attraction can lie hidden in
models that started out repulsive.

 \item Another topic worthy of further study is the coupling $u(-A,A,-A,A)$
where $A = \frac{1}{2}\vec{Q_N} =(\pi /2 , \pi /2 ) $ and lies on the Fermi
surface  for any $r$ see Fig.17.  (The present remarks apply equally well to
the point
$A' = (\pi /2 \  -\pi /2 )$.) This coupling is  a $V$ a $W$ and an $F$
. It receives flow contribution from the ZS' and BCS diagrams and is forward
scattering amplitude which will control particle self-energies.  Is the
distinguished
nature of this point (from the point of view of the RG) related to
why so many investigators ( Sachdev 1989, Trugman 1988, Elser {\em et al}
1990,  Boninsegni and Manousakis 1990)  find holes occurring at  the point A
when the half-filled
systems are doped? Once could pick   the initial value for
the flow the coupling generated by the  nearest neighbor  interaction, follow
the flow and look at the effective theory at very low energies to see if an
answer comes out.

 \item  Consider the problem on a square lattice without nesting, say because
of same sub-lattice hopping. At very large nearest neighbor  repulsion, we can
see that a CDW will result, with more particles in one or the other
sub-lattice.
However, there will be no instability at infinitesimal repulsion. One expects
from continuity that the transition will take place at small coupling for small
nesting violations.  One should then see this phase transition  at weak
coupling from the RG.
\item
 Notice that our analysis depended on the nested Fermi surface. While the
Fermi surface  was nested in the absence of any interaction, don't the
interactions cause it to move? Won't the shape change from perfect nesting
even if we change the chemical potential to sit at half-filling? What happens
to the CDW instability then? First of all this question does not affect our one
loop calculation which uses  the zeroth order propagators with their zeroth
order formulas for $\varepsilon$).  Whether or not the nested surface  will
stay nested at higher orders in the interaction is an open question which could
control the higher terms in the $\beta$-function and
 decide the ultimate destination of the flow. It will not however change the
fact that the free-field fixed point is unstable since that has been
established
close to the fixed point by our one loop calculation. However if the CDW
instability
is to really take place, the flow must keep going till we hit a fixed
point with a gap. Now, it is clear in coordinate space that at strong coupling
in a bipartite lattice, there will be a CDW. For this to come out of the RG,
nesting must be preserved
 as we renormalize. This is however a conjecture and has not been proven.
 \end{itemize}

 Another  interesting question is the following: if we set the hopping
coefficient $r=0$, we seem to decouple the chains. Will we then get a
Luttinger liquid? No. This is because the interaction terms couple the various
decoupled chains. The exact cancellation that took place between the BCS
and CDW instabilities  in a one chain  model with just one coupling will not
repeat itself anymore.
\section{$\ \  $NON-FERMI LIQUIDS IN d =2}
 So far we have seen two means by which the Fermi liquid could be
destroyed: BCS and CDW
 instabilities. In both cases, the flow came about because {\em individual}
Feynman diagrams had logarithmic singularities. Thus the perturbation series
had zero radius of convergence. (There were essential singularities of the
form $e^{-1/u}$ in, say, the CDW order parameter. ) If perturbation theory
can tell us about the instabilities, why follow the   RG  route? The answer,
from the $d=1$ example is that even if individual diagrams diverged, it was
possible for the $\beta$-function to vanish,  producing novel scale invariant
behavior. We are looking for such a state in $d=2$. I see no evidence for it
if
 \begin{itemize}
 \item the Fermi surface  is spherical,
 \item the coupling is weak,
 \item the input interaction is short ranged.
 \item we work in infinite volume from the start.
 \end{itemize}
 In particular, I have examined, together with A.Ruckenstein and H.Schulz
the channel analyzed  by Anderson (1990):
 the incoming particles of opposite spin at  the same momentum on or near the
Fermi surface . We found that there was no flow in this coupling  as the cut-
off went to zero. Indeed there was no singularity in the diagram when the
external (euclidean) frequencies vanished. Setting them to nonzero values did
not help. Of course setting them equal to some {\em real} frequency did
cause singularities, but these correspond to propagating modes and not
instabilities of the ground state. (Recall that the usual instabilities were
seen
at zero external frequency.) But it must be pointed out that this is not in
variance with Anderson's arguments which rely very much on doing things in
finite volume and then carefully taking the infinite volume limit.
Unfortunately
the RG in a finite volume is not an easy prospect and we were hoping that if
(despite Anderson's
 cautionary note about going to infinite volume too quickly)   the effect
showed up in our calculation, it would be additional corroborations to
Anderson's argument, but within the standard  infinite-volume machinery.
Engelbrecht and Randeria (1991) who studied this problem in the  low
density expansion did not find any instability.The same remarks apply to their
calculation also.

 We must therefore relax some of the above conditions. As mentioned above,
Anderson dropped the infinite volume condition.  Another   possibility
(Ruckenstein {\em et al} 1989)  is that at strong coupling a new possibility,
the {\em marginal Fermi liquid liquid} , which has impressive
phenomenological success, arises. The present weak-coupling analysis has
nothing to say about it and surely cannot exclude it. For example, it is
possible for the bare coupling that enters the action of the narrow cut-off
theory  (or $\Gamma^{\omega}$ in the Abrikosov {\em et al} (1963)
treatment) to be singular. Earlier it was stated that this would not happen
because the bare couplings were obtained from the input parameters by
integrating out safe modes.  But this only assures us that the individual
diagrams that add up to give the bare coupling
 are finite. It is certainly possible for the infinite sum to diverge beyond
some
 maximum coupling. This does not contradict  Landau's analysis  since the
basic assumption that the physics in question is a continuation of the
noninteracting problem is invalid. The stability of the marginal fermi liquid
has studied by Zimanyi and Bedell (1991).

  Another possibility is that even before mode integration, the input coupling
is singular.

  Stamp(1992, 1993) has taken a pragmatic approach and considered the
effect of singular interactions, setting aside the question of
 their origin.

 I decided to look at the Coulomb interaction in this light. We cannot simply
say that it gets screened; this is  a picture that makes sense when a subset of
diagrams in standard perturbation theory are resummed in a certain way to
produce the screened propagator for the Coulomb potential. The book-
keeping  is different in   RG :  the coupling that goes into the action has not
been screened by particle-hole pairs at the Fermi surface . It is easy see
(within the sharp cut-off scheme) that at any stage $Q << \Lambda$ is
unscreened. The correct procedure is to follow the evolution of the bare
coupling as the modes are eliminated and see where it ends up when no more
integration is left over. It is shown in  the Appendix  that the final
potential is
screened.
 (This is a smooth cut-off version of a sharp cut-off calculation devised with
G.Murthy.) This analysis  however assumes that the fermion propagator has
the standard  Fermi liquid theory form. Ideally one should let the fermion
propagator also evolve as modes are eliminated and see if we still end up with
a screened interaction.This has not been done.

  Now there {\em are} concrete examples of non-fermi liquid  behavior if we
are willing to consider impurity problems.  Consider for example  the
example provided by
   Affleck and Ludwig (1991a, 1991b, 1992) from  the Kondo problem. More
recently Perakis {\em et al} (1993) have given another example from the
Kondo problem which shows non-fermi liquid behavior for a range of
parameters.

   Although the search for non-fermi liquid  did not yield anything at weak
coupling, it is a worthwhile goal  since the copper-oxides seem to call for
something different. As pointed out by Anderson, these may not be connected
to the fermi liquid  fixed point.  Rather than reach the novel fixed point from
the fermi liquid, one could attempt writing down different fixed points. They
may require additional fields  besides fermions, e.g., gauge boson (Nagaosa
and Lee (1990), Ioffe and Kotliar (1990), Polchinski (1993)) : in a strong
coupling field theory the low energy physics may bear no simple relation to
the microscopic theory.  Another route is to start with one or more  one-
dimensional systems, which can have  Luttinger liquid behavior, and couple
them perturbatively (Wen  1990, Schulz 1991).

 \section{$ \ \ \ $SUMMARY AND OUTLOOK}

 The main aim of this paper was to find a way to apply the RG methods to
interacting nonrelativistic fermions, in particular to understand the various
instabilities for weak perturbations.  In particular we wanted to see if the
system remained gapless. Since the RG was so successful in dealing with
critical phenomena, it was decided to follow a path where one relied heavily
on analogy to this prior application. Since gapless systems corresponded to
critical systems the idea was to use the language of fixed points and their
perturbations.

 We started with a  brief historical review starting from the original
formulation of the RG for use in field theory, and ending with the modern
approach pioneered mainly by Kadanoff and Wilson.
It was pointed out that while the RG has always expressed the invariance of
the theory under a change in cut-off followed by a suitable change in the
parameters, the emphasis has shifted from viewing  the cut-off as an artifact
to be sent to infinity (where it belongs in a continuum theory like quantum
electrodynamics) to viewing it as a dividing line between interesting and
uninteresting degrees of freedom (the slow and fast modes) even in a problem
where the cut-off is finite to begin with.

In Section II we discussed the charged (complex) scalar field in four
dimensions. It was shown how if one wanted to study physics at long
distances, one could work with just the "slow" modes in the functional
integral and how the unwanted fast modes were to be eliminated and the new
couplings deduced. The gaussian fixed point was studied in detail and the
flows to one loop were deduced. It was shown how, upon reducing the cut-off
to very small values, i.e., reducing the phase space to a tiny ball, the
coupling
functions reduce to a handful of coupling constants, which were just the first
few terms in the Taylor series of the coupling functions about the origin.  A
comparison between the modern way to find the flow equations ($\beta$-
functions) using mode elimination and field theory methods  (trying to get rid
of cut-off dependence by proper choice of bare parameters) was provided. It
was shown how although the two had very different book-keeping schemes,
they gave the same answer for relevant and marginal couplings.

In Section III the functional integral method for fermions was introduced.
First a few simple problems in the thermodynamics and dynamics of
fermionic oscillators were solved using operator methods. Then the
Grassmann integral formulation was introduced and the same results were
regained. The rules for calculating correlation functions (Wick's theorem for
fermions) were  derived and used in the calculations.

The stage was now set for dealing with the nonrelativistic fermions. The
strategy would be the following.  We would first start with noninteracting
fermions and write a hamiltonian that faithfully described the physics near the
Fermi surface, i.e., within a cut-off $\Lambda$ on either side of it. The logic
was that the questions we were interested in,  were decided by these modes at
least at weak coupling. The corresponding functional integral would then be
written. A mode elimination process that reduced the cut-off and rescaled
momenta and fields and left the action invariant would be found. Given the
RG and its fixed point, it would  then be  possible to classify the
perturbations
as relevant irrelevant or marginal. While this closely paralleled the scalar
field
problem one major difference was anticipated: since we are renormalizing
towards the fermi surface and not the origin, the remaining phase  space will
be  infinitesimal perpendicular the Fermi surface, but of fixed size in the
tangential direction. Thus the fixed point would be characterized by coupling
functions that depended nontrivially on the angles used to parametrize the
Fermi surface.
 The case $d=1$ was clearly exceptional since the Fermi surface consisted of
just two points. This made the problem very similar to continuum field
theories wherein the phase space for bosons and fermions is a ball centered at
the origin.

The study was limited to spinless fermions. At this point the reader can surely
see  that the  inclusion of spin  really is straightforward. Since all the
interesting flows were due to special properties of the shape of the Fermi
surface, and since spinless fermions could display these shapes, they were the
clear pedagogical choice. To keep the discussion concrete, a nearest neighbor
spinless fermion problem at half-filling was frequently invoked. The question
was whether it developed a CDW gap at arbitrarily small repulsion.

 The RG scheme worked remarkably well in the $d=1$ warmup, described in
Section IV.  We took two slices (of width $2\Lambda$) near each Fermi
point (L/R) and found a mode elimination scheme that left the action of the
free theory invariant. We then turned on a quartic interaction. At tree level
it
was found that only the frequency and  momentum $(k = K - K_F$)
independent part of this coupling was marginal, the rest were irrelevant, as
were all couplings with six or more fields. The marginal coupling did depend
on the discrete internal index $L/R$ but due to the Pauli principle, reduced to
just one independent number. We had to go to one loop to resolve the fate of
this marginal coupling. The  $\beta$-function route  avoided the CDW and
BCS mean-field instabilities  by playing them against each other and giving
the correct answer: a scale-invariant system, the  Luttinger liquid.  (This
correctly describes   the nearest neighbor model, which can be solved
exactly.) We also discussed the quadratic terms that were induced by the
quartic term upon mode elimination. These meant the Fermi surface  was
moving to take into account the interactions. There were two options. We
could add counter term, determined order by order, to keep $K_F$ fixed, or
we could let the surface  move to the new $K_F$. The latter would be found
by doing renormalized perturbation theory: we split the original $\mu$ into
two parts, one which gives the correct $K_F$ in the propagators and a
counter-term, which keeps it there. The division would be revealed to us
order by order. It is a remarkable fact that Fermi systems, unlike Bose
systems can remain critical for a range of $\mu$; they do this by using the
fact that their fixed point is characterized by a whole surface, which can
wiggle around as interactions are turned on.  This idea applies with minor
modifications in higher $d$, and we will not speak of it further.

 The scheme was then extended in Sections V and VI to rotationally invariant
Fermi surfaces  in $d=2$ and $d=3$. In $d=2$   an annulus of width
$2\Lambda$ and radius $K_F$ was used. The RG transformation  was
essentially the same as in $d=1$ except for the fact that we had an integral
over the "internal" variable $\theta $ rather than a sum over the Fermi points
L and R.  At tree level  only two couplings, $F(\theta )$ and $V(\theta )$,
which corresponded to forward scattering  and Cooper-pair interactions
survived. They had no dependence on external $k$ or $\omega$. Going to
one loop we saw that there was no flow of  $F$. This was because of the
kinematics of $d >1 $. There was however flow in  $V$. We decoupled the
flow into an infinite number of equations, one for each angular momentum
$l$. The flow was marginally irrelevant  for repulsion (corresponding to the
findings of  Morel-Anderson) and marginally relevant for attraction,
corresponding to the BCS instability.  We then formally set $V=0$  and
identified $F$ with Landau's F function. Thus RG had led us automatically to
Fermi liquid theory. The situation in $d=3$ was essentially the same, except
for one difference that was significant: while in $d=2$ only forward scattering
amplitudes survived, in $d=3$ nonforward ones did also. However the latter
did not have any effect on low energy, low-momentum transfer physics.

 Having identified the Fermi liquid  as the fixed point of our RG, we next
asked why it is solvable, i.e., why, despite the quartic interactions, one is
able
to calculate a lot response functions. It was pointed out that as we eliminate
modes, a $1/N$ expansion emerges, with $N = K_F /\Lambda$.
 Landau theory is the $N= \infty$ limit. (By using certain collective
coordinates, this limit can be made to correspond to a saddle point which
gives the  exact answer.) This is the first example where $N$ really is large
in
the original problem. However  a two stage assault was needed to ensure this:
first use RG to bring $\Lambda /K_F$ to a small value and then use $1/N$.
Of course the $1/N$ is not a new way to do Landau theory, but just a new
way to understand it: it has always been known that bubble graphs (ZS loops)
dominate the low momentum response. However, after  the application of the
RG there was nothing left but ZS bubble graphs due to the kinematics. The
bubble graphs were ones in which the loop angle runs over all values.
However these  need not have amounted to much since the width of $k$
integration was going to zero. But the graphs did survive since the integrand
had a $\delta $-function singularity on the Fermi  surface  and  the integral
was oblivious to the cut-off. (This is why the graph survived but did not
contribute to the $\beta$-function, which probes the sensitivity to cut-off.)

  An RG version of the Kohn-Luttinger instability was given. The details were
relegated to the Appendix.

 We briefly looked at three effects in  Fermi liquid theory since each taught
us
something and also gave the readers familiar with RG but not Fermi liquid
theory,  some instant gratification for their efforts.

 By embedding   Fermi liquid theory in the framework of RG and $1/N$ we
not only automate the process that Landau had to finesse with his genius, we
also prepare ourselves better to study variants of the problem he  attacked,
such as the problem with impurities which is a big field (Lee and
Ramakrishnan 1985).

 We then studied generic nonrotationally invariant problems. We found that
we were led to $F$ and $V$ which no longer depended on just the difference
between their arguments. The flow equations for $V$ were derived. By going
to a Fermi surface  with no time-reversal invariance, we could eliminate the
BCS instability and have a real  Fermi liquid theory at $T= 0 $. The new
feature here was the changing shape of the Fermi surface  as we
renormalized.

 Moving on to the case of a nested surface  in $d=2$ , we found that a new
coupling $W$ corresponding to momentum transfer $\vec{Q_N} = (\pi\ \pi)$
survived at tree level. At one loop this coupling began to flow. For the case
of
the perturbation corresponding to the nearest neighbor  interaction (truncated
to modes within the cut-off) , we saw the CDW instability. It was pointed out
that the points $(\pm \pi /2\ \pm \pi /2 )$ where holes seemed to appear upon
doping the half-filled system are exceptional from the point of view of the
RG: the are forward scattering amplitudes that  flow due to BCS and CDW
diagrams. Several problems were proposed for further study.

 Finally we discussed the possibility of singular Landau parameters and non-
fermi liquids. This was surely a possibility at strong coupling, but did not
seem to happen at weak coupling within the scheme we were employing.

Although we used the modern Kadanoff--Wilson approach to renormalization
we frequently made contact with the old field theoretic scheme for computing
the $\beta$-functions. The main
difference between the two schemes was that in the former, all loop momenta
were at the cut-off, while in the latter, one was at the cut-off and the rest
at or
below it. In section II it was pointed out that in the case of the scalar
field,
the difference did not show up in the flow of the marginal couplings at one
loop,  since we could set all external momenta to zero and momentum
conservation meant that if one propagator was at the cut-off, so was the other.
It was pointed out in the fermion problem even if we set all $k$'s to zero,
there could still be large momentum transfers of order $K_F$. Yet we saw
repeatedly that the two approaches gave the same one loop flow for marginal
couplings. Let us recall why. First, $F$ never flowed in either scheme since it
got its contribution at the Fermi surface and did not know about $\Lambda$ in
either scheme.
As for $V$, when we set the incoming momenta on the Fermi surface, at
opposite angles, the total momentum vanished, just as in the scalar field
theory, and the two propagators in the loop
were equal and opposite and at the cut-off in the BCS diagram. The other two
diagrams did have large momentum transfers.  They were however
suppressed in both schemes by kinematics, but not equally: the suppression
was by $d\Lambda /K_F$in the modern scheme and by  $\Lambda /K_F$ in
the field theory scheme. This difference did not matter at the fixed point
since
both factors vanished. Finally $W$ flowed due to the ZS diagram. Even
though the momentum transfer was large ($\vec{Q}_N$ )
it was such that if one propagator was at the cut-off so was the other due to
the condition
$E(K) = - E(K+Q_N)$.  The other two diagrams were suppressed by phase
space.

The field theory scheme was the best choice for studying screening and the
Kohn-Luttinger effect discussed in the Appendix.

 There are many possible extensions. Inclusion of spin will produce new
effects such as spin-density waves, but no new formalism is required to
incorporate it. Inclusion of bosons is very interesting, but not discussed here
due to lack of space.  Ye and Sachdev (1991) have used the RG ideas
espoused here to study a boson-fermion system describing the metal -
superconductor transition. Polchinski discussed phonons in his TASI article
(1992) and in a more recent preprint (1993).

 We can study disordered systems using the replica trick. Finally we can go to
finite T. The fixed point discussed here will describe the crossover to
$T=\infty$ in the early stages.

 To conclude,  the analysis in this paper has shown that the RG, which has
proven its worth in critical phenomena, chaos etc., is just as effective in
helping us understand interacting fermions. Conversely, the Fermi systems,
with their novel phase space and fixed point structure, offer us a far from
ordinary manifestation of the RG at work.

 With various phenomena such as  Fermi liquid theory , BCS ,  CDW  and
SDW instabilities, screening, and  the Kohn-Luttinger phenomena all under
the auspices  of the RG, we have a better chance of solving extensions and
generalizations of these problems.

 \section*{ACKNOWLEDGEMENTS}
 During the past two years I have had  many occasions to discuss this material
with colleagues everywhere. I have profited immensely from their generosity,
interest and criticism. It is however practically impossible to list all of
them.
As they read the paper they will no doubt find parts which they helped
clarify. I must however acknowledge a few whose ears I bent beyond the
linear regime: I. Affleck, S.Chakravarty, D.H.Lee, A.Ludwig, G.Murthy,
J.Polchinski, A.Ruckenstein, H.Schulz,  and D.Vollhardt.  I thank
E.Trubowitz for explaining some of the rigorous works to me.
 My  thanks to my colleagues at Yale - A.Chubukov, G.Moore, J.Ye and
especially  N.Read and S.Sachdev with whom I have discussed virtually
every topic on countless occasions.

 Last but not the least, I thank John Wilkins for coaxing me into   improving
this paper in many ways and making it more accessible to a wider audience
through his very detailed comments and suggestions. I welcome the
opportunity to acknowledge his help.

 This research was supported  in part by  a grant from the  Donors of the
Petroleum Research Fund    and  a  Grant  DMR 9120525    from the National
Science Foundation. This support is gratefully acknowledged. I finally thank
for the Aspen Institute for Physics for a very stimulating summer of  1992
when many of the above mentioned discussions took place.

 \newpage

 \appendix
 \section{APPENDIX}
 Here we will use the  field theory approach to study two phenomena:
screening of the Coulomb potential and the Kohn-Luttinger effect. A smooth
cut-off will be used on loop momenta.

 {\bf \large Coulomb Screening}\\

   We know that the instantaneous Coulomb interaction, before any mode
elimination, is given by
   $\frac{4\pi e^2 K^{2}_{F}}{Q^2}$. The extra $K_{F}^{2}$ comes from
the way our fields are normalized. The corresponding bare vertex is
   \beq
   V(4321) = 4\pi e^2 K^{2}_{F} \left[ \frac{1}{(K_3 -K_1)^2} -
\frac{1}{(K_4 - K_1)^2} \right] .\label{barecoul}
   \eeq
    Since we will be focusing on $K_3 \simeq K_1$, we will drop the second
term.

   It is generally agreed that the Coulomb potential gets screened.  Should we
be using a screened version here? No! Screening comes from organizing
diagrammatic perturbation theory in a certain way by first summing a class of
(RPA) diagrams. In RG the organization is different. Since the Coulomb
potential or bare coupling in the action  is unambiguously known before any
mode elimination, we must see what it evolves into as we carry out the RGT.
In the  field theory approach, using a smooth cut-off $e^{-\alpha |k|}$ where
$\alpha = 1/\Lambda$,  we have to second order :
   \beq
   \Gamma (Q) = V(Q\alpha ) + V^2(Q\alpha )\int \frac{dk d\omega
d\Omega}{(2\pi )^4}
   \frac{e^{-\alpha |k|}e^{-\alpha |k'|}}{(i\omega - E(K))(i\omega - E(K'))}
   \label{ field theory approach}
   \eeq
   where $K' = K + Q$, and where k and $k' = |K'| - K_F$ run from $-\infty$
to $\infty$. We have not shown the ZS' and BCS diagrams since they do not
dominate as the ZS diagram does at small Q. Our plan is to find the $\beta$-
function by setting the $\alpha$-derivative of $\Gamma$ to zero. But first let
us evaluate the former more explicitly. Doing the $\omega$ integral we obtain
two equal contributions from processes where  hole gets promoted to a
particle and vice versa and end up with
   \beq
   \Gamma (Q) = V(Q\alpha ) - V^2(Q\alpha )\int_{-\infty}^{0} \int_{-
1}^{1}\frac{dk dz}{2\pi^2} \frac{2m
   e^{-\alpha |k|}e^{-\alpha |k'|}\theta (k')}{(Q^2 + 2KQz)}
   \eeq
    We can now do the k and z integrals. In doing so we  replace K by $K_F$
wherever appropriate and use the fact that k, k' and Q are all much smaller
than $K_F$.
    For example we set
    \beqr
    k' &= &|K + Q | - K_F\simeq k + \frac{Q^2 + 2K_FQz}{2K_F}
    \eeqr
   and so on.
    The result is
    \beqr
    \Gamma (Q) &= &V - \frac{V^2 m}{K_F \pi^2} \frac{1 - e^{-\alpha
Q}}{\alpha Q}\\
    &\simeq & V - \frac{V^2 m}{K_F \pi^2} \frac{1}{1 + \alpha Q}
    \eeqr
    where in the last equation we have used a simpler function with the same
limits at small and large values to facilitate the analysis.
    The $\beta$-function is now computed to be
    \beq
    \alpha \frac{ dV}{d\alpha} = - \frac{V^2 m}{K_F \pi^2} \frac{\alpha
Q}{(1 + \alpha Q)^2}
    \label{coulbeta}
    \eeq
    Notice the flow is strongest at $\alpha Q \simeq 1$ i.e., $Q \simeq
\Lambda$, which makes sense in the sharp cut-off: we need a minimum
momentum $Q \simeq 2\Lambda$ to knock a particle from the shell at $-
\Lambda$ to the shell at $\Lambda$. When $\vec{Q}$ is too different from
this range, the flow is essentially nil. (Had we used sharp cut-off the $\beta$
function would have had a string of $\theta$ functions, which is why we do
not.)  Integrating the flow from $\alpha = 0$ to $\alpha = \alpha $
    \beq
    V(Q\alpha ) = \frac{1}{\frac{1}{V(Q0)} +  \frac{ m}{K_F \pi^2}
\frac{\alpha Q}{1 + \alpha Q}}
    \label{coulbeta2}
    \eeq
    where $V(Q0)$ is the input potential before mode elimination. The final
answer, in terms of $\Lambda$ is that
    \beqr
    V(Q\alpha ) &=& \frac{4\pi e^2 K^{2}_{F}}{Q^2 +
\frac{Q}{Q+\Lambda}\frac{4e^2mK_F}{\pi}}\\
			 &\equiv& \frac{4\pi e^2 K^{2}_{F}}{Q^2 +
\frac{Q}{Q+\Lambda}\Theta^2}
    \eeqr
    where $\Theta$ is the inverse of the Thomas-Fermi screening length.

   This  is the bare potential that goes into the action when the cut-off is
$\Lambda$. To regain the potential before mode elimination, we must set
$\Lambda = \infty$. You may ask how we can have a $\Lambda > K_F$. The
point is
that $\Lambda$ is defined as the inverse of the $\alpha$ in Eqn.(\ref{
field theory approach}) and not as the real cut-off for the integral. Indeed
with
$\alpha = 0$,
the integral is still finite because it is limited by $Q$: for very
small $Q$, we cannot knock states far below $K_F$ to states above it and
only such processes contribute to the $\omega$-integral.

    Now screening refers not to the bare coupling in the action but to the full
physical four point function. It is however clear that at fixed Q if we send
$\alpha $ to $\infty$ we kill all loops and the bare coupling itself gives the
full
answer, which is
    \beqr
    \Gamma (Q) &=& V(Q,\infty)\nonumber \\
    &= &\frac{4\pi e^2 K^{2}_{F}}{Q^2 + \frac{4e^2mK_F}{\pi}}.\\
	      &=& \frac{4\pi e^2 K^{2}_{F}}{Q^2 + \Theta^2}.\label{TF}
    \eeqr

    Let us examine what we have above. If we fix $Q$ at some value and
lower $\Lambda$, this is what happens to the bare charge in the action. First,
at $\Lambda = \infty$, we have the unscreened potential $4\pi e^2
K^{2}_{F} /Q^2$.
   It has this form as long as $\Lambda >> \Theta^2 / Q$. When $Q <<
\Theta^2 / Q$, it goes as
   $4 \pi e^2 K^{2}_{F} \Lambda /(\Theta^2 Q)$. Finally when $\Lambda <<
Q$, we get the screened  form
   in Eqn.(\ref{TF}).

  We must be clear about what is done here: we tried to understand how
screening takes place in the RG scheme as we eliminate modes.  Since we
assumed the fermion propagator had the  Fermi liquid theory form throughout,
we haven't really verified that non- Fermi liquid theory is ruled out. In other
words, we must study the evolution of the fermion propagator as modified by
the Coulomb potential as we go along. This has not been done; we have
assumed Fermi liquid behavior.

   {\bf \large The Kohn-Luttinger Effect}\\

   Years ago Kohn and Luttinger pointed out that in principle any system will
face the BCS instability at low temperature, even if the initial coupling is
repulsive. Let us recall their argument with no reference to the RG. Consider
the BCS amplitude to one loop as shown in Figure 18 :
   \beqr
   \Gamma (-K_3\ K_3 \ -K_1\ K_1 ) \equiv \Gamma (Q = K_1 -K_3 ) &=&
   V (Q) + BCS + ZS + ZS'
   \eeqr
   Let us compute the coefficients of the Legendre expansion
   \beq
   \Gamma_{l} = \int_{-1}^{1} P_{l} (z)\Gamma (z) dz
   \eeq
   where $z = \val{\Omega}{1} \cdot \val{\Omega}{3}$ and $l$ will be odd
due to the Pauli principle.

   The bare potential will make a contribution $V_l$,  assumed to be positive.
Since $V(z)$
is assumed to be analytic in the interval $-1 \le z \le 1$, we must
have $V_{l} \simeq e^{-l} $ in order that the infinite sum over polynomials
and all derivatives converge. Now look at the one loop corrections. They are
nominally smaller, being of second order. However their dependence on $l$
is very interesting.
This is because  the  ZS  and ZS' graphs  have singularities
when $Q = K_1 - K_3 = 2K_F$ or $Q' = K_1 - K_4 = 2K_F$ respectively,
which correspond to $z = \mp 1$. Let us focus on just ZS, since the Pauli
principle will determine ZS' for us later on. Due to the singularity, the
Legendre expansion coefficients fall as
   \beq
   \delta \Gamma_{l} \simeq \frac{V_{\pi}^2}{l^4}
   \eeq
   where $V(\pi )$, which  enters both vertices is the backward scattering
amplitude.
 (Let us see why. At the left vertex 1 gets knocked into $3 = -1$, so that we
have  a momentum transfer $2K_F$ in the direction of 1.  The loop
momentum $K$ must be nearly $-1$ getting knocked into $+1$ if it is to lie
on or near the Fermi surface and obey momentum conservation.   A similar
argument applies at the other vertex.) {\em It follows that if we hold $V$
fixed and look at large $l$, the second term, which is attractive (for odd $l$,
which is all we have) will dominate over the first, which is falling
exponentially. } There is no question of hoping that $V(\pi ) = 0$, since it is
(up to an overall minus  sign) the sum of all the $V_l$'s,  all assumed non-
negative.

 We are trying to reproduce this in the RG language. The procedure will be
just as in the screening calculation, except now we do not assume $Q$ is
much smaller than $K_F$, indeed it is nearly $2K_F$.  Now we get, upon
doing the $\omega$ integral and setting all unimportant  factors to unity but
paying attention to the sign,
 \beq
 \Gamma (Q)  = V(Q)  - V^2 (\pi )\int_{0}^{\infty} dk\int_{-1}^{1}dz
\frac{\theta (k') e^{-\alpha k}e^{-\alpha k'}}{Q^2 + 2KQz}
 \label{kl1}
 \eeq
 where, once again $\alpha = \frac{1}{\Lambda}$, and  the primed quantities
refer to $K' = K + Q$.
 Let us define
 \beq
 x = 2K_F - Q.
 \eeq
  By drawing a sketch of the Fermi surface  you may verify that since all the
action is near $Q = 2K_F$,
  \beq
  e^{-\alpha k'}\simeq e^{-\alpha (k - x)}
  \eeq
  and that $-1 \le z \le 1$ for
  $k > x$ while $z_m \le z \le 1$ for $k < x$ with
$z_{m}$ being the point where $k' =0$. Putting all this in

  \beqr
  \Gamma &=& V - V^2 (\pi ) \left[  e^{\alpha x} \int_{x}^{\infty}
\frac{dk}{2KQ} \ln (\frac{Q + 2K}{Q - 2K} ) e^{-2\alpha k}  \right.
\nonumber \\
  & & \left. + e^{\alpha x} \int_{0}^{x} \frac{dk}{2KQ} \ln (\frac{Q^2 +
2KQ}{2K_F k}) e^{-2\alpha k} \right].
  \eeqr

 By subtracting the second integral from $0$ to $\infty$, which does not
affect singularity in question,
and shifting the origin  and rescaling k, we get
 \beq
 \Gamma = V - V^2 (\pi ) e^{-\alpha x}\int_{0}^{\infty} e^{-\alpha y}\ln
(\frac{y + 2x}{y + x} )dy.
 \eeq
 We will drop the $e^{-\alpha x}$ since it does not modify the singularity at
$x=0$. Next we approximate as follows:
 \beqr
 x &=& 2K_F - Q \nonumber \\
  &=& 2K_F (1 - \sin \theta_{13} /2 )\nonumber \\
  &\simeq & K_F (1 - \sin \theta_{13} /2 ) (1 + \sin \theta_{13} /2
)\nonumber \\
  &=& K_F ( 1 + z)
  \eeqr
  Hereafter we will use $K_F = 1$. We now do an integration by  parts, throw
out the surface  term and obtain
  \beq
  \Gamma = V - \frac{V^2(\pi )}{\alpha}\int_{0}^{\infty}e^{-\alpha y}\left[
\frac{1}{y +1+z} - \frac{2}{2y + 1 + z} \right] dy
  \eeq
  Now do the angular momentum transform using

    \beq
  Q_{l} (z_0 ) = \frac{1}{2} \int_{-1}^{1} \frac{P_{l} (z) dz}{z_0 - z}
  \eeq
  to obtain

\beq
   \Gamma = V + \frac{V^2(\pi )}{\alpha}     \int_{1}^{\infty} Q_{l} (y)
\left[ e^{\alpha /2}e^{-\alpha y /2} - e^{\alpha} e^{-\alpha y} \right] dy.
   \eeq

    Next we use the result that as  $ l \rightarrow \infty $,
    \beqr
    Q_{l} (y)  \rightarrow    \sqrt{2/ \pi X} e^{-X}\\
    X &=& l \sqrt{y^2 -1}
    \eeqr
    to obtain for large $l$ and $\alpha$
    \beqr
    \Gamma &= &V + \frac{V^2( \pi )}{\alpha l^2} \int_{0}^{\infty} e^{-X}
\sqrt{X} \left[ e^{-\alpha X^2 /2l^2} - e^{-\alpha X^2 /4l^2} \right] dX \\
    &\equiv & V + L.
    \eeqr
   Note that
   \beqr
   L ( \alpha /l \rightarrow
   \infty ) &\simeq & -\frac{c}{l^{1/2}\alpha^{7/4}}\\
   L ( l/ \alpha \rightarrow  \infty )&\simeq& \frac{-c'}{l^4}.
   \eeqr
   We fit $L$ with a simpler function with the same limits:
   \beqr
   \Gamma &=& V -  \frac{V^2( \pi )}{\alpha l^2} R(\alpha /l^2 )\\
   R(x) &=& \frac{x}{1 + x^{7/4}}.
    \eeqr
    If we calculate the $\beta$-function
    (including a factor of 2 due to the ZS'
diagram) and also include the usual contribution from the BCS diagram we
get the result quoted in the text:
     \beq
  \frac{dV_l}{dt} = - \frac{V_l^2}{4\pi}  - \frac{V^2 (\pi )
\lambda^{7/4}}{l^{15/2} \left[ \lambda^{7/4} + l^{-7/2} \right]^2}.
  \eeq
  \newpage
  \section{REFERENCES}
  Abrikosov, A. A. ,  L. P. Gorkov,  and I. E. Dzyaloshinski,  1963,  {\em
Methods of Quantum Field Theory in Statistical Mechanics},  Dover,  New
York.  \\
Anderson,  P. W. ,  and G. Yuval, 1970,  Phys.  Rev.  Lett.  {\bf23},  89. \\
Anderson,  P. W. ,  1984,  {\em Basic Notions of Condensed Matter
Physics},  Benjamin-Cummings,  Menlo Park. \\
Anderson, P. W. ,  1990,  Phys. Rev. Lett.  {\bf 64},  1839.      \\
Baranov, M. A. ,  A. V. Chubukov and M. Kagan,  1992,  Int.  J.  Mod.
Phys,   {\bf 6},   2471.   \\
Baym, G. ,   and C. Pethick,  1991,  {\em Landau Fermi Liquid Theory},
Wiley,  New York. \\
Benfatto,  G. ,   and G. Gallavotti,  1990,  Phys. Rev. {\bf B42},  9967. \\
Berezin, F. A. ,  1966 {\em The Method of Second Quantization},  Academic
Press. New York. \\
Boninsegni, M. ,  and E. Manousakis,  1990,  Phys. Rev.  1991,  {\bf B43},
10353. \\
Bourbonnais, C. ,  and L. G. Caron, 1991,  Int. J. Mod. Phys,  {\bf B5},
1033. \\
Brazovskii, S.A., 1975 Zh. Eksp.Teor. Fiz {\bf 68}, 175. (1975 ,
Sov.Phys.JETP {\bf 41} 85.\\
Callan, C.G., 1970, Phys. Rev. {\bf D2}, 1541.\\
Castro Neto, A.H., and E. Fradkin, 1993, {\em Bosonization of the Low
Energy excitations of Fermi Liquids}, University of Illinois preprint, March
1993.\\
De Castro, C. ,  and W. Metzner,  1991,  Phys. Rev. Lett.  {\bf 67},  3852. \\
Elser, V. ,  D. Huse,  B. Shraiman and E. D. Siggia,  1990,  Phys. Rev.  {\bf
B41},  6715. \\
Engelbrecht, E. ,  and M. Randeria,  1991,  Phys. Rev.  Lett.  {\bf   67}, \\
Feldman, J. ,  and E. Trubowitz,  1990,  Helv. Phys. Acta,  {\bf 63},  157. \\
Feldman, J. ,  and E. Trubowitz,  1991,  Helv. Phys. Acta,  {\bf 64}, 213. \\
Feldman, J, .  J. Magnan, V. Rivasseau,  and E. Trubowitz,  1992,  {\bf 65},
679. \\
Fisher. M.E., 1974,  Rev. Mod. Phys. {\bf 4}, 597.\\
Fisher, M. E. ,  1983,  {\em Critical Phenomena}, F. W. J. Hahne,  Editor,
Lecture Notes Number 186, Springer-Verlag,  Berlin, (1983).  \\
Gell-Mann, M. ,  and F. E. Low,  1954,  Phys. Rev. {\bf 95},  1300. \\
Goldenfeld, N. ,  1992 {\em Lectures on Phase Transitions  and the
Renormalization Group. }, Addison Wesley. \\
Haldane, F. D. M. ,  1981,  J. Phys. {\bf C14},  2585. \\
Hertz, 1976,  Phys. Rev.  {\bf B14},  1165. \\
Houghton, A. ,  and J. B. Marston,  1992,  {\em Bosonization and Fermion
Liquids in Dimensions Greater Than One},  Brown University preprint. \\
Ioffe, L. B. ,  and G. Kotliar,  Phys. Rev.  1990,  {\bf B42},  10348. \\
Itzykson, C. ,  and J. B. Zuber,  1980,  {\em Quantum Field Theory},  Mc
Graw Hill, New York. \\
Itzykson, C. ,  and J. M. Drouffe,  1989,  {\em Statistical Field Theory Vol.
I},  Cambridge University Press. \\
Kadanoff, L. P. ,  1965,  Physica {\bf 2} 263. \\
Kadanoff, L.P., 1977, Rev.Mod. Phys., {\bf 49}, 267.\\
Kogut, J.B., 1979,  Rev. Mod. Phys., {\bf 51}, 659.\\
Kohn, W. ,   and J. Luttinger,  1965,   Phys. Rev. Lett. {\bf 15},  524. \\
Krishna-murthy, H.R., J.W.Wilkins and K.G.Wilson, 1980, Phys. Rev. {\bf
B21}, 1008, 1024.\\
Landau L. D. ,  1956,  Sov.  Phys.  JETP {\bf 3},  920. \\
Landau L. D. ,  1957,  Sov.  Phys.  JETP {\bf 8},  70. \\
Landau L. D. ,  1959,  Sov.  Phys.  JETP {\bf 5},  101. \\
Le Bellac, M. ,  1991,  {\em Quantum and Statistical Field Theory},  Oxford
University Press,  New York. \\
Lee, P. A. ,  and T. V. Ramakrishnan,  1985,   Rev. Mod. Phys. {\bf 57},
287.  \\
Leggett, A. J. ,  1975,   Rev.  Mod. Phys,  {\bf 47},  331.  \\
Lifshitz, E.M,  and L.P.  Pitayevskii,  1980,  {\em Statistical Physics II},
Pergamon Press,  Oxford. \\
Ludwig, A. W. W. ,  and I. Affleck (1991a),  Phys. Rev. Lett.  {\bf 67},
3160. \\
Ludwig, A. W. W. ,  and I. Affleck (1991b),  Nucl. Phys. {\bf B360 FS},
641. \\
Ludwig, A. W. W. ,  and I. Affleck (1992),  Phys. Rev. Lett.  {\bf68},  1046.
\\
Luttinger, J. M. ,  1960,  Phys. Rev. {\bf 119},  1153. \\
Luttinger, J. M. ,  1961,   Phys. Rev.  {\bf 121},  942. \\
Ma, S.K., 1976, {\em Modern Theory of Critical Phenomena},
W.A.Benjamin.\\
Mahan, G. D. ,  1981,  {\em Many-body Physics},  Plenum,  New York. \\
Mermin, D. ,  1967,  Phys. Rev.  {\bf  159},  161. \\
Morel, P. ,  and P. W. Anderson,  1962,  Phys. Rev. {\bf 125},  1263. \\
Nagaosa, N. ,  and P. A. Lee,  Phys. Rev. Lett.  1990,  {\bf 64},  2450. \\
Negle, J. F. ,  and H. Orland,  1988,  {\em Quantum Many-particle Systems},
Addison Wesley,New York. \\
Nozieres, P. , 1974,  J. Low. Temp. Phys, {\bf 17},  31. \\
Perakis, I. ,   C. M. Varma,  and A. E. Ruckenstein,  1993,  Phys. Rev. Lett.
{\bf 70},  3467. \\
Pines and P. Nozieres,  1966,  {\em The Theory of Quantum Liquids},
Addison-Wesley,  Menlo Park. \\
Plischke, M. ,  and B. Bergerson,  1989,  {\em Equilibrium Statistical
Physics},  Prentice Hall,  New Jersey. \\
Polchinski, J. ,   1992,  {\em Effective Field Theory and the Fermi Surface},
preprint NSF-ITP-92-XX UTTG-20-92,  to appear in TASI proceedings. \\
Polchinski, J., 1993, {\em Low Energy Dynamics of the Spinon-Gauge
System}, Preprint  NSF-ITP-93-93, UTTG-09-93. \\
Ruckenstein, A. E. ,  S. Schmidt-Rink and C. M. Varma,  1989,  Phys. Rev.
Lett. {\bf  63},  1996. \
Sachdev, S. ,  1989,  Phys. Rev.  {\bf B39},  12232. \\
Schulz, H. J. ,  1991,  Int.  J. Mod. Phys. {\bf B43},  10353.          \\
Schwinger, J. ,  {\em Quantum Kinematics and Dynamics},  W. A. Benjamin.
\\
Shankar, R. ,  1991,  Physica {\bf A177},  530. \\
Solyom, J. ,  1979,  Advances in Physics, {\bf 28},  201.        \\
Stamp, P. C. E. ,  1992,  Phys. Rev. Lett.  {\bf 68},  2180. \\
Stamp, P, C. E. ,  1993,  J. de. Physique, {\bf 3},  625. \\
Stuckelberg, E. C. G. ,  and A. Peterman,  (1953),  Helv. Phys. Acta,  {\bf
26},  499. \\
Swift,J., and P.C. Hohenberg, 1977, Phys. Rev. {\bf A15}, 319.\\
Symanzik, K., 1970, Comm. Math. Phys. {\bf 18}, 227. \\
Trugman, S. ,  1988,  Phys. Rev. {\bf B 37},  1597. \\
Vollhardt, D. ,   and P. Wolfle,  1990,  {\em Superconducting Phases of
$He^3$},  Taylor - Francis. \\
Weinberg, S., 1986, Prog.Theo.Phys., (Supplement), {\bf 86}, 43.\\
Weinberg, S. ,  1993,  {\em Effective Action and Renormalization Group
Flow of Anisotropic  Superconductors},  University of Texas preprint UTTG-
18-93. \\
Wen, X. G. ,  1990,  Phys. Rev.  {\bf B42},  6623. \\
Wilson. K.G., 1971, Phys. Rev. {\bf B4},3174, 3184.\\
Wilson, K. G. ,  Rev. Mod. Phys. , 1975,  {\bf 47},  773. \\
Wilson, K. G. ,  and J. B. Kogut,  1974,  Physics Reports,  {\bf12},   7. \\
Wilson, K. G. ,  and M. E. Fisher,
1977,   Phys. Rev. Lett.  {\bf 28},  240. \\
Yang, C. N. ,  and C. P. Yang,  1976,  Phys. Rev.  {\bf 150},  321.  \\
Ye, J. ,  and S. Sachdev,  1991,   Phys. Rev.  {\bf 44},  10173. \\
Zimanyi, G. ,  and K. Bedel,  1991,  Phys. Rev. Lett.  {\bf 66},  228. \\
Zinn-Justin,  J. , 1989,  {\em Quantum Field Theory and Critical
Phenomena},  Clarendon Press.

  \newpage
  \section{TABLES}
  \setlength{\baselineskip}{0.375in}
\begin{tabular}{||l|l ||} \hline
\multicolumn{2}{||c||}{Table I Definitions of terms frequently used in
connection with the RG} \\ \hline
Symbol & Meaning  \\ \hline
RG & Renormalization Group \\
Z & Classical partition function or Feynman's path integral for  quantum
problem. \\
$\Lambda$ & The cut-off, the maximum allowed value of momentum $k$.\\
s and t & The parameter in the RG: $\Lambda = \Lambda_0 /s= \Lambda_0
e^{-t}$, where $\Lambda_0$ is fixed.\\
Slow modes $\phi_<$ & Modes to be retained.\\
Fast modes $\phi_>$ & Modes to be integrated out.\\
$\beta$-function & The rate of change of couplings with $t$, the logarithm of
the cut-off. \\
$S$ & The action or hamiltonian.The Boltzmann weight is $e^{S}$.\\
$\cal{S} $ & The space of all hamiltonians. Each axis is used to measure one
parameter. \\
$S_A $ or $S_B$ & Any two actions or hamiltonians describing two different
systems. Points in $\cal{S}$. \\
Critical system & A system tuned to be at a phase transition. Has power law
correlations.\\
Critical exponents & Exponents for the above power laws. Are universal. \\
Critical surface & The locus of all actions or hamiltonians that describe
critical systems. \\
$S^{*}$ & The fixed point of the RG transformation in the space of
hamiltonians. \\
Relevant variable & Any deviation from $S^{*}$ which gets amplified under
the RG action. \\
Irrelevant variable & Any deviation from $S^{*}$ which gets renormalized to
zero.\\
Marginal variable & Any deviation from $S^{*}$ which remains fixed under
the action of the RG.
\\ \hline
\end{tabular}
\newpage
\setlength{\textheight}{9in}
\setlength{\headsep}{-.5in}
\setlength{\baselineskip}{0.375in}
\setlength{\oddsidemargin}{-.75in}
\begin{tabular}{||l|l  ||} \hline
\multicolumn{2}{||c||}{Table II  Fixed point couplings and flows: a summary}
\\ \hline
Item & Comments and relationships\\ \hline
$\phi $ ,$\phi^{*} $& Complex scalar field and its conjugate. \\
$<\phi^{*}(k_2) \phi (k_1 )> \equiv <\overline{2}1>$& Two point function.
\\
$G(k) = 1/k^2$ & Propagator.  $<\phi^{*}(k) \phi (k' )> = (2\pi )^4 \delta^4
(k -k') G(k)$.\\
Wick's Theorem & Gives N- point functions in terms of 2-point function\\
 &  in gaussian model. \\
$<\overline{4} \overline{3}21>$ $ =<\overline{4}2><\overline{3}1> +
<\overline{4}1><\overline{3}2>$& Four point function evaluated using
Wick's theorem. \\
$S_0 = -\int_{0}^{\Lambda}\deek  \phi^{*} (k) k^2 \phi (k)$ & Gaussian
model action with cut-off $\Lambda$.\\
$\delta S = -\int_{0}^{\Lambda}\deek  \phi^{*} (k)r(k) \phi (k)$ & Quadratic
perturbation.\\
$r(k) = r_0 + r_2 k^2 + \ldots $& Quadratic coupling function. Only $r_0$ is
relevant;\\
 & $r_2$ is marginal and the  rest are irrelevant. \\
$\delta S = -\frac{1}{2!2!}\int_{\Lambda}\phi^{*} (4) \phi^{*}(3) \phi^(2)
\phi (1) u(4321)$ & Quartic perturbation in schematic form. \\
$u(4321) = u_0 + O(k)$ & Quartic coupling function. $u(4321) = u(3421) =
u(4312)$. \\
 & Only $u_0$ is marginal, higher Taylor coefficents are irrrelevant. \\
  RG action& Reduce $\Lambda$ by $s$ by integrating out fast modes. \\
 & Rewrite result
 in terms of $\phi' (k') = s^{-3} \phi (k)$, where $k' = sk$. \\
$\zeta $ & Field rescaling factors
defined by $\phi' (k') = \zeta \phi (k)$. \\
 & Chosen to make gaussian action the fixed point. \\
Cumulant expansion & $<e^{\Omega}> = e^{<\Omega> + \frac{1}{2}
(<\Omega^2> - <\Omega>^2) +\ldots }$. \\
Tree graphs & Graphs with no closed loops. \\
Loop graphs & Graphs in which there are closed loops.\\
 &  One works to a given number of loops.\\
Connected graphs& Graphs in which there are no disjoint parts. \\
Tree level RG & Calculation with zero loops in  Feynman diagrams.
Reduced   \\
  & to  ignoring  fast modes
       and reexpressing  the perturbation \\  & in terms  of new momenta and
fields. \\
       & \\
  RG flow to one loop  & $\frac{dr_0}{dt} = 2r_0 + au_0$ and
$\frac{du_0}{dt} = -bu^{2}_{0}$, $a,b >0$.\\
   & \\
\\ \hline
\end{tabular}
\newpage

\begin{tabular}{||l|l ||} \hline
\multicolumn{2}{||c||}{Table III Summary of the Fermion Oscillators} \\
\hline
Symbol & Comments or definitions  \\ \hline
$\Psi , \Sid $ & Fermion destruction and creation operators.\\
 & $\{ \Sid , \Psi \} =  \Sid \Psi + \Psi \Sid =1
 \ \ \Sid^2 = \Psi^2 =0.$ \\
 $N$ = $\Sid \Psi $ & Number operator. $N=0\ or 1$.\\
 $|0>$ & State with $N=0$. \\
 $|1>$ & State with $N=1.$\\
$ H_0 = \Omega_0 \Sid \Psi$ & Oscillator hamiltonian.\\
$\mu$ & Chemical potential.\\
$\beta = 1/T$ & Inverse temperature.\\
$H = H_0 - \mu N$ & Free energy operator. Also called hamiltonian.\\
$Z= Tr e^{-\beta (H_0 - \mu N)}$ & Grand partition function.\\
$<\Omega > = \frac{Tr\Omega e^{-\beta (H_0 - \mu N)}}{Tr e^{-\beta (H_0
- \mu N)}} $ & Average of operator $\Omega$. \\
$A $& Free energy. Defined by $Z= e^{-\beta A(\mu , \beta )}$.\\
"Hubbard model" & $H_0 = \Omega_0 N + \frac{U}{2}N(N-1)$\\
$<N>$ & Average occupation. At $\beta \rightarrow \infty $ this becomes \\
 & $\theta (\mu - \Omega_0 )$  \ for single oscillator,\\
  & $ \theta (\mu - \Omega_0 ) + \theta (\mu - \Omega_0 -U)$ \ for "Hubbard
model".\\
$T(\Psi (\tau ) \Sid (0) )$  & $\theta (\tau ) \Psi (\tau ) \Sid (0)
 - \theta (-\tau )  \Sid (0) \Psi (\tau ) $.\ \ \\
 $G(\tau )$ & $<T(\Psi (\tau ) \Sid (0) )>$. The Green's function.\\
  $-G(0^{-})$ & $N$ \\
 $G(\omega )$ &$ \int_{-\infty}^{\infty} G(\tau ) e^{i\omega \tau } d\tau  =
\frac{1}{\Omega_0 - \mu - i\omega }$\\
 $G(\tau ) $& $\int_{-\infty}^{\infty} G(\omega ) e^{-i\omega \tau }
\frac{d\omega}{2\pi}$ \\
 &
\\ \hline
\end{tabular}
\newpage

\begin{tabular}{||c||} \hline
\multicolumn{1}{||c||}{Table IV Fermion Path Integrals: Useful Relations} \\
\hline \\
$\psi$ and $\sib$ are Grassmann numbers.  They anticommute with each
other
and with \\ fermion creation and destruction operators. Their differentials are
also Grassmann numbers.\\
$|\psi > = |0> - \psi |1> $ \\
$<\sib | = <0| - <1|\sib $\\
$\Psi |\psi > = \psi |\psi > $\\
$<\sib | \Sid = <\overline{\psi} |\overline{\psi} $\\
 $<\sib | \psi >  = e^{\sib \psi}$\\
 $<\sib | H(\Sid , \Psi )|\psi > = H(\sib , \psi )$\\
 $\int \psi d\psi = 1$\\
$\int 1 d\psi = 0 $ \\
$ \int\overline{\psi} d\sib = 1 $ \\
$\int 1 d\sib = 0 $ \\
$\int\overline{\psi} \psi
d\psi d\sib = 1 = - \int\overline{\psi} \psi  d\sib d\psi
$\\ \\
$\int e^{-\overline{\psi}_iM_{ij}\psi_j}
[\prod_{i} d\overline{\psi}_i d\psi_i
] = det M $\\
\\
$\left< \overline{\psi}_i
\psi_j \right>_0 = \frac{\int\overline{\psi}_i \psi_j
e^{\sum_{k} a_k\overline{\psi}_k \psi_k}\prod_{k} d\overline{\psi}_k
d\psi_k}{
 \int e^{\sum_{k} a_k\overline{\psi}_k \psi_k}\prod_{k} d\sib_k d\psi_k} =
\frac{\delta_{ij}}{a_i} \equiv <\overline{i} j>$\\
 \\
 $\left< \sib_i\overline{\psi}_j  \psi_k\psi_l \right>_0 =
\frac
{\delta_{il}\delta_{jk}}{a_ia_j} - \frac{\delta_{ik}\delta_{jl}}{a_ia_j}
\equiv <\overline{i} l><\overline{j} k> - <\overline{i} k><\overline{j}l>$
(Wick's\ theorem)\\
  \\
 $I = \int |\psi ><\overline{\psi}
 | e^{-\overline{\psi} \psi}d\overline{\psi}
d\psi$\\
  \\
$Tr \Omega = \int d\overline{\psi}
d\psi <-\overline{\psi} | \Omega | \psi >
e^{-\overline{\psi} \psi}$\\
 \\
$Z_{oscillator} = \int [d\overline{\psi} (\tau ) d\psi (\tau)
]e^{\int_{0}^{\beta}\overline{\psi}(\tau )
( - \frac{\partial}{\partial \tau} -
\Omega_0 + \mu )\psi (\tau ) d\tau} $\\
 \\
$Z_{oscillator, \beta = \infty} =
\int [d\overline{\psi} (\omega )d\psi (\omega
) ]e^{ \intom \overline{\psi} (\omega ) ( i\omega - \Omega_0 + \mu ) \psi
(\omega ) } $\\
 \\
$<\overline{\psi} (\omega_1 ) \psi(\omega_2 ) > = \frac{2\pi \delta
(\omega_1 - \omega_2 )}{i\omega_1 - \Omega_0 + \mu }$\\ \\
$<\overline{\psi}
(\omega ) \psi (\omega ) >= \frac{2\pi \delta (0)}{i\omega -
\Omega_0 + \mu } $\\
\\ $2\pi \delta (0) = \beta $\\ \\
$<N>= \frac{1}{\beta Z} \frac{\partial Z}{\partial \mu} =\int_{-
\infty}^{\infty}\frac{d\omega}{2\pi} \frac{e^{i\omega o^{+}}}{i\omega -
\Omega_0 + \mu } $\\

\\ \hline
\end{tabular}
\newpage

\begin{tabular}{||l||} \hline
\multicolumn{1}{||l||}{Table V Spinless Fermions in $d=1$: Summary of
symbols and formulae.} \\ \hline
\normalsize
CDW \ \ Charge density wave.\\
$H
     = -\frac{1}{2} \sum_n \sid
     (n+ 1) \psi (n) + h.c.  + U_0 \sum _n (\sid (n)
\psi (n) -1/2  ) ( \sid (n + 1) \psi (n+1) -1/2 )$ \\
     $n_j = \sid_j \psi_j$ \ \ Particle number at site $j$. \\
     $\mu = U_0$ The
     chemical potential that ensures half-filling for repuslion
$U_0$. \\
     $n_j =  \frac{1}{2} + \frac{1}{2}(-1)^j \Delta $ The mean field ansatz.
$H$ appears as follows in terms of $\Delta$:\\
     $H = -\frac{1}{2} \sum_n \sid (n+ 1) \psi (n) + h.c  +U_0 [
\frac{1}{4}\sum_j \Delta^2 - \Delta \sum_j (-1)^j n_j ] + U_0 \left[ \sum_j
:n_j::n_{j+1}: \right] $\\
     $:n_j:$ = charge density at site $j$ with mean value subtracted.\\
 $U_0 \sum_j :n_j::n_{j+1}:$
 \ \  Part neglected in mean-field calculation.\\
$\Delta $ \ \ CDW order parameter.\\
$E(k) = -\cos K $ \ \ Dispersion relation for free fermions. \\
$K_F = \pi /2$ \ \ The  Fermi momentum at half filling. \\
$K$ \ \  Momentum measured from origin.\\
$k = |K| - K_F$ \ \ Momentum measured relative to the Fermi
surface(points).\\
L and R: Names for left and right Fermi points $K = \mp \pi /2$. \\
$H_0 = - \intK \sid (K) \psi (K) \cos K $ \ \ Free field hamiltonian. \\
 $H_0 = \sum_i \intk \psi^{\dag}_{i} (k) \psi_i (k) k $ \ \ Free  field
hamiltonian linearized near $K_F$. \\
 \\
$ Z_0 =\int \prod_{i=L,R}\prod _{|k|<\Lambda} d\psi_{i} (\omega k)
d\sib_{i} (\omega k)e^{S_0}$ Free fermion partition function.\\ \\
 $ S_0        = \sum_{i = L, R} \intk \intom \sib_{i} (\omega  k) (i\omega -
k)\psi_{i} (\omega k) $ Free fermion action.\\ \\
 $ k'=sk\ \ \omega' = s\omega \ \ \psi_{i} '(k'\omega ') = s^{-3/2} \psi_{i}
(k\omega  ) $ \ \ RG transformation.\\ \\
$ \delta S_2 = \sum_i \intk \intom \mu (k\omega  )\sib_{i} (\omega  k)
\psi_{i} (\omega k)$ \ \ Quadratic perturbation.\\ \\
$\mu (k, \omega) = \mu_{00} + \mu_{10} k + \mu_{01} i\omega + \cdots  $
Taylor expansion of qudratic coupling.\\
 Only $\mu_{00}$ is relevant. The rest are marginal or irrelevant. \\ \\
 $\delta S_4 =
 \frac{1}{2!2!}\int_{K\omega} \sib (4) \sib (3) \psi (2) \psi (1)
u(4, 3, 2 ,1)$ \ \ Quartic  perturbation in schematic form.\\ \\
 $\int_{K \omega} \! \! =\! \!  \left[ \prod_{i=1}^{4}\int_{-
\pi}^{\pi}\frac{dK_i}{2\pi}\int_{-
\infty}^{\infty}\frac{d\omega_i}{2\pi}\right] \left[ 2\pi
\overline{\delta} (K_1 + K_2 - K_3 - K_4) 2\pi \delta (\omega_1 + \omega_2
- \omega_3 - \omega_4 )\right] $ \\ \\
$\overline{\delta}$: \ \  Delta function for momentum conservation modulo
$2\pi$. \\
$Umklapp \ process$ \ \ A process where momentum is changed by a multiple
of $2\pi$.\\ \\
$u(4,3,2,1) = U_0 \sin (\frac{K_1 - K_2}{2} ) \sin (\frac{K_3 -
K_4}{2})\cos (\frac{K_1 +K_2 - K_3 -K_4}{2})$ \ \ Nearest neighbor
model.\\ \\
$u'_{i_4 i_3 i_2 i_1}(k_i', \omega_i'  )
=u_{i_4 i_3 i_2 i_1}(k_i'/s, \omega_i'
/s ) $ \ \ Renormalization
of quartic coupling at tree level. \ \ $i  = L \  , R$.\\ \\
$u_0 = u_{LRLR}         = u_{RLRL}
	  = -u_{RLLR}
	    = -u_{LRRL}$\  Symmetries of marginal coupling constant.\\
	    $ \frac{du_0}{dt} = 0$\ \ Flow at one loop.\\

\\ \hline
\end{tabular}
\newpage

\setlength{\baselineskip}{0.425in}
\begin{tabular}{||l||} \hline
\multicolumn{1}{||l||}{Table VI Phase space for various fixed points:
Summary of symbols and formulae.} \\ \hline
One particle phase space is as follows:\\
$d=1$ :
The Fermi surface is a pair of points, called L(left) and R (right).\\
phase space = $\sum_{i=L}^{R} \intk   \intom$, where $ k = |K| -K_F$.\\ \\
$d=2$:
Circular case. The Fermi surface is parametrized by angle $\theta$.\\
phase space = $  \iT\intk \intom $, where $k = |K| - K_F$. \\ \\
$d=2:$ Noncircular case. The Fermi surface is parametrized by an angle
$\theta$ if connected \\ and by an additional label $\alpha = \pm 1$ if it has
two branches.\\
phase space = $\sum_{\alpha} \iT J(\theta ) \int_{-\Lambda}^{\Lambda}
\frac{d\varepsilon}{2\pi} \intom $,\ where $J$ is the Jacobian on the Fermi
surface \\ and $\varepsilon $  measures the energy relative to the Fermi
surface. \\  \\
$d=3$: Spherical case. The Fermi surface is parametrized by $z = \cos \theta$
and $\phi$. \\
phase space = $\int \frac{d\Omega}{4\pi^2} \intk \intom $ where $d\Omega
= d\phi dz = d\phi d\cos \theta$.\\  \\
\hline
In all cases the RG transformation reduces $\Lambda$ (limit on $k$ or
$\varepsilon$) by a factor $s = e^{-t}$. \\
\hline
The quartic interaction is written in schematic form as \\ \\
 $  \delta S_4 = \int_{\Lambda} \sib (4) \sib (3) \psi (2) \psi (1) u(4321)
  $ \\ \\

where  the labels $1$ to $4$ stand for all the attributes of the fields:\\
momentum, frequency and spin if included.\\
The coupling obeys $u(4321) = - u(3421) = -u(4312) = u(3412)$ due to
Fermi statistics.\\
$\int_{\Lambda}$ stands for the integral over the above phase space  for each
of the four\\
fields,  restricted by  delta functions that impose  $\omega$ and   momentum
\\ conservation (modulo $2\pi$ if appropriate).\\
\hline

\\ \hline
\end{tabular}
\newpage

\begin{tabular}{||l  |  l ||} \hline
\multicolumn{2}{||c||}{Table VII  Fixed point couplings and flows: a
summary} \\ \hline
\multicolumn{2}{|| c ||}{The marginal part of coupling $u$ depends only on
the angles that parametrize the Fermi surface} \\ \hline
\multicolumn{2}{||c||}{Subscript "NN" denotes values in nearest neighbor
model} \\ \hline
Fermi Surface & Couplings marginal at tree Level  and their  flow at one
loop\\ \hline
$A$: d=2:  & $u(\theta_4 =
\theta_2 ,\theta_3 = \theta_1 , \theta_2 , \theta_1 )
=-u(\theta_4 = \theta_1 ,\theta_3 = \theta_2 , \theta_2 , \theta_1 )$
\\
Circular Fermi surface
& = $F(\theta_1 , \theta_2 ) = F(\theta_1 - \theta_2 )
\equiv F(\theta_{12})$  \\
   & $ F_{NN} =  U_0 (1- \cos \theta_{12})$  \\
   & $\frac{dF}{dt} =0.$\\
    & \\
    & $u(\theta_4 = -\theta_3 ,\theta_3 ,
    \theta_2 = - \theta_1 ,  \theta_1 )
=V(\theta _3 - \theta_1) \equiv V(\theta_{13} )$ \\

      & $\frac{dV(\theta_{1} - \theta_3) }{dt} = -\frac{1}{8\pi^2} \iT
V(\theta_{1} - \theta )
 V(\theta - \theta_{3}) $
\\
  & $V_{NN}= U_0 \cos \theta_{13} $ \\
   & $\frac{dV_l}{dt} =
   -\frac{V^{2}_{l}}{4\pi}$ ,\ $V_l = \iT e^{il\theta }
V(\theta )$. \ (BCS instability if $V_l < 0 $)\\
    & \\
 \hline
  & \\
$B$: d=2: Noncircular
& $F$ same as in $A$, but $F(\theta_1 , \theta_2 )\ne
F(\theta_1 - \theta_2 ). $
 \ \    $\frac{dF}{dt} =0.$\\
Time reversal invariant.
& $V$ same as in $A$  but $V(\theta_3 , \theta_1
)\ne V(\theta_3 - \theta_1 ) $ \\
Asumed connected & $ \frac{dV(\theta_{1} ; \theta_3) }{dt} = -
\frac{1}{8\pi^2} \iT  V(\theta_{1} ; \theta )
 V(\theta ; \theta_{3}) J(\theta )   $   \\ \\
 \hline
  & \\
$C$: d=2: Non circular  & $F$ is as in $B$. \ \   $\frac{dF}{dt} =0.$\\
 No time-reversal  & No $V$. No BCS instability \\
 invariance  & \\
 \hline
 & \\

 $D$: d=2: nested
 & $F$ defined as in $B$     and  $\frac{dF}{dt} =0$      \\
 $\alpha$ labels branch & $F_{NN} = U_0 (\sin^2 (\frac{\theta_1 -
\theta_2}{2}) + \frac{1}{2} (1 - \frac{\cos \theta_1 \cos \theta_2}{r^2} -
\frac{\alpha_1
\alpha_2}{r^2} \sqrt{r^2 - \cos^2 \theta_1}\sqrt{r^2 - \cos^2
\theta_2}))$ \\ \\
$r$: hopping anisotropy
&    $V$  defined as in $B$.   Same flow as in $B$
\\
    &    $
    V_{NN} (\theta_3 \alpha_3 \ \theta_1 \alpha_1 ] = U_0 [ \sin
\theta_1
\sin \theta_3 + \frac{\alpha_1 \alpha_3}{r^2}\sqrt{r^2 - \cos^2
\theta_1} \sqrt{r^2 - \cos^2 \theta_3}    ]
    $ \\
     & \\
 & $W[\theta_2 \alpha_2 ;
 \theta_1 \alpha_1 ] = u[ \theta_2 + \pi , -\alpha_2
;\theta_1 + \pi
, -\alpha_1 ; \theta_2 \alpha_2 ; \theta_1 \alpha_1 ]  \equiv
u(\theta^{'}_{2} ,
\alpha^{'}_{2}, \theta_{1}^{'} \alpha_{1}^{'} ; \theta_2
\alpha_2, \theta_1 \alpha_1 ) $ \\
 \\
    & $\frac{dW\left[
    \alpha_2 \theta_2\ \alpha_1 \theta_1 \right]}{dt} = - \int
\sum_{\alpha}
W\left[ \alpha_2 \theta_2 \ \alpha'\ \theta' \right] W\left[ \theta
\alpha \ \theta_1
\alpha_1 \right] \frac{J(\theta )d\theta }{(2\pi )^2}\ \
$
\\
 & \\
 & $W_{NN} =
 -U_0 ( \sin^2
 (\frac{\theta_1 - \theta_2}{2}) + \frac{1}{2} ( 1 - \frac{\cos
\theta_1
 \cos \theta_2}{r^2} - \frac{\alpha_1 \alpha_2 }{r^2} \sqrt{
 r^2 - \cos^2 \theta_1 }\sqrt{r^2 - \cos^2 \theta_2} ) )$ \\
  & \\
\hline
 & \\
$E$:  d=3: Spherical &
$u(\vec{\Omega_4},\vec{\Omega}_3,\vec{\Omega}_2,\vec{\Omega}_1)|
_{\vec{\Omega}_2 \cdot \vec{\Omega}_1 = \vec{\Omega_4} \cdot
\vec{\Omega}_3}
 = F(\vec{\Omega}_1 \cdot \vec{\Omega}_2; \phi_{12;34} )\equiv F(z,\phi ).
$ \ \   $\frac{dF}{dt} =0.$ \\
$\vec{\Omega_i}$ is direction on sphere.  & $F(z,0) $ is Landau's $F$
function. \\
$\phi_{12;34}$ is angle between   & $F_{NN} = U_0 (1-z)\cos \phi$. \\
 $1-2$ and $3-4$ planes   & $u(-\vec{\Omega}_3,\vec{\Omega}_3, -
\vec{\Omega}_1,\vec{\Omega}_1) = V(\vec{\Omega}_3 \cdot
\vec{\Omega}_1 \equiv V(z_{13})$ \\
    & $V_{NN} = U_0  z_{13} = U_0 \vec{\Omega}_1 \cdot
\vec{\Omega}_3.$\\
     & $\frac{dV_l}{dt} = -c V^{2}_{l} $ where $c>0$ and  $V_l = \int
V(z)P_l (z) dz$ \\
       \\
\hline
\end{tabular}
\newpage

  \section{FIGURE CAPTIONS}
  {\bf Figure 1}  Four types of diagrams that appear
in the cumulant expansion to lowest order, corresponding
to the $2^4 = 16$ choices for the legs on the vertex to be
fast (F) or slow (S).
Fig1.(a) corresponds to all slow modes, denoted by $S$. This
a {\em tree graph} with no integration over fast (F) modes. Fig.1b
typifies graphs that vanish since they involve an odd number of fast
lines. Figure 1c is the {\em tadpole} graph. Here two fast lines have been
joined  to form a loop. This term with two external slow lines will
renormalize the quadratic term. Figure 1d is a two loop graph. It
comes from a vertex with all fast lines  upon joining them in
pairs. Its contribution  is a constant as far as the slow modes
are concerned. \\
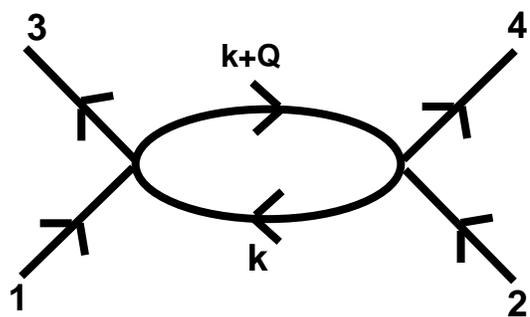
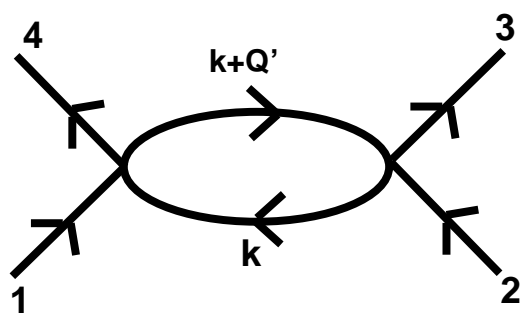
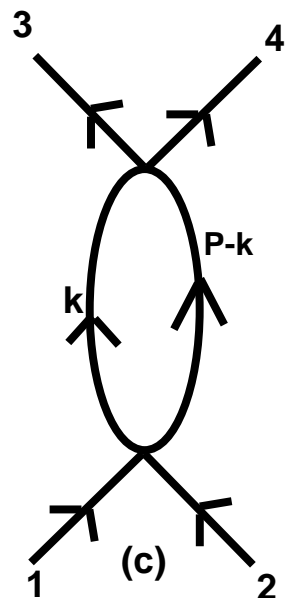
{\bf Figure 2} The three one loop graphs that renormalize
the quartic coupling of the scalar field. The names ZS, ZS'
and BCS are used to label the topology of the graphs and do not
imply the corresponding phenomena (like BCS  superconductivity).
In part (a) lines $1$ and $3$ meet at a vertex, in (b) $1$ and
$4$ meet while in the last lines $1$ and $2$ meet. All loop momenta
lie in the shell being eliminated. The external momenta can be chosen
to vanish if we want the renormalization of the marginal part of the
quartic coupling. When this choice is made, both propagators
have the same momentum. This is true for all three graphs. \\
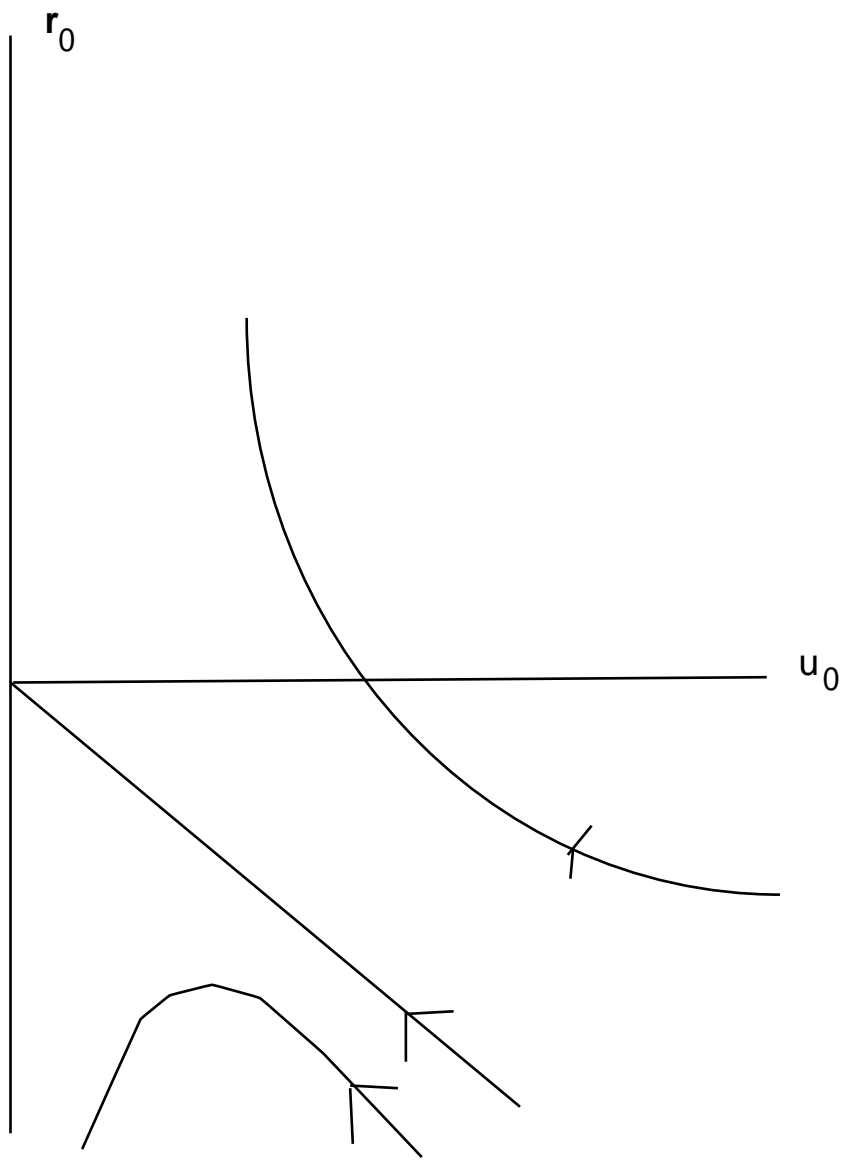
{\bf Figure 3} One loop flow of the quadratic ($r_0$ and quartic $u_0$
couplings in the scalar field in $d=4$. Notice that although $u_0$ is
irrelevant, a point on the $u_0$ axis does not flow to the gaussian
fixed point at the origin. For this this to  happen, we must tune the
parameters to lie on the critical surface, shown by the arrow flowing
into the origin. \\
{\bf Figure 4}  Graphical representation of the perturbative calculation
of the type 1 fermionic propagator in the toy Hubbard model. Figure 4a
shows the two terms we kept in the analysis to one loop. Figure 4b shows
the corrections that are graphically disconnected and ignorable since
they get cancelled by the partition function that comes in the
denominator of all averages. Fig 4c denotes a correction which
embellishes the type $2$ propagator in the loop we did consider.
The effect of such graphs is to turn the free propagator in the loop
to the exact propagator for $2$ which in turn means the integral
over the loop equals the exact density $N_2$. The last part shows a two loop
contribution that vanishes upon to $\omega$ integration.  We have
not shown the iterates of  the connected diagrams which  are part
of the  geometric series which we assumed and summed in the text. \\
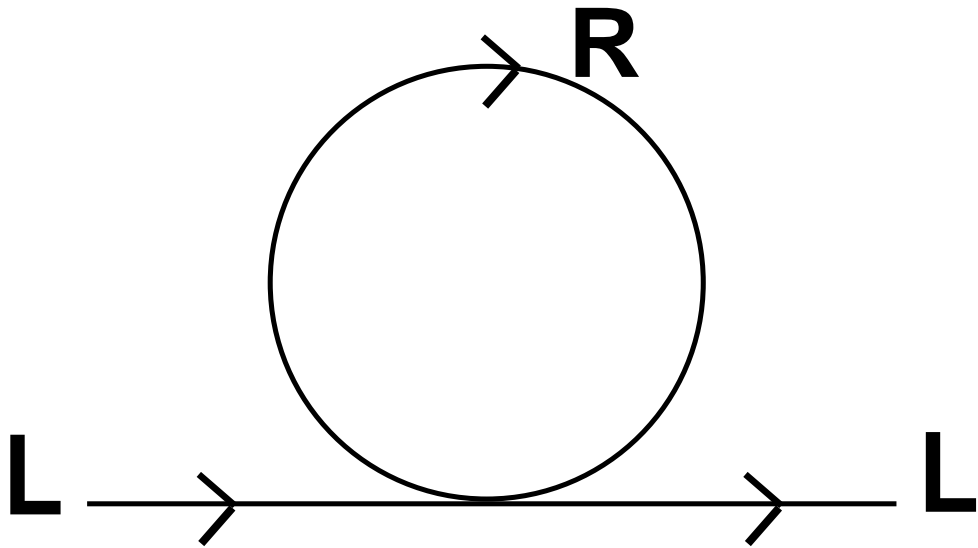
{\bf Figure 5} The tadpole graph which renormalizes the fermion at
one loop. It has no dependence on $k$, the deviation of the external
momentum from $K_F$  or $\omega$ We have used this freedom to set
both these to zero on the external legs. The effect of this graph may be
neutralized
by a counter-term corresponding to a change in chemical potential. One may
do this
if one wants to preserve $K_F$.\\
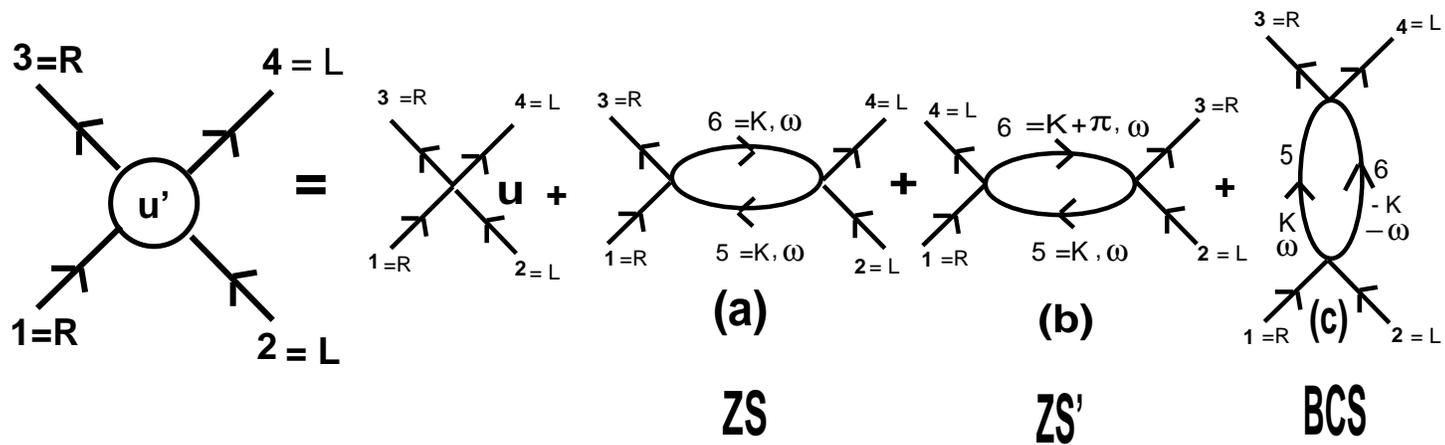
 {\bf Figure 6} The one loop graphs for $\beta (u)$ for the spinless
 fermions in $d=1$.  The loop  momenta  lie in the shell of width
 $d\Lambda$ being eliminated. The external frequencies being all
 zero, the loop frequencies are equal for ZS and ZS' graphs and
 equal and opposite for the BCS graph.  The ZS graph does not
 contribute since both loop momenta are equal (the momentum
 transfer $Q$  at the left vertex is $0$) and  lie a distance
 $\Lambda$ above or below  the Fermi surface and the $\omega$
 integration vanishes  when the poles  lie on the same half-plane.
 The ZS' graph has momentum transfer  $\pi$ at the left and right
 ends. This changes the sign of the  energy of the line entering
 left the vertex. The $\omega$ integral is nonzero, the poles being on
 opposite half-planes. The BCS graph (c) also survives since the
 momenta
 loop momenta  are equal and opposite (since the incoming momentum
 is zero) and this again makes the poles go to the opposite half-planes
 because the lines have {\em opposite }frequencies. The labels $1\ldots 6$
 refer to the master Eqn.(312).\\
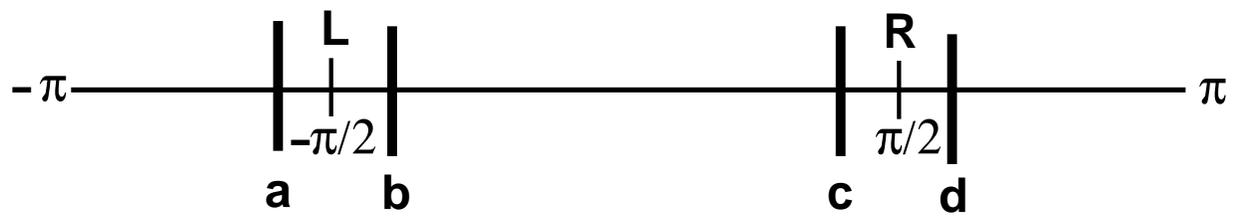
   {\bf Figure 7} The figure shows the regions
   of momentum space being integrated out in the $d=1$ spinless fermion
   problem. The thick lines stand for the slices of width
   $d\Lambda$. They lie a distance $\Lambda$
   from the Fermi points L and R. In the ZS graph which has zero momentum
transfer, both lines lie on the same
   slice  and the $\omega$-integral gives zero. In the ZS' graph,  the
   momentum transfer  $\pi$, connects  $a$ and  $c$ (which have opposite
energies)
   and $b$ and $d$ similarly related. In the BCS diagram the loop momenta
are
   opposite and correspond to $a,d$ or $b,c$.\\
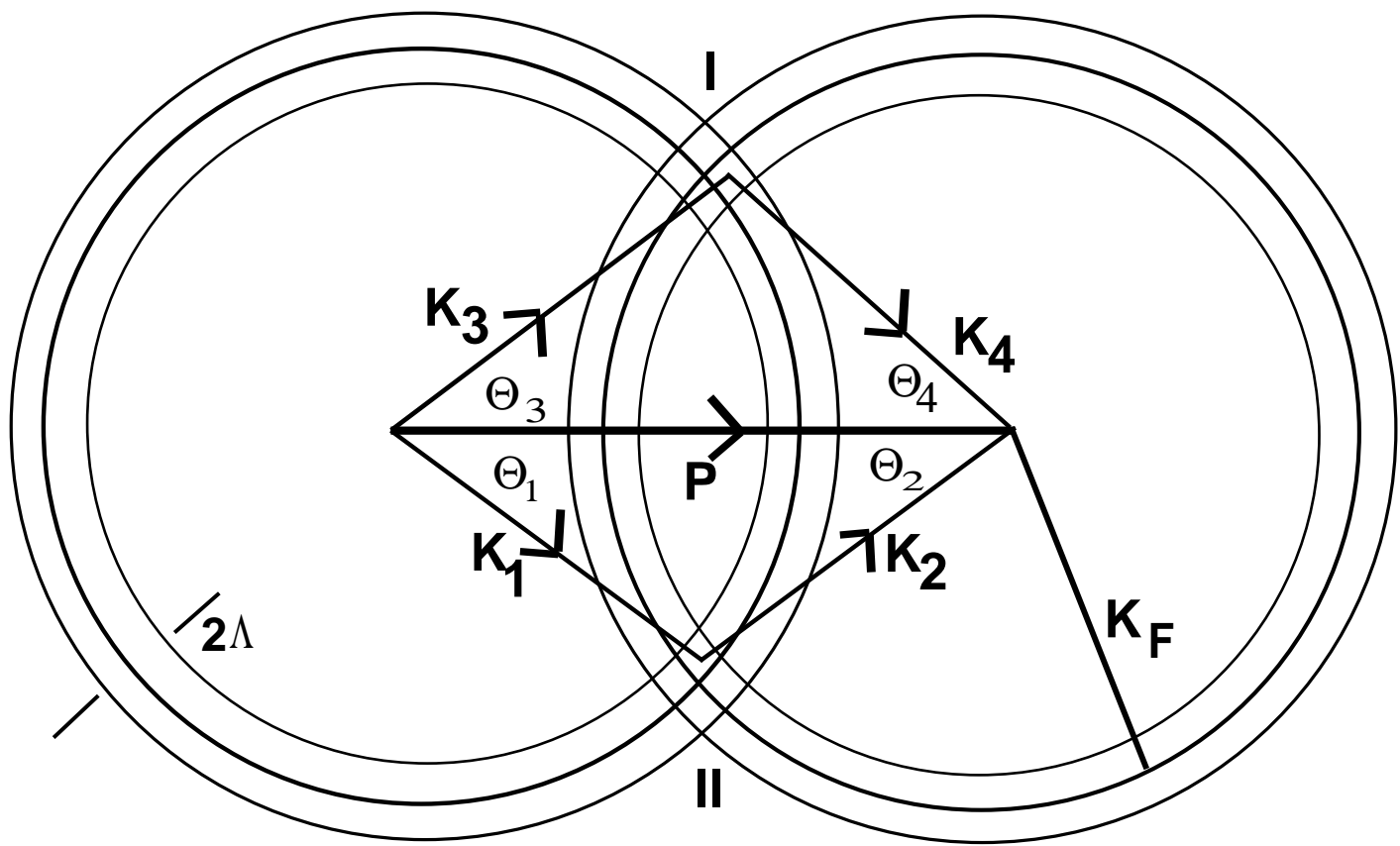
{\bf Figure 8} The geometric construction for determining the allowed
values of momenta. If 1 and 2 add up to P, 3 and 4 are constrained as
shown, if they are to  add up to P and lie within the cut-off. Note
that both  the magnitude of $3$ {\em  and  its direction } $\theta_3$
are constrained to
lie within $\simeq \Lambda /K_F$  of $\theta_1$.  Had we chosen a
$K_3$ that terminated in intersection II, $\theta_3$ would have been
within $\Lambda /K_F$ of $\theta_2$.  At the fixed point, the angels become
equal
pairwise. If the incoming momenta $1$ and $2$ are
equal and opposite,the two shells coalesce and $3$ and $4$ are free to point
in all
directions as long as they are equal and opposite.
\\
{\bf Figure 9} The diagrams that renormalize  the marginal quartic couplings
for a
$d=2$ circular Fermi surface. All external frequencies are chosen to vanish
and
all external momenta are on the Fermi surface at angles
 given by $\vec{\Omega}_i$. The  marginal couplings obey
 $\vec{\Omega}_3 = \vec{\Omega}_1$ and  $
\vec{\Omega}_4 = \vec{\Omega}_2$ ($F$ coupling), or    the above with
$4\leftrightarrow 3$
( $-F$, by Fermi statistics) or $\vec{\Omega}_2 =- \vec{\Omega}_1$
and $\vec{\Omega}_4 =- \vec{\Omega}_3$, (coupling $V$ ). All loop
momenta
must lie within a shell of width $d\Lambda$.  In the first case (F),
there is no momentum transfer at the left vertex ($Q=0$) of the ZS graph.
No matter what direction we choose for $K$, the other loop
momentum $K+Q = K$ will also lie in the shell.  However  the two
propagators
have the same energy and
the
 $\omega$-integral
gives zero.
In the  other two graphs  the loop angle is restricted to
to lie in a region of width $d\Lambda /\Lambda$.
Figure 10 shows this for the ZS' graph and Fig.8 can be used to show this for
the BCS graph if we replace the annuli of thickness $\Lambda$ with  shells of
thickness
$d\Lambda$.
Thus none of the diagrams cause a flow of $F$ at the fixed point. The flow of
$V$, receives no contribution from the ZS and ZS' diagrams because
there is no correlation between incoming and outgoing momenta and both
$Q$ and $Q'$  will be of order $K_F$ and the diagrams will be down by
$d\Lambda /\Lambda$.
The BCS diagram however does cause a flow of $V$ exactly as in $d=1$: the
loop
frequencies are equal and opposite but the loop energies are equal
due to time-reversal invariance.   \\
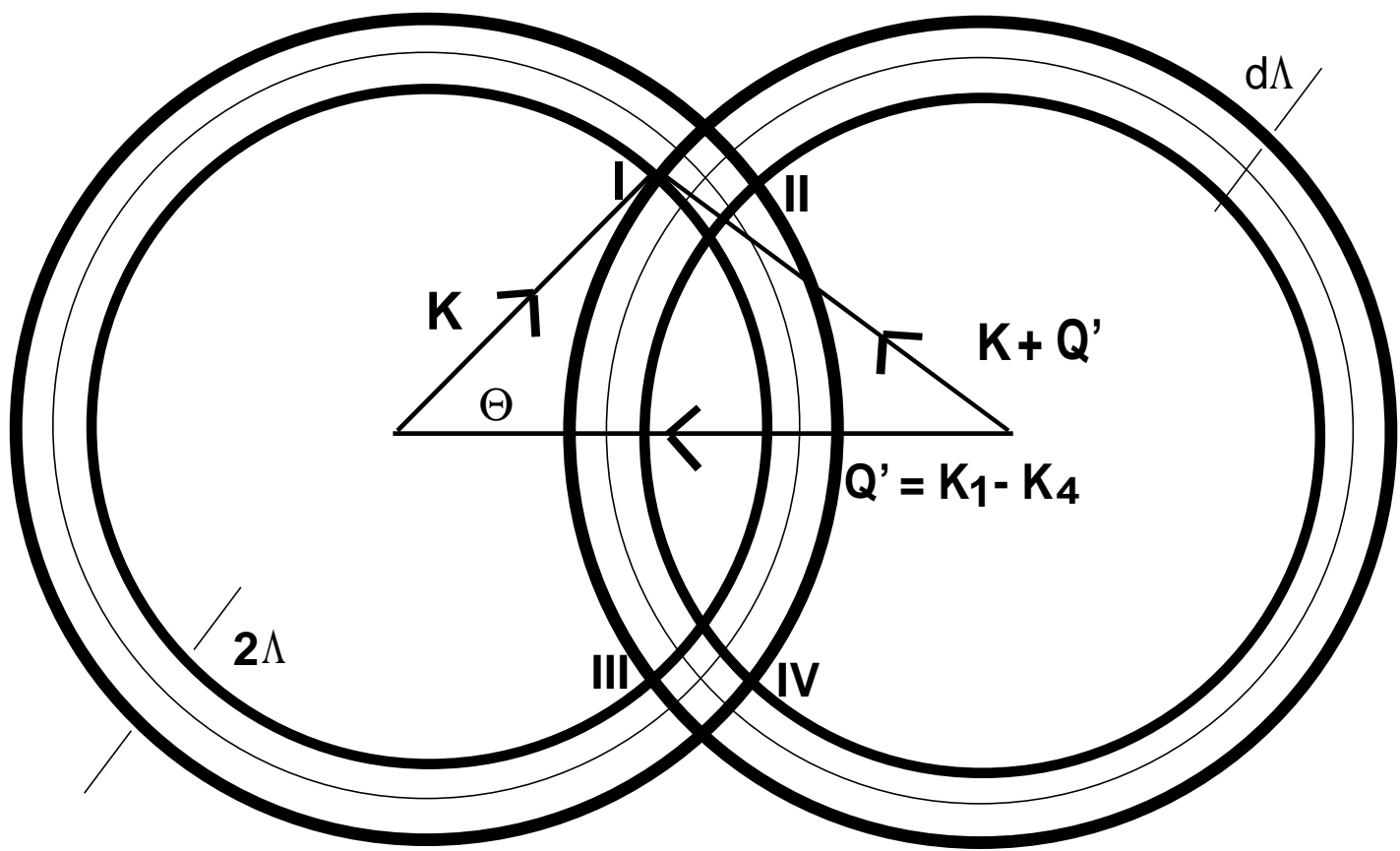
 {\bf Figure 10}  Construction for determining the allowed values of
loop momenta in  ZS'.The requirement that the loop momenta come from the
shell  and
differ by $Q'$ forces them to lie in one of the eight intersection regions of
width
$(d\Lambda )^2$. The diagram does cause a flow of $F$ or $V$ for this
kinematical reason. (In the field theory  approach, the suppression factor will
be $\Lambda /K_F$ instead of
$d\Lambda /K_F$ and the diagram is once again unimportant at the fixed
point.)\\
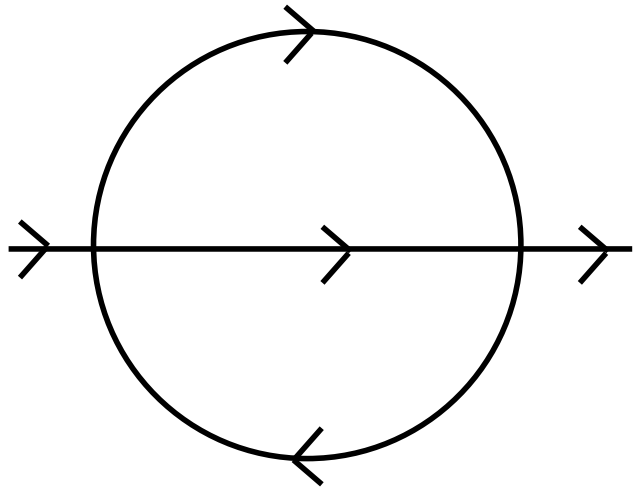
{\bf Figure 11} The sunrise diagram contribution to self-energy. This has two
loops
and is suppressed by a factor $\Lambda /K_F$, or in the
language of the $1/N$ expansion, by a factor  $1/N$. \\
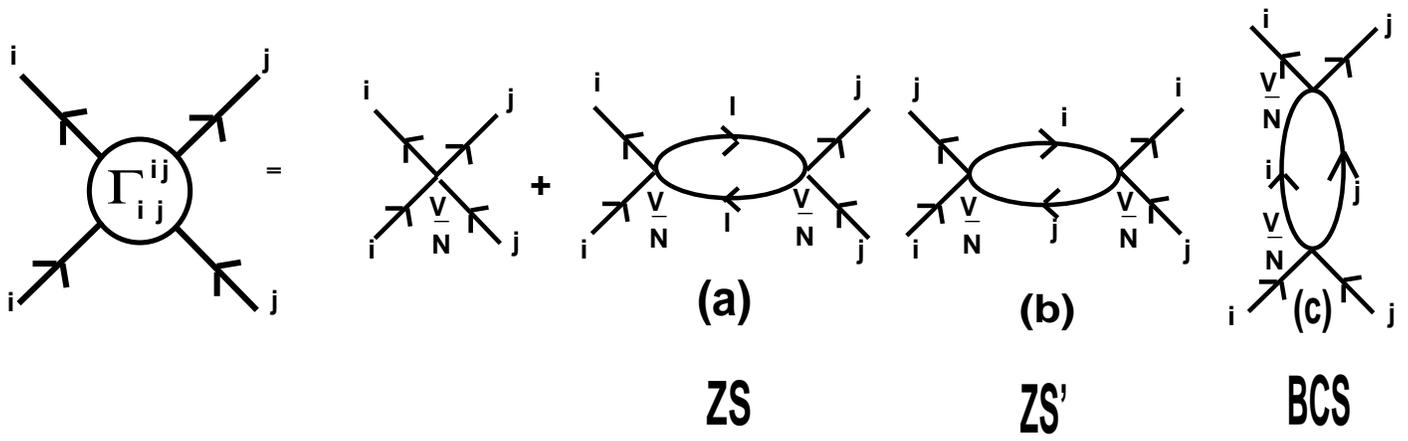
{\bf Figure 12} Flow of indices in the $1/N$ analysis for the problem with
an $F$ type interaction. The tree level
graph is order $1/N$. The ZS graph is of the same order since the sum
over $l$ pays for the extra vertex. The other two graphs are order $1/N^2$.\\
If we consider interactions of the $V$ type, the BCS diagram and its iterates
will be favored.\\
{\bf Figure 13} Role of the counter-term (shown by a cross)
in $G^{-1}$. The tadpole (order $u$) is cancelled by the next diagram with
just a cross.
The value of the cross is found by first evaluating the tadpole and then
arranging the
cancellation. The next one with two loops is cancelled
exactly by the one following it, where the cross still stands for its order $u$
value.
The sunrise diagram (last) will require a change in the counter-term, but this
will
be of relative order $1/N$ and hence ignorable in the large $N$ or small
$\Lambda /K_F$ limit.\\
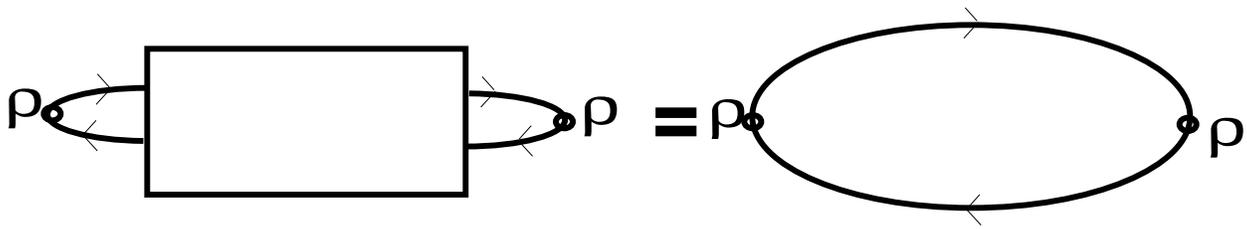
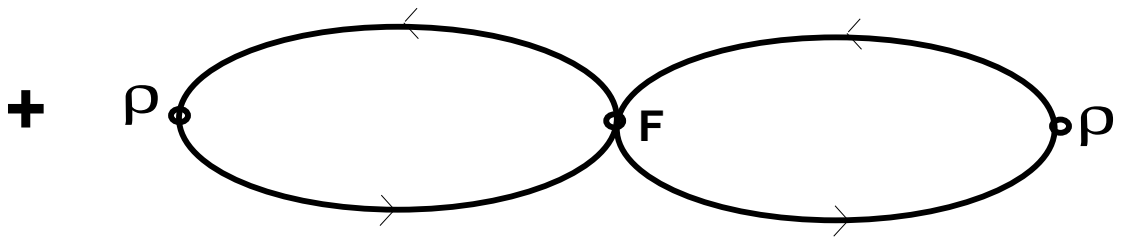
{\bf Figure 14} The compressibility at small
$Q,\omega$, expressed as a  correlation function of two densities. It is
assumed we have
renormalized down to a very small cut-off $\Lambda$ which is still larger
than
probe momentum.
The $\rho$'s stand for
the external
probes or fields  coupling to the particle density. The first graph
on the
right hand
side is what one gets in free-field theory. The rest of the graphs are
corrections due to interactions. The bubble sum, i.e., the iterates of the ZS
loop are all of the same order in $1/N$. If we try to introduce any large
scattering amplitudes
(these are marginal in $d=3$)
into the graph, (say across one of the ZS bubbles),  there will be a
suppression by powers of
$\Lambda /K_F$.
Only the ZS bubbles
survive the
kinematical restrictions that all propagators lie within the narrow
cut-off.\\
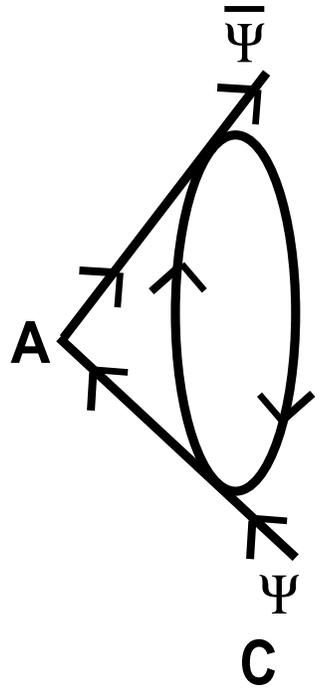
 {\bf Figure 15} Ward identity at work. The idea is show that
 the coupling of the charge density to a field  $A$ is unaffected by
 interactions in the limit of vanishing probe frequency and momenta.  Figure
15a represents the
 renormalization of the quadratic coupling by the addition of the self-energy
 $\Sigma ( k, \omega )$. (The loop momenta are all fast.)If we expand
$\Sigma$ in a power series in $i\omega$, the first derivative term will
  modify the coefficient of the $i\omega$ term in the
 quadratic piece by $\Sigma' = \frac{\partial \Sigma}{\partial i\omega}$.
 The field must then be rescaled to neutralize this.
 The graph in  Figure 15b represents $\Sigma'$, the
 (external) frequency derivative of that in Figure 15a.  The derivative
introduces
 an extra propagator. The corresponding diagram represents  the change in the
field rescaling. However the graph in Fig.15c, which reflects the fact
 that the slow fields can couple indirectly to the external field $A$ via fast
modes
 denoted by $\psi_F$,
   equals the  graph in Figure 15b when external momentum and
  frequency of probe vanish. Consequently the field rescaling and
modification of the coupling to the external field exactly cancel for these two
graphs.  \\
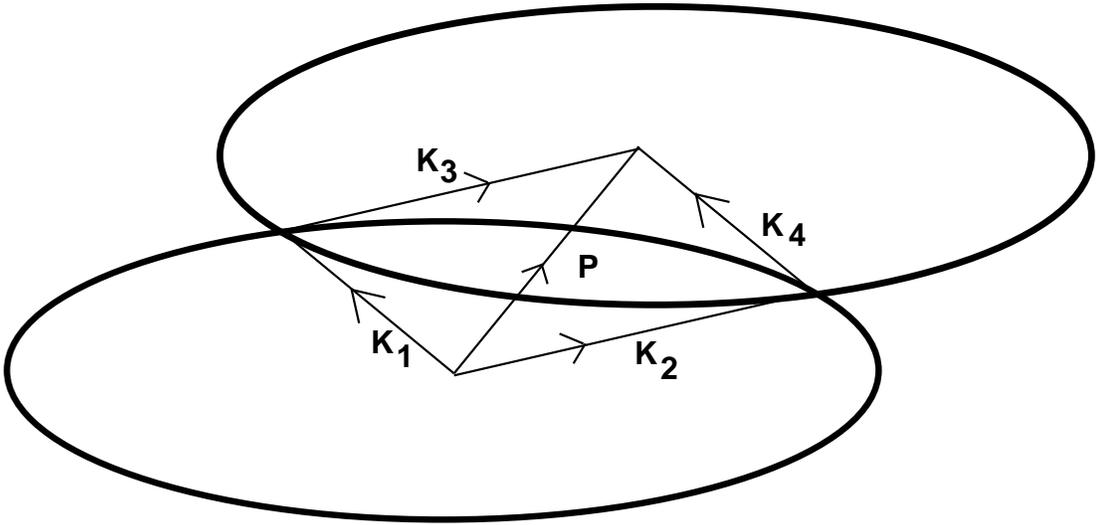
 {\bf Figure 16} A Fermi surface with time-reversal symmetry but no
rotational invariance. The dark line may be taken to represent a very thin
shell
left
after a lot of  renormalization. The figure shows that once again the set of
initial
vectors  $1$ and $2$ coincides with the set of final vectors  as the cut-
off goes to zero. All the results of the circular case hold except for one
thing:
$F(\theta_1 , \theta_2 )$ and $V(\theta_3 , \theta_1)$ are no longer functions
of the differences of their arguments. The only instability is the BCS
instability which only requires time-reversal invariance.  Indeed the
construction in the figure  assumes this symmetry: the solutions it gives for
$\vec{K}_3$ and $\vec{K}_4$ actually point from the surface to the center
of the second ellipse instead of the other way around. However time-reversal
invariance assures us that if we continue these vectors past the origin of the
second ellipse, by an equal amount, we will hit the surface, so that
$\vec{K}_3$ and $\vec{K}_4$ as shown are acceptable solutions.
 {\em In general,  to find the final set using the initial set, we must draw
the
time-reversed version of the second surface and displace its center by
$\vec{P}$
relative to the first.} This did not matter so far due to time-reversal
invariance of the surfaces. If this condition were forgotten, one would
erroneously conclude that when $\vec{P}=0$, the two surfaces will coincide
and the Cooper pairs can roam over all angles,  leading to the BCS instability.
The correct construction would show in this case that the  two surfaces, even
with $\vec{P}=0$ would intersect only in a few places of area $\simeq
\Lambda^2$.\\
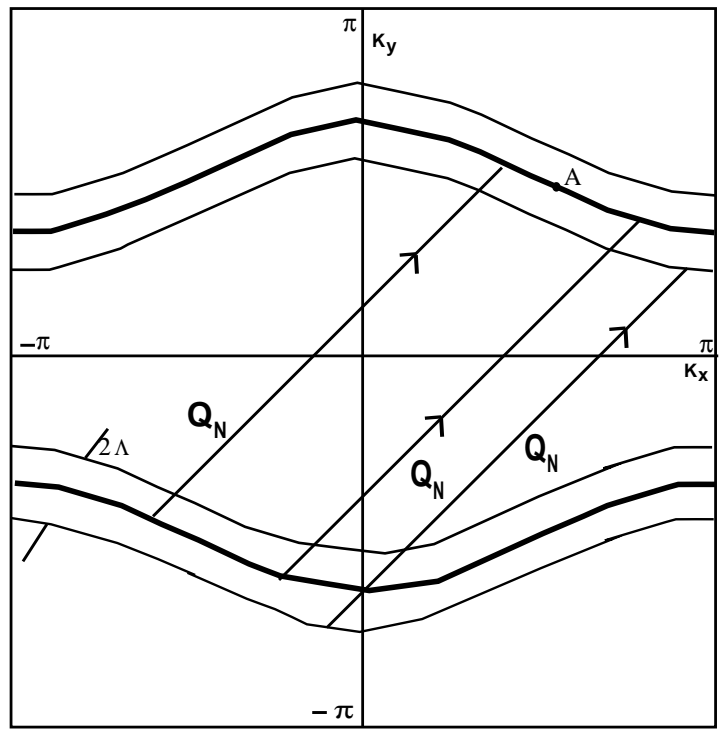
 {\bf Figure 17}  A nested Fermi surface with $\vec{Q}_N = (\pi , \pi )$ and
hopping anisotropy $r>1$. The filled states go from the origin to the dark
lines, the two branched Fermi surface. Any point on Fermi surface (defined
here by zero energy $\varepsilon =0$)   goes to another on the surface
 upon addition of $\vec{Q}$ which reverses the energy. Consequently  any
point just (above) below the filled sea  goes to a point just above (below)
the
sea, leading to the failure of perturbation theory at second order when a
perturbation of momentum $\vec{Q}_N$ is introduced and to the flow of the
coupling $W$ through the graph where the momentum transfer is
$\vec{Q}_N$.    The thin lines are equal-energy contours at $\varepsilon =
\pm \Lambda$ and stand for the infinitesimal shells being integrated out in the
RG program.
 The point A at $(\pi /2 \ \pi /2 ) $ is privileged. It lies on the Fermi
surface for
all $r$, scatters into minus itself under the addition  of $\vec{Q}_N$. The
coupling for $A, -A \rightarrow A, -A$ is an $F$, a $V$ and a $W$ and
flows for more than one reason. It is conjectured that this  is probably why
holes are found at these points (and two more, obtained by reflecting on the
y-axis) upon doping.  \\
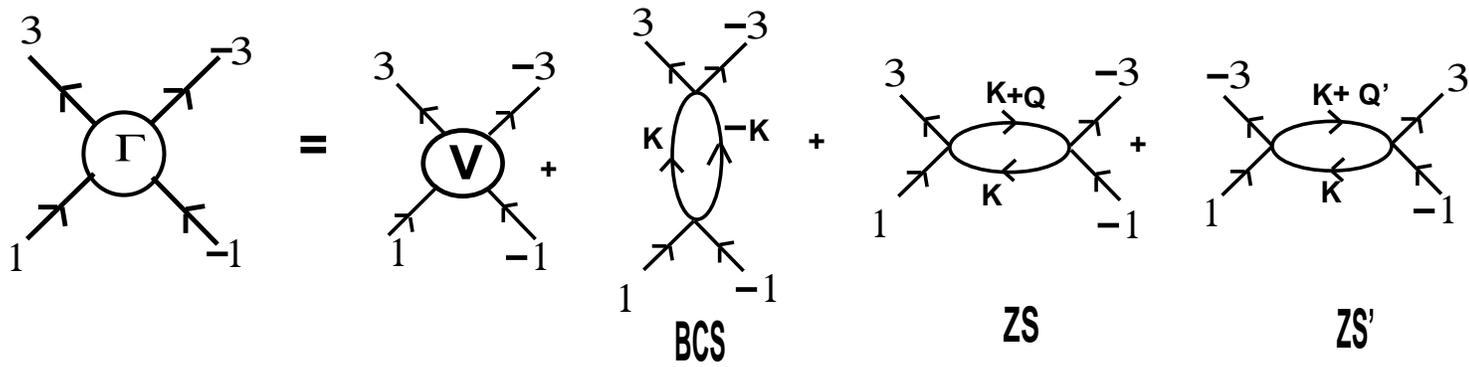
 {\bf Figure 18} The Kohn-Luttinger diagrams. $\Gamma$ is the full
 BCS amplitude and $V$ is the bare vertex. The ZS and ZS' diagrams have
singularities when the momentum transfer equals $2K_F$ whereupon  $K
\simeq -1$ in ZS scatters into $K + Q \simeq 1$. Thus $V = V(\pi )$.
\newpage
\section{FIGURES}

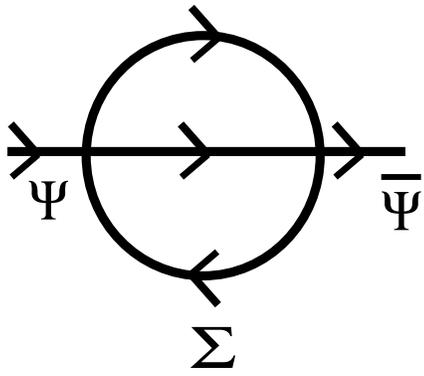
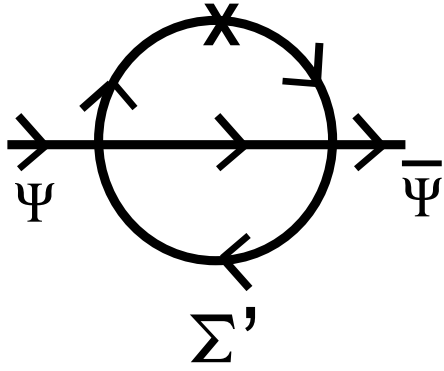
\end{document}